%% file: main.tex
\documentclass[Royal,sagev,doublespace]{sagej_edit}
\usepackage{moreverb,url}
\usepackage[colorlinks,bookmarksopen,bookmarksnumbered,citecolor=red,urlcolor=red]{hyperref}
\usepackage{pdflscape}
\usepackage[linesnumbered,ruled,vlined]{algorithm2e}
\usepackage{multirow}
\usepackage{commath}  
\usepackage{amsfonts}  
\usepackage{hyperref}       
\usepackage{url}   
\usepackage{mathtools}
\usepackage{amsmath}
\usepackage{graphicx}
\usepackage{multirow}
\usepackage{array}
\usepackage{mathrsfs} 
\usepackage{caption}
\usepackage{subcaption}
\usepackage{tikz}
\usepackage{color}
\usepackage{ulem}
\usepackage{xspace} 
\usepackage{diagbox}
\usepackage{nicematrix}
\usetikzlibrary{shapes,decorations,arrows,calc,arrows.meta,fit,positioning}
\tikzset{
    -Latex,auto,node distance =1 cm and 1cm,semithick,
    state/.style ={ellipse, draw, minimum width = 0.7 cm},
    point/.style = {circle, draw, inner sep=0.04cm,fill,node contents={}},
    bidirected/.style={Latex-Latex,dashed},
    el/.style = {inner sep=2pt, align=left, sloped}
}

\newtheorem{decision}{Decision}
\newtheorem{theorem}{Theorem}

\usepackage[colorinlistoftodos]{todonotes}
\newcommand\BibTeX{{\rmfamily B\kern-.05em \textsc{i\kern-.025em b}\kern-.08emT\kern-.1667em\lower.7ex\hbox{E}\kern-.125emX}}
\def\volumeyear{2023}
\begin{document}

\runninghead{}

\title{Estimating dynamic treatment regimes for ordinal outcomes with household interference: Application in household smoking cessation}

\author{Cong Jiang\affilnum{1}, Mary Thompson\affilnum{2}, Michael Wallace\affilnum{2} }
\affiliation{\affilnum{1} Faculty of Pharmacy, Université de Montréal, Canada
\\
\affilnum{2} Department of Statistics and Actuarial Science, University of Waterloo, Canada
}

\corrauth{Mary Thompson, Department of Statistics and Actuarial Science, University of Waterloo, Waterloo, ON N2L 3G1, Canada}

\email{methomps@uwaterloo.ca}

\begin{abstract}
The focus of precision medicine is on decision support, often in the form of dynamic treatment regimes (DTRs), which are sequences of decision rules. At each decision point, the decision rules determine the next treatment according to the patient's baseline characteristics, the information on treatments and responses accrued by that point, and the patient's current health status, including symptom severity and other measures. However, DTR estimation with ordinal outcomes is rarely studied, and rarer still in the context of interference - where one patient's treatment may affect another's outcome. In this paper, we introduce the weighted proportional odds model (WPOM): a regression-based, approximate doubly-robust approach to single-stage DTR estimation for ordinal outcomes. This method also accounts for the possibility of interference between individuals sharing a household through the use of covariate balancing weights derived from joint propensity scores. Examining different types of balancing weights, we verify the approximate double robustness of WPOM with our adjusted weights via simulation studies. We further extend WPOM to multi-stage DTR estimation with household interference, namely dWPOM (dynamic WPOM). Lastly, we demonstrate our proposed methodology in the analysis of longitudinal survey data from the Population Assessment of Tobacco and Health study, which motivates this work. Furthermore, considering interference, we provide optimal treatment strategies for households to achieve smoking cessation of the pair in the household. 
\end{abstract}


\maketitle

\section{Introduction}
\input{Sections/Introduction}
\section{Methodology} \label{Sec2}
\input{Sections/Methodology}

\section{Simulation Studies} \label{Sec3}
\input{Sections/Simulation}

\section{Population Assessment of Tobacco and Health Study} \label{Sec4}
\input{Sections/PATH}

\section{Conclusion and Discussion}
\input{Sections/Conclusion}

\normalem
\bibliography{Reference}%
\section*{Declaration of conflicting interests}
The author(s) declared no potential conflicts of interest with respect to the research, authorship, and/or
publication of this article.

\section*{Supporting information}

Additional supporting information can be found online in the Supporting Information section at the end of this article.
\section*{Data availability statement}
 The data that support the findings of this study come from the Population Assessment of Tobacco and Health (PATH) Study. Restrictions apply to the availability of these data, which were used under license for this study. Data collected in the PATH Study are available from \url{https://doi.org/10.3886/ICPSR36231.v30} with the permission of the Population Assessment of Tobacco and Health Study Restricted-Use Files (ICPSR 36231).

All code for simulated data generation is provided in the online repository \url{https://github.com/CONGJIANG/dWPOM_Rcodes}. 

\newpage
\input{Sections/appendix}
\end{document}


\maketitle

\section{Appendix A: Non-parametric structural equation model}

Suppose we observe $J$ i.i.d. copies of $\boldsymbol{O}=(\boldsymbol{S}, \boldsymbol{D}, \boldsymbol{W}, \boldsymbol{R}, \boldsymbol{A},Y)$ with true distribution $P_0$ which belongs to some model space $\omega$. For a given treatment $k, for~k\in\{1,2,\cdots,14\}$, the hierarchical non-parametric structural equation model (NPSEM) that we assume can be written as:
\begin{align*}
    &\boldsymbol{S}=g_{\boldsymbol{S}}(\varepsilon_{\boldsymbol{S}})\\
    &D^{(k)}=g_D(\boldsymbol{S},\varepsilon_D)\\
    &\{\boldsymbol{W},\boldsymbol{R}\}=g_{\boldsymbol{W},\boldsymbol{R}}(\boldsymbol{S},\varepsilon_{\boldsymbol{W}, \boldsymbol{R}})\\
    &A^{(k)}=
	g_{A}(\boldsymbol{S},\boldsymbol{W},R^{(k)},\varepsilon_{A})\cdot D^{(k)}\\
    &Y=g_Y(A^{(k)},\boldsymbol{S},\boldsymbol{W},\boldsymbol{R},\varepsilon_Y)
\end{align*}
Specifically, we start from the study-level random variables $\boldsymbol{S}$ where the values for observations in the same study are equal. The treatment availability $D$ for each study is generated conditioning on $\boldsymbol{S}$. We assume the individual-level variable $\boldsymbol{W}$ and antimicrobial resistance indicator $\boldsymbol{R}$ are jointly distributed based on $\boldsymbol{S}$. The treatment assignment $A$ of each subject is jointly distributed conditioning on  $\boldsymbol{S},\boldsymbol{W},\boldsymbol{R}$. Note that $fg_{A^{(k)}|A^{(k^*)}}$ is degenerate with all mass at zero when $D^{(k)}=0$. Finally, we can generate the outcome on the basis of all the variables except the treatment availability. We also assume that all the errors $\varepsilon_{\boldsymbol{S}},\varepsilon_{D},\varepsilon_{\boldsymbol{W},\boldsymbol{R}},\varepsilon_{A}$ are independent. In this NPSEM model, we assume there are no unmeasured covariates affecting both the outcome and the treatment assignment and/or availability.\\

We use lower case letters with indices $ij$ or $j$ to represent the observed realizations of individual patient variables and study-level variables. Let $C_j$ be the set of indices of subjects in study $j$. Then, we can write the data probability density function (pdf) of observed data as 
$$\prod_{j=1}^Jf_{\boldsymbol{D}}(\boldsymbol{d}_{j}|\boldsymbol{s}_j)f_{\boldsymbol{S}}(\boldsymbol{s}_{j})\prod_{i\in C_j}\biggl\{f_{\boldsymbol{W},\boldsymbol{R}}(\boldsymbol{w}_{ij},\boldsymbol{r}_{ij}|\boldsymbol{s}_{j})f_{\boldsymbol{A}}(\boldsymbol{a}_{ij}|\boldsymbol{s}_{j},\boldsymbol{w}_{ij},\boldsymbol{r}_{ij},\boldsymbol{d}_j)f_Y(y_{ij}|\boldsymbol{a}_{ij},\boldsymbol{s}_{j},\boldsymbol{w}_{ij},\boldsymbol{r}_{ij})\biggr\}$$
\\
Then we separate the relevant components of the density into the parts specific to the treatment $k$ and other treatments $k^*$. And we assume that the distribution of treatment $k$ is not affected by other treatment availability and resistance. We define $f_{\boldsymbol{A}^{(k^*)}}$ as the counterfactual pdf of treatment $k^*$ such that $k^*\neq k$ under the intervention $\boldsymbol{d}^{(k^*)}=1$.  Then the pdf can be written as
\begin{align*}
    &\prod_{j=1}^Jf_{D^{(k)}}(d_{j}^{(k)}|\boldsymbol{s}_j)f_{\boldsymbol{D}^{(k^*)}}(\boldsymbol{d}_{j}^{(k^*)}|\boldsymbol{s}_j)f_{\boldsymbol{S}}(\boldsymbol{s}_{j})\prod_{i\in C_j}\biggl\{f_{\boldsymbol{W},\boldsymbol{R}}(\boldsymbol{w}_{ij},\boldsymbol{r}_{ij}|\boldsymbol{s}_{j})\\
&f_{\boldsymbol{A}^{(k^*)}}(\boldsymbol{a}_{ij}^{(k^*)}|\boldsymbol{s}_{j},\boldsymbol{w}_{ij},\boldsymbol{r}_{ij}^{(k^*)},\boldsymbol{d}_j^{(k^*)})f_{A^{(k)}}(a_{ij}^{(k)}|\boldsymbol{s}_{j},\boldsymbol{w}_{ij},r_{ij}^{(k)},d_j^{(k)},\boldsymbol{a}_{ij}^{(k^*)})f_Y(y_{ij}|\boldsymbol{a}_{ij},\boldsymbol{s}_{j},\boldsymbol{w}_{ij},\boldsymbol{r}_{ij})\biggr\}
\end{align*}
\\
We define $f_Y^{(k)}$ as the counterfactual pdf of outcome given treatment $k$. We assume the counterfactual pdf under the intervention settings $d^{(k)}=1$ and $a^{(k)}=1$. Note that $a^{(k)}=1$ implies $d^{(k)}=1$. Then the pdf can be written as
\begin{align*}
    &\prod_{j=1}^Jf_{\boldsymbol{D}^{(k^*)}}(\boldsymbol{d}_{j}^{(k^*)}|\boldsymbol{s}_j)f_{\boldsymbol{S}}(\boldsymbol{s}_{j})
    \prod_{i\in C_j}\biggl\{f_{\boldsymbol{W},\boldsymbol{R}}(\boldsymbol{w}_{ij},\boldsymbol{r}_{ij}|\boldsymbol{s}_{j})\\
&f_{\boldsymbol{A}^{(k^*)}}(\boldsymbol{a}_{ij}^{(k^*)}|\boldsymbol{s}_{j},\boldsymbol{w}_{ij},\boldsymbol{r}_{ij}^{(k^*)},\boldsymbol{d}_j^{(k^*)})f_Y^{(k)}(y_{ij}(d_j^{(k)}=1,a_{ij}^{(k)}=1)|\boldsymbol{a}_{ij}^{(k^*)},\boldsymbol{s}_{j},\boldsymbol{w}_{ij},\boldsymbol{r}_{ij})\biggr\}\\
 =&\prod_{j=1}^Jf_{\boldsymbol{D}^{(k^*)}}(\boldsymbol{d}_{j}^{(k^*)}|\boldsymbol{s}_j)f_{\boldsymbol{S}}(\boldsymbol{s}_{j})
    \prod_{i\in C_j}\biggl\{f_{\boldsymbol{W},\boldsymbol{R}}(\boldsymbol{w}_{ij},\boldsymbol{r}_{ij}|\boldsymbol{s}_{j})\\
&f_{\boldsymbol{A}^{(k^*)}}(\boldsymbol{a}_{ij}^{(k^*)}|\boldsymbol{s}_{j},\boldsymbol{w}_{ij},\boldsymbol{r}_{ij}^{(k^*)},\boldsymbol{d}_j^{(k^*)})f_Y^{(k)}(y_{ij}(a_{ij}^{(k)}=1)|\boldsymbol{a}_{ij}^{(k^*)},\boldsymbol{s}_{j},\boldsymbol{w}_{ij},\boldsymbol{r}_{ij})\biggr\}
\end{align*}
\\

\section{Appendix B: Proof of identifiability of $\psi(\boldsymbol{V}^{(k)})$}

We aim to estimate the parameters of interest on a global population based on the observed data. Under the assumptions defined in Section 2.2 in the main manuscript, we can rewrite the CATE as:\\
\begin{align*}
  &\psi\{\boldsymbol{V}^{(k)}\}&\\
    =&\mathbb{E}[Y\{a^{(k)}=1\}|\boldsymbol{V}^{(k)}]-E[Y\{a^{(k)}=0\}|\boldsymbol{V}^{(k)}]&\\
=&\mathbb{E}(\mathbb{E}[Y\{a^{(k)}=1\}|\boldsymbol{X}^{(k)}]|\boldsymbol{V}^{(k)})-E(E[Y\{a^{(k)}=0\}|\boldsymbol{X}^{(k)}]|\boldsymbol{V}^{(k)})&\\
=&\mathbb{E}(\mathbb{E}[Y\{a^{(k)}=1,d^{(k)}=1\}|\boldsymbol{X}^{(k)}]|\boldsymbol{V}^{(k)})-\mathbb{E}(\mathbb{E}[Y\{a^{(k)}=0\}|\boldsymbol{X}^{(k)}]|\boldsymbol{V}^{(k)})&\\
=&\mathbb{E}(\mathbb{E}[Y\{a^{(k)}=1,d^{(k)}=1\}|\boldsymbol{X}^{(k)},D^{(k)}=1]|\boldsymbol{V}^{(k)})&\\
&-\mathbb{E}(\mathbb{E}[Y\{a^{(k)}=0\}|\boldsymbol{X}^{(k)}]|\boldsymbol{V}^{(k)})&by~assumption (c)\\
=&\mathbb{E}(\mathbb{E}[Y\{a^{(k)}=1,d^{(k)}=1\}|A^{(k)}=1,\boldsymbol{X}^{(k)},D^{(k)}=1]|\boldsymbol{V}^{(k)})&\\
&-\mathbb{E}(\mathbb{E}[Y\{a^{(k)}=0\}|\boldsymbol{X}^{(k)}]|\boldsymbol{V}^{(k)})&by~assumption (d)\\
=&\mathbb{E}[\mathbb{E}\{Y|A^{(k)}=1,\boldsymbol{X}^{(k)},D^{(k)}=1\}|\boldsymbol{V}^{(k)}]-\mathbb{E}[\mathbb{E}\{Y|A^{(k)}=0,\boldsymbol{X}^{(k)}\}|\boldsymbol{V}^{(k)}]&by~assumption (a)\\
=&\mathbb{E}[\mathbb{E}\{Y|A^{(k)}=1,\boldsymbol{X}^{(k)}\}|\boldsymbol{V}^{(k)}]-\mathbb{E}[\mathbb{E}\{Y|A^{(k)}=0,\boldsymbol{X}^{(k)}\}|\boldsymbol{V}^{(k)}]
\end{align*}

Both the two terms can be estimated from the observed data, so the CATE is identifiable.\\

\section{Appendix C: Efficient Influence Function for $\boldsymbol{\beta_V}$ }

We derive the EIF for the coefficients $\boldsymbol{\beta_{\boldsymbol{V}}}$ in a working i.i.d. setting. Depending on the geometry of influence function, the EIF is given by $\prod(\mathcal{D}|T_Q)$ where $\mathcal{D}$ is an arbitrary influence function, $T_Q$ is the tangent space which is tangential to model space $\mathcal{M}$ and has the same dimension with $\mathcal{M}$. $\prod(\mathcal{D}|T_Q)$ refers to the projection of $\mathcal{D}$ onto space $T_Q$ \cite{semipara2006}. Here, we use the score function for IPTW as $\mathcal{D}_{IPTW}$ which is acturally an EIF up to a normalizing matrix and decompose the tangent space $T_Q$ to the direct sum of $T_Y$ and $T_{\boldsymbol{X}}$. As discussed by Tsiatis \cite{semipara2006}, we have
        \begin{align}
            \prod(\mathcal{D}_{IPTW}|T_Q)=&\prod(\mathcal{D}_{IPTW}|T_Y\oplus T_{\boldsymbol{X}}) \nonumber\\
            =&\underbrace{\mathbb{E}(\mathcal{D}_{IPTW}|\boldsymbol{X},A,Y)}_{(\textit{i})}-\underbrace{\mathbb{E}(\mathcal{D}_{IPTW}|\boldsymbol{X},A)}_{(\textit{ii})}+\underbrace{\mathbb{E}(\mathcal{D}_{IPTW}|\boldsymbol{X})}_{(\textit{iii})}
        \end{align}
 where
   $$ \mathcal{D}_{IPTW}=\biggl[\biggl\{\frac{I_{A=1}}{g(1|\boldsymbol{X})}-\frac{I_{A=0}}{g(0|\boldsymbol{X})}\biggr\}Y-\boldsymbol{V}^\intercal\boldsymbol{\beta_V}\biggr]\boldsymbol{V}$$
Note that $\boldsymbol{V}$ is a $(p+1)$-dimensional random covariate vector and the symbol $\intercal$ indicates a transpose.\\

Obviously, for the term $(\textit{i})$, $\mathbb{E}(\mathcal{D}_{IPTW}|\boldsymbol{X},A,Y)=\mathcal{D}_{IPTW}$. And the term $(\textit{ii})$,
    $$\mathbb{E}(\mathcal{D}_{IPTW}|\boldsymbol{X},A)
    =\biggl[\biggl\{\frac{I_{A=1}}{g(1|\boldsymbol{X})}-\frac{I_{A=0}}{g(0|\boldsymbol{X})}\biggr\}\mathbb{E}(Y|\boldsymbol{X},A)-\boldsymbol{V}^\intercal\boldsymbol{\beta_V}\biggr] \boldsymbol{V}$$
Then for the term $(\textit{iii})$,
\begin{align*}
    \mathbb{E}(\mathcal{D}_{IPTW}|\boldsymbol{X})
    =&\mathbb{E}\biggl(\biggl[\biggl\{\frac{I_{A=1}}{g(1|\boldsymbol{X})}-\frac{I_{A=0}}{g(0|\boldsymbol{X})}\biggr\}Y-\boldsymbol{V}^\intercal\boldsymbol{\beta_V}\biggr] \boldsymbol{V}\big{|}\boldsymbol{X}\biggr)\\
    =&\mathbb{E}\biggl(\biggl[\biggl\{\frac{I_{A=1}}{g(1|\boldsymbol{X})}-\frac{I_{A=0}}{g(0|\boldsymbol{X})}\biggr\}Y\big{|}\boldsymbol{X}\biggr]-\boldsymbol{V}^\intercal\boldsymbol{\beta_V} \biggr)\boldsymbol{V}\\
   =&\biggl[\biggl\{\frac{\mathbb{E}(Y,I_{A=1}|\boldsymbol{X})}{g(1|\boldsymbol{X})}-\frac{\mathbb{E}(Y,I_{A=0}|\boldsymbol{X})}{g(0|\boldsymbol{X})}\biggr\}-\boldsymbol{V}^\intercal\boldsymbol{\beta_V} \biggr]\boldsymbol{V}\\
    =&\biggl\{\mathbb{E}(Y|A=1,\boldsymbol{X})-\mathbb{E}(Y|A=0,\boldsymbol{X})-\boldsymbol{V}^\intercal\boldsymbol{\beta_V}\biggr\}\boldsymbol{V}
\end{align*}

Finally, according to equation $(2)$, we can combine all three terms as
\begin{align*}
    \mathcal{D}(Y,A,\boldsymbol{X},\boldsymbol{V})=&\prod(\mathcal{D}_{IPTW}|T_Q)
    =\mathbb{E}(\mathcal{D}_{IPTW}|\boldsymbol{X},A,Y)-\mathbb{E}(\mathcal{D}_{IPTW}|\boldsymbol{X},A)+\mathbb{E}(\mathcal{D}_{IPTW}|\boldsymbol{X})\\
    =&\biggl[\mathbb{E}(Y|A=1,\boldsymbol{X})-\mathbb{E}(Y|A=0,\boldsymbol{X})+\\
    &\biggl\{\frac{I_{A=1}}{g(1|\boldsymbol{X})}-\frac{I_{A=0}}{g(0|\boldsymbol{X})}\biggr\}\{Y-\mathbb{E}(Y|\boldsymbol{X},A)\}-\boldsymbol{V}^\intercal\boldsymbol{\beta_V}\biggr]\boldsymbol{V}\\
     =&\biggl[\biggl\{\frac{I_{A=1}}{g(1|\boldsymbol{X})}-\frac{I_{A=0}}{g(0|\boldsymbol{X})}\biggr\}\{Y-\overline{Q}(A,\boldsymbol{X}\}+\overline{Q}(1,\boldsymbol{X})-\overline{Q}(0,\boldsymbol{X})-\boldsymbol{V}^\intercal\boldsymbol{\beta_V}\biggr]\boldsymbol{V}
\end{align*}
\\
$\mathcal{D}(Y,A,\boldsymbol{X},\boldsymbol{V})$ is a $(p+1)$-dimensional random vector. We define a normalizing matrix $M$ 
 $$M=-\mathbb{E}\biggl\{\frac{\partial }{\partial \boldsymbol{\beta_{V}}}\mathcal{D}(Y,A,\boldsymbol{X},\boldsymbol{V})\biggr\}=\mathbb{E}(\boldsymbol{V}^\intercal\boldsymbol{V})$$
 \\
The dimension of $M$ is $(p+1)\times(p+1)$. The efficient influence function for coefficients $\boldsymbol{\beta_V}$ is $\mathcal{D}_{\boldsymbol{\beta_{V}}} =M^{-1}\mathcal{D}(Y,A,\boldsymbol{X},\boldsymbol{V})$, i.e.
\begin{equation}
    \mathcal{D}_{\boldsymbol{\beta_{V}}} 
    =M^{-1}\biggl[\biggl\{\frac{I_{A=1}}{g(1|\boldsymbol{X})}-\frac{I_{A=0}}{g(0|\boldsymbol{X})}\biggr\}\{Y-\overline{Q}(A,\boldsymbol{X})\}+\overline{Q}(1,\boldsymbol{X})-\overline{Q}(0,\boldsymbol{X})-\boldsymbol{V}^\intercal\boldsymbol{\beta_V}\biggr]\boldsymbol{V}\nonumber\\
\end{equation}
An equivalent result was also given by Rosenblum and van der Laan \cite{Rosenblum}.\\

\section{Appendix D: Augmented inverse probability of treatment weighted estimator (A-IPTW)}

In simple single-treatment single-dataset settings, A-IPTW involves a transformation of the outcomes using two components: 1) an outcome regression conditional on treatment and covariates, and 2) weights comprising of the inverse of the propensity score, where the propensity is the probability of treatment conditional on covariates \cite{van2014}. A-IPTW allows for doubly-robust and locally efficient estimation of causal parameters \cite{Kang}. \\

The A-IPTW estimator can be defined as a linear regression of the transformed outcome~\cite{Asma,van2014} on the set of covariates $\boldsymbol{V}$. We can construct the doubly robust  transformed outcome as:
$$Y^{*}=\frac{2A-1}{g_n(A|\boldsymbol{X})}\{Y-\overline{Q}_n(A,\boldsymbol{X})\}+\overline{Q}_n(1,\boldsymbol{X})-\overline{Q}_n(0,\boldsymbol{X})$$
Here, a subscript $n$ is used to denote a fitted value of the propensity score or outcome regression quantities. The coefficient estimates in the linear regression of $Y^{*}$ are denoted as $\hat{\boldsymbol \beta}^{AIPTW}_{\boldsymbol V}$, and are the A-IPTW estimates of $\boldsymbol{\beta}_{\boldsymbol{V}}$.\\

Note that the A-IPTW estimator satisfies the equation 
$\sum _{j=1}^J \sum _{i\in C_j} \mathcal{D}_{{ij,n}}({{\hat{\boldsymbol \beta}^{AIPTW}_{\boldsymbol V}}})=0$
where $\mathcal{D}_{{ij,n}}({{\hat{\boldsymbol \beta}^{AIPTW}_{\boldsymbol V}}})$ is the empirical influence curve evaluated at  $\overline{Q}_n(a_{ij},\boldsymbol{x}_{ij})$ and $g_n(a_{ij}|\boldsymbol{x}_{ij})$, with parameter estimate $\hat{\boldsymbol \beta}_{V}$.   The estimator is doubly robust, meaning that if either the propensity scores or the outcome regression estimates converge to their true values,  $\hat{\boldsymbol \beta}^{AIPTW}_{\boldsymbol V}$ converges to ${\boldsymbol \beta}_{\boldsymbol{V}}$ \cite{van2014} as the sample sizes of both the in-study populations and the number of studies increase. If both of the estimators $\overline{Q}_n$ and $g_n$ are consistent and converge overall at  sub-parametric $n^{-1/4}$ rates under regularity conditions, the A-IPTW is an asymptotically linear estimator. To estimate the variance of the parameter of interest, it suffices to consider the influence function of the estimator \cite{semipara2006}. \\

 \section{Appendix E: Simulation results}
 
 \begin{center}
\begin{table}[!htbp]
\caption{The simulation data generating mechanism for $J\in
\{10,30,50\}$ studies and $n_j=300~(for~j=1,2,\cdots,J)$ subjects in each study. $C_j$ is the set of indices of subjects in study $j$.}
\label{table:simulation}

\centering
 \begin{tabular}{c c}
 \\[-1.2em]
\toprule\toprule\\[-1.2em]
\textbf{Variable} & \textbf{Generating Mechanism}\\ \\[-1.2em]
\midrule\midrule\\[-1.2em]
$S_{1,ij}$    & 
\begin{tabular}[c]{@{}l@{}}$S_{1,j}\sim N(mean=0.5,sd=0.8,n=J)$\\ Set $S_{1,ij}=S_{1,j}$ for all i in $C_j$\\\\[-1.2em]
\end{tabular}       \\
\midrule\\[-1.2em]
$S_{2,ij}$    & 
\begin{tabular}[c]{@{}l@{}}$S_{2,j}\sim N(mean=0.5S_{1,ij}+0.1,sd=0.5,n=J)$\\ Set $S_{2,ij}=S_{2,j}$  for all i in $C_j$\\\\[-1.2em]
\end{tabular} \\
\midrule\\[-1.2em]
$W_{1,ij}$    & $W_{1,ij}\sim N(mean=0.3S_{1,j},sd=0.7,n=n_j)$\\
$W_{2,ij}$    & $W_{2,ij}\sim
Bin(logit(p)=0.5+0.1S_{1,j},n=n_j)$\\
$W_{3,ij}$    & $W_{3,ij}\sim
Bin(logit(p)=0.4+0.2S_{1,j},n=n_j)$\\\\[-1.2em]
\midrule\\[-1.2em]
$D_{ij}^{(k)}$    & 
\begin{tabular}
[c]{@{}l@{}}$D_j^{(1)}\sim Bin(logit(p)=0.8+0.5S_{1,ij},n=J)$\\  $D_j^{(2)}\sim Bin(logit(p)=0.9+0.4S_{1,ij},n=J)$\\  
$D_j^{(3)}|(D_j^{(1)},D_j^{(2)})$\\
\[~~\left\{\begin{array}{lr}
	= 1 &if~d_j^{(1)}=d_j^{(2)}=0\\
	\sim Bin(logit(p)=0.2+0.3S_{1,ij},n=J)&otherwise \\
\end{array}\right.\]\\
Set $D_{ij}^{(k)}=D_j^{(k)}$  for all $i$ in $C_j$
\\[0.4em]
\end{tabular} \\
\midrule\\[-1.2em]
$A_{ij}^{(k)}$    &
\begin{tabular}
[c]{@{}l@{}}$A_{ij}^{(1)}\sim Bin(logit(p)=-2+1.6S_{1,ij}+2.5W_{1,ij}+1.2W_{2,ij},n=n_j)$\\  
$A_{ij}^{(2)}\sim Bin(logit(p)=0.6+0.3S_{1,ij}+0.4W_{3,ij},
n=n_j)$\\  
$A_{ij}^{(3)}\sim Bin(logit(p)=0.5+0.3S_{1,ij}+0.2W_{1,ij},n=n_j)$\newline\\ \\[-1em]
\end{tabular} \\
\midrule\\[-1.2em]
$Y_{ij}$ &
\begin{tabular}[c]{@{}l@{}}
Without random effects:\\
$Y_{ij}\sim \{2+0.3W_{1,ij}+0.6W_{2,ij}+0.9W_{3,ij}+0.8S_{1,ij}-0.4S_{2,ij}$\newline\\
$~~~~~+0.5A_{ij}^{(1)}+A_{ij}^{(2)}+0.5A_{ij}^{(3)}+A_{ij}^{(1)}(0.65W_{1,ij}+0.35W_{3,ij})$\newline\\
$~~~~~+A_{ij}^{(2)}(0.8W_{1,ij})+A_{ij}^{(3)}(W_{2,ij})+\epsilon, n=n_j\}$\newline\\
$where~\epsilon\sim N(mean=0, sd=1)$\\\\[-1.2em]
With random effects:\\
$Y_{ij}\sim \{2+0.3W_{1,ij}+0.6W_{2,ij}+0.9W_{3,ij}+0.8S_{1,ij}-0.4S_{2,ij}$\newline\\
$~~~~~+0.5A_{ij}^{(1)}+A_{ij}^{(2)}+0.5A_{ij}^{(3)}+A_{ij}^{(1)}(0.65W_{1,ij}+0.35W_{3,ij})$\newline\\
$~~~~~+\underline{{A}_{ij}^{(1)}(0.7S_{2,ij})}+A_{ij}^{(2)}(0.8W_{1,ij})+A_{ij}^{(3)}(W_{2,ij})+\epsilon, n=n_j\}$\newline\\
$where~\epsilon\sim N(mean=0, sd=1)$\\\\[-1.2em]
\end{tabular}\\
 \bottomrule\bottomrule
\end{tabular}
\vspace{1.8mm}

\end{table}
\end{center}

 \begin{landscape}
 \begin{table}[!htbp]
 \fontsize{10.5pt}{11.25pt}\selectfont
\centering
\caption{Estimated effect modification coefficients and their corresponding non clustered (noCl), clustered (Cl) and Monte Carlo (MC) standard errors based on TMLE under four scenarios for three sample sizes. The simulated data are generated without random effects. The four scenarios are as follows: Scenario 1 - both $Q$ and $g$ models are correct; Scenario 2 - $Q$ model is correct, $g$ is null; Scenario 3 - $Q$ model is null, $g$ model is correct; Scenario 4 - both $Q$ and $g$ models are null.}
\label{table:3}
 \begin{tabular}{c c c c c c c c c c c c c c c}
\toprule\toprule
& & & \multicolumn{4}{c}{\makecell{N=3000, no. of cluster=10}}& \multicolumn{4}{c}{\makecell{N=9000, no. of cluster=30}}& \multicolumn{4}{c}{\makecell{N=15000, no. of cluster=50}}
\\
\textbf{Scenarios} & \textbf{Var} & \textbf{$\beta_V$} & \textbf{$\hat{\beta}_V$} & \textbf{$SE_{noCl}$} & \textbf{$SE_{Cl}$} & \textbf{$SE_{MC}$} & \textbf{$\hat{\beta}_V$} & \textbf{$SE_{noCl}$} & \textbf{$SE_{Cl}$} & \textbf{$SE_{MC}$} &\textbf{$\hat{\beta}_V$} & \textbf{$SE_{noCl}$} & \textbf{$SE_{Cl}$} & \textbf{$SE_{MC}$}\\
\midrule\midrule\\[-1em]
\multirow{5}{*}{Scenario 1} &$W_1$ & \textbf{0.65} & \textbf{0.65} & 0.15 & 0.14 & 0.16 & \textbf{0.65} & 0.10 & 0.10 & 0.10 & \textbf{0.65} & 0.09 & 0.09 & 0.08\\ 
 &$W_2$ & \textbf{0.00} & \textbf{0.00} & 0.18 & 0.16 & 0.17 & \textbf{0.00} & 0.12 & 0.11 & 0.11 & \textbf{0.00} & 0.10 & 0.10 & 0.09 \\
 &$W_3$ & \textbf{0.35} & \textbf{0.35} & 0.16 & 0.15 & 0.14 & \textbf{0.35} & 0.11 & 0.10 & 0.09 & \textbf{0.35} & 0.09 & 0.09 & 0.08 \\ 
 &$A^{(2)}$ & \textbf{0.00} & \textbf{0.01} & 0.16 & 0.15 & 0.17 & \textbf{0.00} & 0.11 & 0.11 & 0.11 & \textbf{0.00} & 0.09 & 0.09 & 0.09 \\ 
 &$A^{(3)}$ & \textbf{0.00} & \textbf{0.00} & 0.16 & 0.16 & 0.20 & \textbf{0.00} & 0.10 & 0.11 & 0.11 & \textbf{0.00} & 0.09 & 0.09 & 0.09 \\ 
\midrule\\[-1em]
\multirow{5}{*}{Scenario 2} &$W_1$ & \textbf{0.65} & \textbf{0.65} & 0.06 & 0.08 & 0.08 & \textbf{0.65} & 0.03 & 0.06 & 0.05 & \textbf{0.65} & 0.02 & 0.05 & 0.04 \\ 
&$W_2$ & \textbf{0.00} & \textbf{0.00} & 0.08 & 0.08 & 0.09 & \textbf{0.00} & 0.04 & 0.05 & 0.05 & \textbf{0.00} & 0.04 & 0.04 & 0.04 \\ 
&$W_3$ & \textbf{0.35} & \textbf{0.35} & 0.08 & 0.08 & 0.08 & \textbf{0.35} & 0.05 & 0.05 & 0.05 & \textbf{0.35} & 0.04 & 0.04 & 0.04 \\ 
&$A^{(2)}$ & \textbf{0.00} & \textbf{0.00} & 0.08 & 0.10 & 0.12 & \textbf{0.00} & 0.04 & 0.07 & 0.06 & \textbf{0.00} & 0.04 & 0.06 & 0.05 \\ 
&$A^{(3)}$ & \textbf{0.00} & \textbf{0.00} & 0.08 & 0.11 & 0.14 & \textbf{0.00} & 0.05 & 0.07 & 0.08 & \textbf{0.00} & 0.04 & 0.06 & 0.06 \\ 
\midrule\\[-1em]
\multirow{5}{*}{Scenario 3} &$W_1$ & \textbf{0.65} & \textbf{0.65} & 0.17 & 0.19 & 0.22 & \textbf{0.65} & 0.13 & 0.15 & 0.17 & \textbf{0.66} & 0.11 & 0.13 & 0.14 \\ 
&$W_2$ & \textbf{0.00} & \textbf{-0.02} & 0.20 & 0.19 & 0.20 & \textbf{-0.01} & 0.14 & 0.14 & 0.14 & \textbf{-0.02} & 0.12 & 0.12 & 0.11 \\ 
&$W_3$ & \textbf{0.35} & \textbf{0.34} & 0.18 & 0.16 & 0.17 & \textbf{0.35} & 0.13 & 0.13 & 0.12 & \textbf{0.35} & 0.11 & 0.10 & 0.09 \\ 
&$A^{(2)}$ & \textbf{0.00} & \textbf{0.08} & 0.18 & 0.23 & 0.28 & \textbf{0.04} & 0.13 & 0.18 & 0.19 & \textbf{0.03} & 0.10 & 0.15 & 0.15 \\ 
&$A^{(3)}$ & \textbf{0.00} & \textbf{0.01} & 0.18 & 0.23 & 0.30 & \textbf{0.02} & 0.12 & 0.18 & 0.20 & \textbf{0.02} & 0.10 & 0.16 & 0.17 \\ 
\midrule\\[-1em]
\multirow{5}{*}{Scenario 4} &$W_1$ & \textbf{0.65} & \textbf{0.62} & 0.06 & 0.19 & 0.17 & \textbf{0.61} & 0.04 & 0.13 & 0.10 & \textbf{0.61} & 0.03 & 0.11 & 0.08 \\ 
&$W_2$ & \textbf{0.00} & \textbf{-0.05} & 0.09 & 0.12 & 0.12 & \textbf{-0.05} & 0.05 & 0.08 & 0.07 & \textbf{-0.06} & 0.04 & 0.06 & 0.05 \\ 
&$W_3$ & \textbf{0.35} & \textbf{0.31} & 0.09 & 0.08 & 0.09 & \textbf{0.31} & 0.05 & 0.05 & 0.06 & \textbf{0.31} & 0.04 & 0.04 & 0.05 \\ 
&$A^{(2)}$ & \textbf{0.00} & \textbf{0.51} & 0.09 & 0.21 & 0.27 & \textbf{0.52} & 0.05 & 0.15 & 0.16 & \textbf{0.52} & 0.04 & 0.12 & 0.13 \\ 
&$A^{(3)}$ & \textbf{0.00} & \textbf{0.03} & 0.09 & 0.22 & 0.30 & \textbf{0.05} & 0.05 & 0.16 & 0.17 & \textbf{0.05} & 0.04 & 0.13 & 0.14 \\ 
\bottomrule\bottomrule
\\[-1.6em]
\end{tabular}

\end{table}
\end{landscape}

 \begin{landscape}
 \begin{table}[!htbp]
 \fontsize{10.5pt}{11.25pt}\selectfont
\centering
\caption{Estimated effect modification coefficients and their corresponding non clustered (noCl), clustered (Cl) and Monte Carlo (MC) standard errors based on TMLE under four scenarios for three sample sizes. The data are generated with random effects. The four scenarios are as follows: Scenario 1 - both $Q$ and $g$ models are correct; Scenario 2 - $Q$ model is correct, $g$ is null; Scenario 3 - $Q$ model is null, $g$ model is correct; Scenario 4 - both $Q$ and $g$ models are null.}
\label{table:3}
 \begin{tabular}{c c c c c c c c c c c c c c c}
\toprule\toprule
& & & \multicolumn{4}{c}{\makecell{N=3000, no. of cluster=10}}& \multicolumn{4}{c}{\makecell{N=9000, no. of cluster=30}}& \multicolumn{4}{c}{\makecell{N=15000, no. of cluster=50}}
\\
\textbf{Scenarios} & \textbf{Var} & \textbf{$\beta_V$} & \textbf{$\hat{\beta}_V$} & \textbf{$SE_{noCl}$} & \textbf{$SE_{Cl}$} & \textbf{$SE_{MC}$} & \textbf{$\hat{\beta}_V$} & \textbf{$SE_{noCl}$} & \textbf{$SE_{Cl}$} & \textbf{$SE_{MC}$} &\textbf{$\hat{\beta}_V$} & \textbf{$SE_{noCl}$} & \textbf{$SE_{Cl}$} & \textbf{$SE_{MC}$}\\
\midrule\midrule\\[-1em]
\multirow{5}{*}{Scenario 1} &$W_1$ & \textbf{0.77} & \textbf{0.75} & 0.15 & 0.14 & 0.17 & \textbf{0.76} & 0.11 & 0.11 & 0.11 & \textbf{0.76} & 0.09 & 0.09 & 0.08 \\
&$W_2$ &  \textbf{0.02} & \textbf{0.01} & 0.18 & 0.16 & 0.16 & \textbf{0.02} & 0.12 & 0.12 & 0.11 & \textbf{0.02} & 0.10 & 0.10 & 0.08 \\ 
 &$W_3$ & \textbf{0.38} & \textbf{0.38} & 0.16 & 0.15 & 0.14 & \textbf{0.38} & 0.11 & 0.11 & 0.09 & \textbf{0.38} & 0.09 & 0.09 & 0.07 \\ 
 &$A^{(2)}$ & \textbf{0.08} & \textbf{0.08} & 0.16 & 0.19 & 0.24 & \textbf{0.08} & 0.11 & 0.14 & 0.14 & \textbf{0.08} & 0.09 & 0.11 & 0.11 \\ 
 &$A^{(3)}$ & \textbf{0.02} & \textbf{0.02} & 0.16 & 0.20 & 0.26 & \textbf{0.02} & 0.10 & 0.14 & 0.14 & \textbf{0.03} & 0.09 & 0.12 & 0.11 \\ 
\midrule\\[-1em]
\multirow{5}{*}{Scenario 2} 
&$W_1$ & \textbf{0.77} & \textbf{0.75} & 0.06 & 0.07 & 0.11 & \textbf{0.76} & 0.03 & 0.04 & 0.06 & \textbf{0.77} & 0.02 & 0.04 & 0.04 \\ 
&$W_2$ & \textbf{0.02} & \textbf{0.02} & 0.08 & 0.08 & 0.09 & \textbf{0.02} & 0.05 & 0.04 & 0.05 & \textbf{0.02} & 0.04 & 0.04 & 0.04 \\ 
&$W_3$ & \textbf{0.38} & \textbf{0.38} & 0.08 & 0.08 & 0.08 & \textbf{0.38} & 0.05 & 0.05 & 0.05 & \textbf{0.38} & 0.04 & 0.04 & 0.04 \\ 
&$A^{(2)}$ & \textbf{0.08} & \textbf{0.07} & 0.08 & 0.15 & 0.19 & \textbf{0.08} & 0.04 & 0.10 & 0.10 & \textbf{0.08} & 0.04 & 0.08 & 0.08 \\ 
 &$A^{(3)}$ & \textbf{0.02} & \textbf{0.02} & 0.08 & 0.17 & 0.22 & \textbf{0.02} & 0.05 & 0.11 & 0.12 & \textbf{0.02} & 0.04 & 0.09 & 0.09 \\ 
\midrule\\[-1em]
\multirow{5}{*}{Scenario 3} 
&$W_1$ & \textbf{0.77} & \textbf{0.73} & 0.18 & 0.20 & 0.25 & \textbf{0.75} & 0.14 & 0.17 & 0.18 & \textbf{0.76} & 0.11 & 0.14 & 0.15 \\ 
&$W_2$ & \textbf{0.02} & \textbf{-0.01} & 0.22 & 0.21 & 0.22 & \textbf{0.00} & 0.16 & 0.15 & 0.15 & \textbf{0.00} & 0.13 & 0.13 & 0.12 \\ 
&$W_3$ & \textbf{0.38} & \textbf{0.36} & 0.20 & 0.18 & 0.18 & \textbf{0.38} & 0.15 & 0.14 & 0.13 & \textbf{0.38} & 0.12 & 0.12 & 0.10 \\ 
&$A^{(2)}$ & \textbf{0.08} & \textbf{0.15} & 0.20 & 0.29 & 0.37 & \textbf{0.12} & 0.14 & 0.22 & 0.25 & \textbf{0.10} & 0.12 & 0.19 & 0.19 \\ 
 &$A^{(3)}$ & \textbf{0.02} & \textbf{0.04} & 0.20 & 0.30 & 0.40 & \textbf{0.05} & 0.14 & 0.23 & 0.25 & \textbf{0.05} & 0.11 & 0.20 & 0.21 \\ 
\midrule\\[-1em]
\multirow{5}{*}{Scenario 4} 
&$W_1$ & \textbf{0.77} & \textbf{0.64} & 0.06 & 0.23 & 0.17 & \textbf{0.63} & 0.04 & 0.16 & 0.11 & \textbf{0.63} & 0.03 & 0.13 & 0.09 \\ 
&$W_2$ & \textbf{0.02} & \textbf{-0.06} & 0.09 & 0.14 & 0.13 & \textbf{-0.07} & 0.05 & 0.08 & 0.07 & \textbf{-0.08} & 0.04 & 0.07 & 0.06 \\ 
&$W_3$ & \textbf{0.38} & \textbf{0.33} & 0.09 & 0.09 & 0.10 & \textbf{0.34} & 0.05 & 0.06 & 0.06 & \textbf{0.34} & 0.04 & 0.05 & 0.05 \\ 
 &$A^{(2)}$ & \textbf{0.08} & \textbf{0.57} & 0.09 & 0.27 & 0.36 & \textbf{0.57} & 0.05 & 0.20 & 0.22 & \textbf{0.58} & 0.04 & 0.16 & 0.17 \\ 
 &$A^{(3)}$ & \textbf{0.02} & \textbf{0.03} & 0.09 & 0.30 & 0.41 & \textbf{0.07} & 0.05 & 0.22 & 0.25 &\textbf{0.06} & 0.04 & 0.19 & 0.20 \\ 
\bottomrule\bottomrule
\\[-1em]
\end{tabular}

\end{table}
\end{landscape}

 \begin{table}[!htbp]
\centering
\caption{Coverage rates based on clustered standard errors for TMLE under the four scenarios for three sample sizes. $Cov_{10},Cov_{30},Cov_{50}$ represent the coverage rates corresponding to three datasets of 10, 30 and 50 studies with total sample sizes of 3000, 9000, 15000, respectively. The results are summarized over 1000 iterations. The four scenarios are as follows: Scenario 1 - both $Q$ and $g$ models are correct; Scenario 2 - $Q$ model is correct, $g$ is null; Scenario 3 - $Q$ model is null, $g$ model is correct; Scenario 4 - both $Q$ and $g$ models are null.}
\label{table:3}
 \begin{tabular}{c c c c c c c c}
\toprule\toprule
& &  \multicolumn{3}{c}{Without random effects}& \multicolumn{3}{c}{With random effects}
\\
\textbf{Scenarios} & \textbf{Var} & \textbf{$Cov_{10}$} & \textbf{$Cov_{30}$} & \textbf{$Cov_{50}$} & \textbf{$Cov_{10}$} & \textbf{$Cov_{30}$} & \textbf{$Cov_{50}$} \\
\midrule\midrule\\[-1em]
\multirow{5}{*}{Scenario 1} 
&$W_1$ & 0.815 & 0.896 & 0.920 & 0.798 & 0.890 & 0.930 \\ 
&$W_2$ &  0.866 & 0.943 & 0.951 & 0.881 & 0.946 & 0.957 \\ 
 &$W_3$ & 0.893 & 0.961 & 0.950 & 0.901 & 0.948 & 0.967 \\ 
 &$A^{(2)}$ & 0.851 & 0.940 & 0.940 & 0.856 & 0.922 & 0.946 \\ 
 &$A^{(3)}$ & 0.860 & 0.935 & 0.953 & 0.826 & 0.881 & 0.896 \\ 
\midrule\\[-1em]
\multirow{5}{*}{Scenario 2} 
 &$W_1$ & 0.908 & 0.976 & 0.976 & 0.750 & 0.853 & 0.876 \\ 
 &$W_2$  &0.874 & 0.945 & 0.947 & 0.855 & 0.917 & 0.928 \\ 
 &$W_3$  & 0.902 & 0.927 & 0.922 & 0.899 & 0.927 & 0.938 \\ 
 &$A^{(2)}$ &0.908 & 0.973 & 0.961 & 0.883 & 0.949 & 0.962 \\ 
 &$A^{(3)}$ & 0.869 & 0.926 & 0.949 & 0.838 & 0.897 & 0.872 \\ 
\midrule\\[-1em]
\multirow{5}{*}{Scenario 3} 
&$W_1$ &  0.827 & 0.898 & 0.915 & 0.797 & 0.892 & 0.911 \\ 
&$W_2$ & 0.864 & 0.917 & 0.959 & 0.865 & 0.914 & 0.951 \\ 
&$W_3$ & 0.885 & 0.943 & 0.951 & 0.883 & 0.932 & 0.953 \\ 
 &$A^{(2)}$ & 0.842 & 0.922 & 0.930 & 0.847 & 0.913 & 0.935 \\ 
 &$A^{(3)}$ & 0.847 & 0.932 & 0.926 & 0.827 & 0.927 & 0.930 \\ 
\midrule\\[-1em]
\multirow{5}{*}{Scenario 4} 
&$W_1$ & 0.939 & 0.972 & 0.961 & 0.900 & 0.891 & 0.881 \\ 
&$W_2$ & 0.892 & 0.920 & 0.873 & 0.892 & 0.859 & 0.744 \\ 
&$W_3$ & 0.864 & 0.858 & 0.819 & 0.849 & 0.829 & 0.806 \\ 
 &$A^{(2)}$ & 0.294 & 0.119 & 0.034 & 0.401 & 0.305 & 0.182 \\ 
 &$A^{(3)}$ & 0.824 & 0.903 & 0.908 & 0.822 & 0.909 & 0.917 \\ 
\bottomrule\bottomrule
\\
\end{tabular}
\end{table}

 \begin{landscape}
 \begin{table}[!htbp]
 \fontsize{10.5pt}{11.25pt}\selectfont
\centering
\caption{Estimated effect modification coefficients and their corresponding non clustered (noCl), clustered (Cl) and Monte Carlo (MC) standard errors based on A-IPTW under the four scenarios for three sample sizes. The data is generated without random effects. The four scenarios are as follows: Scenario 1 - both $Q$ and $g$ models are correct; Scenario 2 - $Q$ model is correct, $g$ is null; Scenario 3 - $Q$ model is null, $g$ model is correct; Scenario 4 - both $Q$ and $g$ models are null.}
\label{table:3}
 \begin{tabular}{c c c c c c c c c c c c c c c}
\toprule\toprule
& & & \multicolumn{4}{c}{\makecell{N=3000, no. of cluster=10}}& \multicolumn{4}{c}{\makecell{N=9000, no. of cluster=30}}& \multicolumn{4}{c}{\makecell{N=15000, no. of cluster=50}}
\\
\textbf{Scenarios} & \textbf{Var} & \textbf{$\beta_V$} & \textbf{$\hat{\beta}_V$} & \textbf{$SE_{noCl}$} & \textbf{$SE_{Cl}$} & \textbf{$SE_{MC}$} & \textbf{$\hat{\beta}_V$} & \textbf{$SE_{noCl}$} & \textbf{$SE_{Cl}$} & \textbf{$SE_{MC}$} &\textbf{$\hat{\beta}_V$} & \textbf{$SE_{noCl}$} & \textbf{$SE_{Cl}$} & \textbf{$SE_{MC}$}\\
\midrule\midrule\\[-1em]
\multirow{5}{*}{Scenario 1} 
&$W_1$ & \textbf{0.65} & \textbf{0.64} & 0.25 & 0.24 & 0.27 & \textbf{0.65} & 0.14 & 0.13 & 0.14 & \textbf{0.65} & 0.10 & 0.10 & 0.11 \\
&$W_2$ &  \textbf{0.00} & \textbf{0.00} & 0.27 & 0.26 & 0.27 & \textbf{0.00} & 0.14 & 0.14 & 0.14 & \textbf{0.00} & 0.11 & 0.11 & 0.11 \\ 
 &$W_3$ & \textbf{0.35} & \textbf{0.35} & 0.23 & 0.22 & 0.22 & \textbf{0.35} & 0.13 & 0.12 & 0.12 & \textbf{0.35} & 0.10 & 0.10 & 0.10 \\ 
 &$A^{(2)}$ & \textbf{0.00} & \textbf{0.01} & 0.23 & 0.22 & 0.26 & \textbf{0.00} & 0.13 & 0.13 & 0.14 & \textbf{0.0}0 & 0.10 & 0.11 & 0.11 \\ 
 &$A^{(3)}$ & \textbf{0.00} & \textbf{0.00} & 0.23 & 0.23 & 0.26 & \textbf{0.00} & 0.12 & 0.13 & 0.14 & \textbf{0.01} & 0.10 & 0.11 & 0.11 \\ 
\midrule\\[-1em]
\multirow{5}{*}{Scenario 2} 
&$W_1$ & \textbf{0.65} & \textbf{0.65} & 0.06 & 0.09 & 0.09 & \textbf{0.65} & 0.03 & 0.06 & 0.05 & \textbf{0.65} & 0.02 & 0.05 & 0.04 \\ 
&$W_2$ & \textbf{0.00} & \textbf{0.00} & 0.08 & 0.08 & 0.09 & \textbf{0.00} & 0.04 & 0.05 & 0.05 & \textbf{0.00} & 0.04 & 0.04 & 0.04 \\ 
&$W_3$ & \textbf{0.35} & \textbf{0.35} & 0.08 & 0.08 & 0.08 & \textbf{0.35} & 0.05 & 0.05 & 0.05 & \textbf{0.35} & 0.04 & 0.04 & 0.04 \\ 
 &$A^{(2)}$ & \textbf{0.00} & \textbf{0.00} & 0.08 & 0.12 & 0.14 & \textbf{0.00} & 0.04 & 0.07 & 0.06 & \textbf{0.00} & 0.04 & 0.06 & 0.05 \\ 
 &$A^{(3)}$ & \textbf{0.00} & \textbf{0.00} & 0.08 & 0.11 & 0.15 & \textbf{0.00} & 0.05 & 0.07 & 0.08 & \textbf{0.00} & 0.04 & 0.06 & 0.06 \\ 
\midrule\\[-1em]
\multirow{5}{*}{Scenario 3} 
&$W_1$ & \textbf{0.65} & \textbf{0.62} & 0.87 & 0.98 & 0.92 & \textbf{0.62} & 0.52 & 0.67 & 0.59 & \textbf{0.60} & 0.39 & 0.55 & 0.44 \\ 
&$W_2$ & \textbf{0.00} & \textbf{0.03} & 0.89 & 0.90 & 0.87 & \textbf{-0.04} & 0.49 & 0.53 & 0.49 & \textbf{-0.05} & 0.38 & 0.42 & 0.38 \\ 
&$W_3$ & \textbf{0.35} & \textbf{0.34} & 0.82 & 0.86 & 0.83 & \textbf{0.36} & 0.47 & 0.54 & 0.49 &\textbf{ 0.34} & 0.36 & 0.41 & 0.36 \\ 
 &$A^{(2)}$ & \textbf{0.00} & \textbf{0.06} & 0.72 & 0.78 & 0.90 & \textbf{0.04} & 0.43 & 0.56 & 0.58 & \textbf{0.05} & 0.33 & 0.46 & 0.45 \\ 
 &$A^{(3)}$ & \textbf{0.00} & \textbf{0.04} & 0.73 & 0.81 & 0.88 & \textbf{0.05} & 0.42 & 0.57 & 0.60 & \textbf{0.07} & 0.32 & 0.47 & 0.48 \\ 
\midrule\\[-1em]
\multirow{5}{*}{Scenario 4} 
&$W_1$ & \textbf{0.65} & \textbf{0.24} & 0.10 & 0.55 & 0.20 & \textbf{0.23} & 0.06 & 0.34 & 0.13 & \textbf{0.23} & 0.04 & 0.27 & 0.10 \\ 
&$W_2$ & \textbf{0.00} & \textbf{-0.31} & 0.12 & 0.36 & 0.15 & \textbf{-0.31} & 0.07 & 0.21 & 0.08 & \textbf{-0.32} & 0.05 & 0.16 & 0.06 \\ 
&$W_3$ & \textbf{0.35} & \textbf{0.29} & 0.12 & 0.48 & 0.12 & \textbf{0.29} & 0.07 & 0.28 & 0.07 & \textbf{0.29} & 0.05 & 0.22 & 0.06 \\ 
 &$A^{(2)}$ & \textbf{0.00} & \textbf{0.50} & 0.12 & 0.34 & 0.38 & \textbf{0.52} & 0.07 & 0.23 & 0.22 & \textbf{0.51} & 0.05 & 0.19 & 0.18 \\ 
 &$A^{(3)}$ & \textbf{0.00} & \textbf{0.04} & 0.12 & 0.37 & 0.43 & \textbf{0.06} & 0.07 & 0.26 & 0.26 & \textbf{0.05} & 0.05 & 0.21 & 0.21 \\ 
\bottomrule\bottomrule
\\[-1em]
\end{tabular}

\end{table}
\end{landscape}

 \begin{landscape}
 \begin{table}[!htbp]
 \fontsize{10.5pt}{11.25pt}\selectfont
\centering
\caption{Estimated effect modification coefficients and their corresponding non clustered (noCl), clustered (Cl) and Monte Carlo (MC) standard errors based on A-IPTW under four scenarios for three sample sizes. The outcomes are generated with random effects. The four scenarios are as follows: Scenario 1 - both $Q$ and $g$ models are correct; Scenario 2 - $Q$ model is correct, $g$ is null; Scenario 3 - $Q$ model is null, $g$ model is correct; Scenario 4 - both $Q$ and $g$ models are null.}
\label{table:3}
 \begin{tabular}{c c c c c c c c c c c c c c c}
\toprule\toprule
& & & \multicolumn{4}{c}{\makecell{N=3000, no. of cluster=10}}& \multicolumn{4}{c}{\makecell{N=9000, no. of cluster=30}}& \multicolumn{4}{c}{\makecell{N=15000, no. of cluster=50}}
\\
\textbf{Scenarios} & \textbf{Var} & \textbf{$\beta_V$} & \textbf{$\hat{\beta}_V$} & \textbf{$SE_{noCl}$} & \textbf{$SE_{Cl}$} & \textbf{$SE_{MC}$} & \textbf{$\hat{\beta}_V$} & \textbf{$SE_{noCl}$} & \textbf{$SE_{Cl}$} & \textbf{$SE_{MC}$} &\textbf{$\hat{\beta}_V$} & \textbf{$SE_{noCl}$} & \textbf{$SE_{Cl}$} & \textbf{$SE_{MC}$}\\
\midrule\midrule\\[-1em]
\multirow{5}{*}{Scenario 1} 
&$W_1$ & \textbf{0.77} & \textbf{0.74} & 0.23 & 0.22 & 0.26 & \textbf{0.76} & 0.14 & 0.14 & 0.15 & \textbf{0.76} & 0.10 & 0.11 & 0.11 \\ 
&$W_2$ &  \textbf{0.02} & \textbf{0.01} & 0.25 & 0.24 & 0.26 & \textbf{0.02} & 0.14 & 0.14 & 0.14 & \textbf{0.02} & 0.11 & 0.11 & 0.11 \\ 
 &$W_3$ & \textbf{0.38} & \textbf{0.38} & 0.22 & 0.20 & 0.22 & \textbf{0.38} & 0.13 & 0.13 & 0.13 & \textbf{0.38} & 0.10 & 0.10 & 0.10 \\ 
 &$A^{(2)}$ & \textbf{0.08} & \textbf{0.07} & 0.22 & 0.25 & 0.32 & \textbf{0.08} & 0.13 & 0.15 & 0.16 & \textbf{0.08} & 0.10 & 0.12 & 0.13 \\
 &$A^{(3)}$ & \textbf{0.02} & \textbf{0.03} & 0.21 & 0.26 & 0.31 & \textbf{0.02} & 0.12 & 0.15 & 0.16 & \textbf{0.03} & 0.10 & 0.12 & 0.13 \\ 
\midrule\\[-1em]
\multirow{5}{*}{Scenario 2} 
&$W_1$ & \textbf{0.77} & \textbf{0.75} & 0.06 & 0.07 & 0.11 & \textbf{0.76} & 0.03 & 0.04 & 0.06 & \textbf{0.77} & 0.02 & 0.04 & 0.04 \\ 
&$W_2$ & \textbf{0.02} & \textbf{0.02} & 0.08 & 0.08 & 0.09 & \textbf{0.02} & 0.05 & 0.04 & 0.05 & \textbf{0.02} & 0.04 & 0.04 & 0.04 \\ 
&$W_3$ & \textbf{0.38} & \textbf{0.38} & 0.08 & 0.08 & 0.08 & \textbf{0.38} & 0.05 & 0.05 & 0.05 & \textbf{0.38} & 0.04 & 0.04 & 0.04 \\ 
&$A^{(2)}$ & \textbf{0.08} & \textbf{0.07} & 0.08 & 0.17 & 0.24 & \textbf{0.08} & 0.04 & 0.10 & 0.10 & \textbf{0.08} & 0.04 & 0.08 & 0.08 \\ 
&$A^{(3)}$ & \textbf{0.02} & \textbf{0.02} & 0.08 & 0.18 & 0.22 & \textbf{0.02} & 0.05 & 0.11 & 0.12 & \textbf{0.02} & 0.04 & 0.09 & 0.09 \\ 
\midrule\\[-1em]
\multirow{5}{*}{Scenario 3} 
&$W_1$ & \textbf{0.7}7 & \textbf{0.70} & 0.96 & 1.06 & 1.02 & \textbf{0.72} & 0.58 & 0.73 & 0.64 & \textbf{0.70} & 0.44 & 0.59 & 0.48 \\ 
&$W_2$ & \textbf{0.02} & \textbf{0.04} & 1.00 & 1.00 & 0.97 & \textbf{-0.03} & 0.55 & 0.59 & 0.56 & \textbf{-0.03} & 0.43 & 0.47 & 0.43 \\ 
&$W_3$ & \textbf{0.38} & \textbf{0.37} & 0.93 & 0.95 & 0.93 & \textbf{0.40} & 0.54 & 0.59 & 0.55 & \textbf{0.38} & 0.40 & 0.45 & 0.41 \\ 
&$A^{(2)}$ & \textbf{0.08} & \textbf{0.13} & 0.82 & 0.91 & 1.04 & \textbf{0.11} & 0.49 & 0.63 & 0.67 & \textbf{0.13} & 0.37 & 0.52 & 0.52 \\ 
&$A^{(3)}$ & \textbf{0.02} & \textbf{0.07} & 0.83 & 0.93 & 1.01 & \textbf{0.09} & 0.47 & 0.64 & 0.67 & \textbf{0.11} & 0.36 & 0.53 & 0.53 \\ 
\midrule\\[-1em]
\multirow{5}{*}{Scenario 4} 
&$W_1$ & \textbf{0.77} & \textbf{0.25} & 0.10 & 0.60 & 0.21 & \textbf{0.24} & 0.06 & 0.38 & 0.13 & \textbf{0.24} & 0.04 & 0.30 & 0.11 \\ 
&$W_2$ & \textbf{0.02} & \textbf{-0.33} & 0.12 & 0.37 & 0.15 & \textbf{-0.33} & 0.07 & 0.22 & 0.09 & \textbf{-0.34} & 0.06 & 0.17 & 0.07 \\ 
&$W_3$ & \textbf{0.38} & \textbf{0.31} & 0.13 & 0.48 & 0.13 & \textbf{0.32} & 0.07 & 0.28 & 0.08 & \textbf{0.32} & 0.06 & 0.22 & 0.06 \\ 
&$A^{(2)}$ & \textbf{0.08} & \textbf{0.56} & 0.12 & 0.40 & 0.45 & \textbf{0.57} & 0.07 & 0.28 & 0.26 & \textbf{0.58} & 0.05 & 0.22 & 0.21 \\ 
&$A^{(3)}$ & \textbf{0.02} & \textbf{0.06} & 0.13 & 0.44 & 0.52 & \textbf{0.08} & 0.07 & 0.32 & 0.33 & \textbf{0.07} & 0.06 & 0.26 & 0.26 \\ 
\bottomrule\bottomrule
\\[-1em]
\end{tabular}
\end{table}
\end{landscape}

\begin{table}[!htbp]
\centering
\label{table:3}
\caption{Coverage rates based on clustered standard errors for A-IPTW under four scenarios for three sample sizes. $Cov_{10},Cov_{30},Cov_{50}$ represent the coverage rates under 10, 30 and 50 studies with sample sizes 3000, 9000, 15000, respectively. The results summarized are over 1000 iterations. The four scenarios are as follows: Scenario 1 - both $Q$ and $g$ models are correct; Scenario 2 - $Q$ model is correct, $g$ is null; Scenario 3 - $Q$ model is null, $g$ model is correct; Scenario 4 - both $Q$ and $g$ models are null.}
 \begin{tabular}{c c c c c c c c}
\toprule\toprule
& &  \multicolumn{3}{c}{\makecell{Without random effects}}& \multicolumn{3}{c}{\makecell{With random effects}}
\\
\textbf{Scenarios} & \textbf{Var} & \textbf{$Cov_{10}$} & \textbf{$Cov_{30}$} & \textbf{$Cov_{50}$} & \textbf{$Cov_{10}$} & \textbf{$Cov_{30}$} & \textbf{$Cov_{50}$} \\
\midrule\midrule\\[-1.6em]
\multirow{5}{*}{Scenario 1} 
&$W_1$ &0.868 & 0.918 & 0.937 & 0.853 & 0.911 & 0.939 \\ 
&$W_2$ & 0.893 & 0.944 & 0.950 & 0.903 & 0.941 & 0.951 \\ 
 &$W_3$ & 0.911 & 0.951 & 0.940 & 0.913 & 0.940 & 0.948 \\ 
 &$A^{(2)}$ &  0.897 & 0.943 & 0.943 & 0.882 & 0.924 & 0.943 \\ 
 &$A^{(3)}$ &0.893 & 0.939 & 0.947 & 0.884 & 0.933 & 0.950 \\ 
\midrule\\[-1.6em]
\multirow{5}{*}{Scenario 2} 
&$W_1$ &0.913 & 0.977 & 0.977 & 0.747 & 0.856 & 0.881 \\ 
&$W_2$ &0.867 & 0.946 & 0.950 & 0.850 & 0.920 & 0.925 \\ 
&$W_3$ & 0.897 & 0.927 & 0.924 & 0.902 & 0.927 & 0.936 \\ 
 &$A^{(2)}$ &0.924 & 0.972 & 0.964 & 0.891 & 0.952 & 0.963 \\ 
  &$A^{(3)}$ &0.868 & 0.928 & 0.947 & 0.871 & 0.939 & 0.952 \\ 
\midrule\\[-1.6em]
\multirow{5}{*}{Scenario 3} 
&$W_1$ & 0.963 & 0.993 & 0.995 & 0.953 & 0.983 & 0.987 \\ 
&$W_2$ &0.955 & 0.973 & 0.962 & 0.951 & 0.968 & 0.956 \\ 
&$W_3$ &0.954 & 0.989 & 0.985 & 0.947 & 0.983 & 0.984 \\ 
 &$A^{(2)}$ &  0.910 & 0.944 & 0.954 & 0.904 & 0.950 & 0.960 \\ 
 &$A^{(3)}$ & 0.902 & 0.943 & 0.949 & 0.897 & 0.946 & 0.946 \\ 
\midrule\\[-1.6em]
\multirow{5}{*}{Scenario 4} 
&$W_1$ & 0.995 & 0.965 & 0.860 & 0.983 & 0.910 & 0.690 \\ 
&$W_2$ & 0.948 & 0.846 & 0.536 & 0.946 & 0.765 & 0.381 \\ 
&$W_3$ &1.000 & 1.000 & 1.000 & 1.000 & 1.000 & 1.000 \\ 
 &$A^{(2)}$ & 0.533 & 0.377 & 0.213 & 0.611 & 0.509 & 0.356 \\ 
 &$A^{(3)}$ & 0.905 & 0.942 & 0.938 & 0.903 & 0.939 & 0.943 \\ 
\bottomrule\bottomrule
\\[-1.6em]
\end{tabular}
\end{table}

 \begin{table}[!htbp]
\centering
\caption{Bias of TMLE and A-IPTW estimators with and without random effects in the simulation of the outcome. 10, 30 and 50 are the number of studies in each simulation. S1-4 represent the four scenarios.}
\label{table:bias of tmle and aiptw}
\resizebox{\columnwidth}{!}{%
  \begin{tabular}{c c |c c c| c c c | c c c| c c c}
\toprule
& &  \multicolumn{6}{c|}{\makecell{Without random effects}}& \multicolumn{6}{c}{\makecell{With random effects}}
\\\cmidrule{3-8}\cmidrule{9-14}
& &  \multicolumn{3}{c|}{\makecell{TMLE}}& \multicolumn{3}{c|}{\makecell{A-IPTW}}&  \multicolumn{3}{c|}{\makecell{TMLE}}& \multicolumn{3}{c}{\makecell{A-IPTW}}
\\\cmidrule{3-5}\cmidrule{6-8}\cmidrule{9-11}\cmidrule{12-14}
\textbf{Scen} & \textbf{Var}& \textbf{${10}$} & \textbf{${30}$} & \textbf{${50}$} & \textbf{${10}$} & \textbf{${30}$} & \textbf{${50}$} & \textbf{${10}$} & \textbf{${30}$} & \textbf{${50}$} & \textbf{${10}$} & \textbf{${30}$} & \textbf{${50}$} \\
\midrule
\multirow{5}{*}{S1} 
&$W_1$ &0.00 & 0.00 & 0.00 & -0.01 & 0.00 & 0.00 & -0.02 & -0.01 & -0.01 & -0.03 & -0.01 & -0.01 \\ 
&$W_2$ & -0.00 & 0.00 & -0.00 & -0.00 & 0.00 & -0.00 & -0.01 & 0.00 & 0.00 & -0.01 & 0.00 & 0.00 \\ 
 &$W_3$ & 0.00 & 0.00 & 0.00 & 0.00 & 0.00 & 0.00 & 0.00 & 0.00 & 0.00 & 0.00 & 0.00 & 0.00 \\ 
 &$A^{(2)}$ & 0.01 & -0.00 & -0.00 & 0.01 & -0.00 & -0.00 & -0.01 & 0.00 & 0.00 & -0.01 & 0.00 & 0.00 \\ 
 &$A^{(3)}$ & -0.00 & -0.00 & 0.00 & 0.00 & -0.00 & 0.01 & 0.01 & 0.00 & 0.01 & 0.01 & 0.00 & 0.01 \\ 
\midrule
\multirow{5}{*}{S2} 
&$W_1$ &0.00 & 0.00 & 0.00 & 0.00 & 0.00 & 0.00 & -0.02 & -0.01 & 0.00 & -0.02 & -0.01 & 0.00 \\ 
&$W_2$ &-0.00 & 0.00 & 0.00 & -0.00 & 0.00 & 0.00 & -0.01 & 0.00 & 0.00 & -0.01 & 0.00 & 0.00 \\ 
&$W_3$ & 0.00 & 0.00 & 0.00 & 0.00 & 0.00 & 0.00 & 0.00 & 0.00 & 0.00 & 0.00 & 0.00 & 0.00 \\ 
 &$A^{(2)}$ &0.00 & 0.00 & -0.00 & 0.00 & 0.00 & -0.00 & -0.01 & 0.00 & 0.00 & -0.01 & 0.00 & 0.00 \\ 
  &$A^{(3)}$ &0.00 & 0.00 & 0.00 & 0.00 & 0.00 & 0.00 & 0.00 & 0.00 & 0.00 & 0.00 & 0.00 & 0.00 \\ 
\midrule
\multirow{5}{*}{S3} 
&$W_1$ & 0.00 & 0.00 & 0.01 & -0.03 & -0.03 & -0.05 & -0.04 & -0.02 & -0.01 & -0.07 & -0.05 & -0.07 \\ 
&$W_2$ &-0.02 & -0.01 & -0.02 & 0.03 & -0.04 & -0.05 & -0.03 & -0.02 & -0.02 & 0.02 & -0.05 & -0.05 \\ 
&$W_3$ & -0.01 & 0.00 & 0.00 & -0.01 & 0.01 & -0.01 & -0.02 & 0.00 & 0.00 & -0.01 & 0.02 & 0.00 \\ 
 &$A^{(2)}$ & 0.08 & 0.04 & 0.03 & 0.06 & 0.04 & 0.05 & 0.07 & 0.04 & 0.02 & 0.05 & 0.03 & 0.05 \\ 
 &$A^{(3)}$ & 0.01 & 0.02 & 0.02 & 0.04 & 0.05 & 0.07 & 0.02 & 0.03 & 0.03 & 0.05 & 0.07 & 0.09 \\ 
\midrule
\multirow{5}{*}{S4} 
&$W_1$ & -0.03 & -0.04 & -0.04 & -0.41 & -0.42 & -0.42 & -0.13 & -0.14 & -0.14 & -0.52 & -0.53 & -0.53 \\ 
&$W_2$ & -0.05 & -0.05 & -0.06 & -0.31 & -0.31 & -0.32 & -0.08 & -0.09 & -0.09 & -0.35 & -0.35 & -0.36 \\ 
&$W_3$ &-0.04 & -0.04 & -0.04 & -0.06 & -0.06 & -0.06 & -0.05 & -0.04 & -0.04 & -0.07 & -0.06 & -0.06 \\
 &$A^{(2)}$ & 0.51 & 0.52 & 0.52 & 0.50 & 0.52 & 0.51 & 0.49 & 0.50 & 0.50 & 0.48 & 0.50 & 0.50 \\ 
 &$A^{(3)}$ &0.03 & 0.05 & 0.05 & 0.04 & 0.06 & 0.05 & 0.01 & 0.05 & 0.04 & 0.04 & 0.06 & 0.05 \\
\bottomrule
\end{tabular}}
\end{table}

 \begin{figure}[!htbp]
 \centering
 \includegraphics[width=16cm,height=18cm]{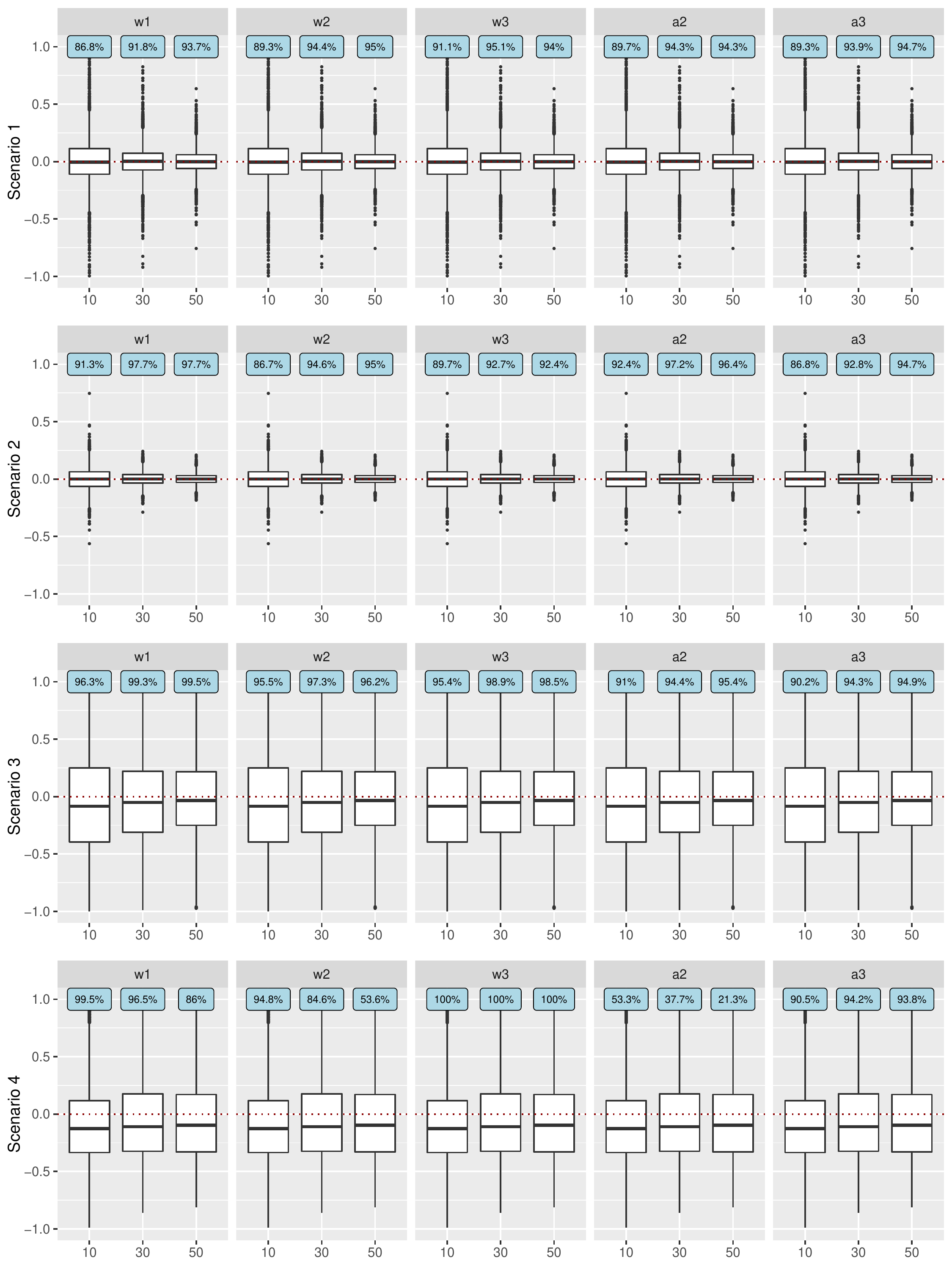}
 \caption{Estimated bias of A-IPTW estimates under four scenarios and three different sample sizes without random effects. $x$-axis represents the number of studies for three sample sizes. Coverage rates based on the clustered sandwich estimators of the standard error are presented in blue boxes. The four scenarios are as follows: Scenario 1 - both $Q$ and $g$ models are correct; Scenario 2 - $Q$ model is correct, $g$ is null; Scenario 3 - $Q$ model is null, $g$ model is correct; Scenario 4 - both $Q$ and $g$ models are null.}
 \end{figure}
 
 \begin{figure}[!htbp]
 \centering
 \includegraphics[width=16cm,height=18cm]{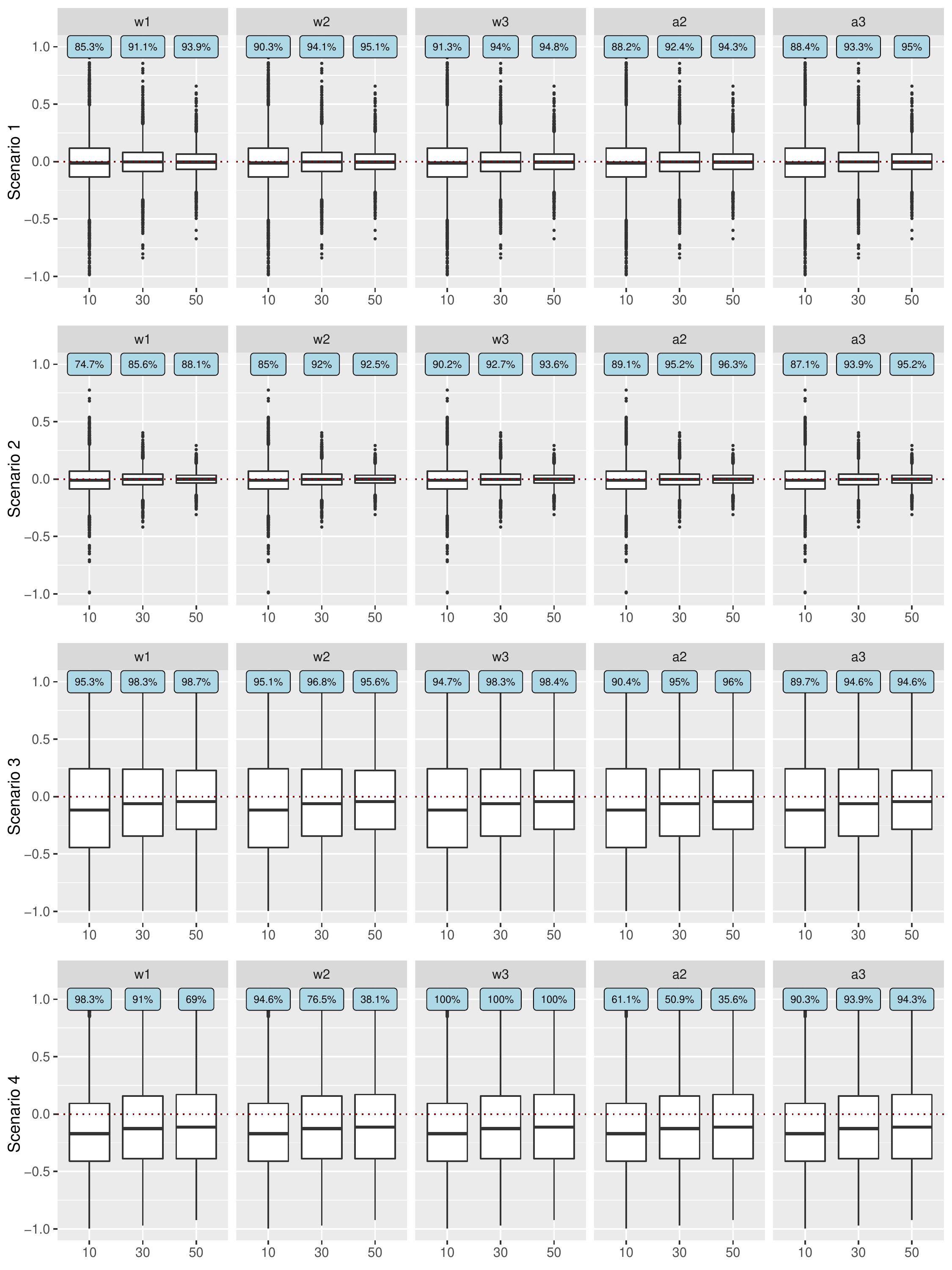}
 \caption{Estimated bias of A-IPTW estimates under four scenarios and three different sample sizes with random effects. $x$-axis represents the number of studies for three sample sizes. Coverage rates based on the clustered sandwich estimators of the standard error are presented in blue boxes. The four scenarios are as follows: Scenario 1 - both $Q$ and $g$ models are correct; Scenario 2 - $Q$ model is correct, $g$ is null; Scenario 3 - $Q$ model is null, $g$ model is correct; Scenario 4 - both $Q$ and $g$ models are null.}
 \end{figure}

 \newpage
 
 \section{Appendix F: MDR-TB results}
 \begin{table}[ht]
\centering
\caption{Sample sizes and the numbers of patients taking each of the 14 medications by study \cite{Ahuja}.}
\resizebox{\columnwidth}{!}{%
\begin{tabular}{lrrrrrrrrrrrrrrr}
  \toprule
  \toprule
 Study & Size & EMB & CAP & CIP & CS & ETO & OFX & PAS & PTO & RIF & SM & PZA & KM/AM & LgFQ & Gp5 \\ 
  \midrule
  \midrule
Ahuja & 722 & 589 &  247 & 346 & 319 & 350 & 270 & 143 & 0 & 619 & 149 & 622 & 193 & 33 & 117 \\ 
  Avendano & 66 & 31 &  0 & 1 & 24 & 13 & 7 & 27 & 0 & 0 & 2 & 24 & 61 & 59 & 66 \\ 
  Burgos & 48 & 48 &  28 & 9 & 29 & 17 & 24 & 3 & 0 & 47 & 28 & 47 & 10 & 8 & 5 \\ 
  Chan & 202 & 88 &  84 & 54 & 163 & 124 & 111 & 133 & 0 & 24 & 29 & 77 & 78 & 10 & 61 \\ 
  Chiang & 125 & 76 &  0 & 0 & 31 & 0 & 122 & 102 & 117 & 27 & 67 & 99 & 42 & 0 & 0 \\ 
  Cox & 77 & 28 &  63 & 0 & 77 & 72 & 77 & 77 & 0 & 2 & 2 & 66 & 22 & 0 & 48 \\ 
  Garcia & 41 & 41 &  0 & 0 & 0 & 0 & 0 & 0 & 0 & 41 & 2 & 41 & 0 & 0 & 0 \\ 
  Granich & 101 & 50 &  65 & 16 & 50 & 45 & 29 & 33 & 0 & 33 & 9 & 70 & 19 & 0 & 0 \\ 
  Koh & 149 & 47 &  1 & 1 & 142 & 0 & 6 & 101 & 138 & 0 & 40 & 80 & 55 & 139 & 56 \\ 
  Leung & 98 & 0 &  0 & 0 & 98 & 0 & 58 & 0 & 0 & 0 & 0 & 57 & 73 & 0 & 0 \\ 
  Migliori & 44 & 12 &  10 & 0 & 7 & 19 & 10 & 15 & 10 & 0 & 4 & 18 & 24 & 30 & 22 \\ 
  Mitnick & 710 & 191 &  349 & 488 & 615 & 478 & 123 & 597 & 232 & 9 & 146 & 279 & 347 & 235 & 557 \\ 
  Narita & 77 & 61 &  11 & 1 & 24 & 26 & 48 & 4 & 0 & 23 & 30 & 55 & 5 & 0 & 24 \\ 
  Palmero & 114 & 38 &  4 & 2 & 82 & 8 & 79 & 52 & 81 & 0 & 15 & 31 & 46 & 0 & 22 \\ 
  Pasvol & 43 & 27 &  0 & 9 & 8 & 0 & 16 & 1 & 12 & 5 & 2 & 29 & 13 & 5 & 2 \\ 
  Pena & 25 & 7 &  0 & 0 & 0 & 0 & 0 & 13 & 0 & 0 & 0 & 0 & 0 & 0 & 7 \\ 
  Perez & 34 & 28 & 0 & 17 & 0 & 0 & 7 & 0 & 9 & 0 & 16 & 17 & 13 & 0 & 24 \\ 
  Quy & 142 & 123 &  0 & 0 & 0 & 0 & 0 & 0 & 0 & 142 & 142 & 142 & 0 & 0 & 0 \\ 
  Riekstina & 1,026 & 407 &  474 & 0 & 888 & 0 & 962 & 552 & 777 & 0 & 26 & 314 & 601 & 6 & 691 \\ 
  Robert & 45 & 28 &  0 & 0 & 13 & 18 & 34 & 4 & 0 & 11 & 6 & 36 & 21 & 0 & 5 \\ 
  Schaaf & 33 & 28 &  0 & 0 & 1 & 28 & 29 & 0 & 0 & 10 & 11 & 29 & 9 & 0 & 9 \\ 
  Seung & 142 & 3 &  0 & 0 & 142 & 0 & 142 & 72 & 142 & 0 & 104 & 71 & 40 & 0 & 0 \\ 
  Shim & 1,364 & 188  & 0 & 0 & 1,199 & 0 & 1,113 & 887 & 1,186 & 149 & 319 & 610 & 477 & 37 & 45 \\ 
  Shin & 608 & 150 & 353 & 0 & 541 & 450 & 545 & 560 & 27 & 0 & 0 & 507 & 254 & 9 & 45 \\ 
  Shiraishi & 61 & 16 &  1 & 1 & 44 & 40 & 0 & 34 & 0 & 3 & 16 & 44 & 36 & 56 & 12 \\ 
  Tabarsi & 43 & 14 &  0 & 0 & 36 & 0 & 43 & 0 & 7 & 0 & 0 & 35 & 41 & 0 & 26 \\ 
  Tupasi & 170 & 45 &  28 & 19 & 160 & 40 & 122 & 140 & 151 & 9 & 71 & 123 & 119 & 72 & 79 \\ 
  Van der Werf & 43 & 39 &  0 & 4 & 5 & 1 & 36 & 15 & 0 & 19 & 0 & 36 & 33 & 0 & 34 \\ 
  Van der Walt& 2,182 & 1,642 & 0 & 0 & 476 & 2,182 & 2,182 & 0 & 0 & 0 & 0 & 2,182 & 2,182 & 0 & 0 \\ 
  Viiklepp & 284 & 62 &  150 & 0 & 263 & 0 & 237 & 211 & 240 & 44 & 41 & 253 & 115 & 0 & 84 \\ 
  Yim & 211 & 81 &  6 & 0 & 192 & 0 & 32 & 161 & 175 & 44 & 89 & 108 & 86 & 167 & 97 \\ 
  \midrule
  Total & 9,030 & 4,188 &  1,874 & 968 & 5,629 & 3,911 & 6,464 & 3,937 & 3,304 & 1,261 & 1,366 & 6,102 & 5,015 & 866 & 2,138 \\ 
   \bottomrule
   \bottomrule
\end{tabular}
}
\end{table}

\begin{landscape}
\begin{table}[htbp]
\centering
  \caption{Summary of proportions of patients with each combination of 6 individual level covariates and medications after removing the missing values. There are 9030 patients in total with 4 missing values for sex (0.3\%), 26 for age (14.4\%), 1304 for HIV (14.4\%), 2345 missing data for cavity (26\%), 519 for past TB (5.7\%) and 1405 for AFB (15.6\%). For each medication, the last two rows represent the number of studies (J) that observed the usage of the given medication and the total sample size corresponding to those J studies.}
  \resizebox{\columnwidth}{!}{%
    \begin{tabular}{llrrrrrrrrrrrrrrr}
    \\[-1.5em]
    \toprule
    \toprule
 \\[-1em]
 & & \textbf{n}&\textbf{EMB} &\textbf{CAP} & \textbf{CIP} & \textbf{CS} & \textbf{ETO} & \textbf{OFX} & \textbf{PAS} & \textbf{PTO} & \textbf{RIF} & \textbf{SM} & \textbf{PZA} & \textbf{KM/AM} & \textbf{LgFQ} & \textbf{Gp5} \\[0.2em]
   \midrule\\[-1em]
    \multirow{2}{*}{\textbf{Sex}} 
    & \textbf{male} &   6,147  & 0.660  & 0.699  & 0.640  & 0.698  & 0.653  & 0.696  & 0.705  & 0.716  & 0.697  & 0.685  & 0.676  & 0.666  & 0.579  & 0.679  \\[0.2em]
    & \textbf{female} &   2,879  & 0.340  &  0.301  & 0.360  & 0.302  & 0.347  & 0.304  & 0.295  & 0.284  & 0.303  & 0.315  & 0.324  & 0.334  & 0.421  & 0.321  \\[0.2em]
    \midrule\\[-1em]
   \multirow{5}{*}{\textbf{Age}}       
   & \textbf{0-17} & 154   & 0.018  & 0.019  & 0.039  & 0.017  & 0.020  & 0.013  & 0.019  & 0.014  & 0.023  & 0.029  & 0.017  & 0.013  & 0.025  & 0.027  \\[0.2em]
   & \textbf{18-25} &   1,283  & 0.129  & 0.157  & 0.186  & 0.153  & 0.170  & 0.136  & 0.164  & 0.137  & 0.059  & 0.119  & 0.137  & 0.161  & 0.232  & 0.175  \\[0.2em]
  & \textbf{26-45} &   4,892  & 0.597  &  0.563  & 0.597  & 0.519  & 0.608  & 0.560  & 0.509  & 0.470  & 0.620  & 0.539  & 0.576  & 0.565  & 0.483  & 0.531  \\[0.2em]
  & \textbf{46-60} &   1,974  & 0.198  & 0.207  & 0.137  & 0.231  & 0.174  & 0.220  & 0.226  & 0.266  & 0.201  & 0.217  & 0.207  & 0.206  & 0.192  & 0.209  \\[0.2em]
  & \textbf{61-91} &      701  & 0.057  & 0.055  & 0.041  & 0.081  & 0.028  & 0.071  & 0.081  & 0.113  & 0.096  & 0.096  & 0.064  & 0.056  & 0.068  & 0.057  \\[0.2em]
\midrule
\multirow{2}{*}{\textbf{HIV}} 
& \textbf{pos.} &   1,159  & 0.261  & 0.124  & 0.285  & 0.075  & 0.266  & 0.147  & 0.029  & 0.011  & 0.397  & 0.089  & 0.208  & 0.179  & 0.024  & 0.055  \\[0.2em]
 & \textbf{neg.} &   6,567  & 0.739  &  0.876  & 0.715  & 0.925  & 0.734  & 0.853  & 0.971  & 0.989  & 0.603  & 0.911  & 0.792  & 0.821  & 0.976  & 0.945  \\[0.2em]
\midrule\\[-1em]
\multirow{2}{*}{\textbf{Cavity} }     
& \textbf{yes} &   4,568  & 0.739  & 0.738  & 0.874  & 0.659  & 0.797  & 0.666  & 0.649  & 0.607  & 0.537  & 0.672  & 0.707  & 0.732  & 0.855  & 0.791  \\[0.2em]
& \textbf{no} &   2,117  & 0.261  & 0.262  & 0.126  & 0.341  & 0.203  & 0.334  & 0.351  & 0.393  & 0.463  & 0.328  & 0.293  & 0.268  & 0.145  & 0.209  \\[0.2em]
\midrule\\[-1em]
\multirow{2}{*}{\textbf{past TB}}
& \textbf{yes} &   6,380  & 0.713  & 0.732  & 0.612  & 0.756  & 0.847  & 0.796  & 0.789  & 0.740  & 0.309  & 0.680  & 0.750  & 0.819  & 0.747  & 0.755  \\[0.2em]
& \textbf{no} &   2,131  & 0.287  & 0.268  & 0.388  & 0.244  & 0.153  & 0.204  & 0.211  & 0.260  & 0.691  & 0.320  & 0.250  & 0.181  & 0.253  & 0.245  \\[0.2em]
\midrule\\[-1em]
\multirow{2}{*}{\textbf{AFB}} 
& \textbf{pos.} &   5,683  & 0.737  & 0.769  & 0.741  & 0.751  & 0.736  & 0.730  & 0.758  & 0.720  & 0.797  & 0.805  & 0.744  & 0.727  & 0.761  & 0.755  \\[0.2em]
& \textbf{neg.} &   1,942  & 0.263  & 0.231  & 0.259  & 0.249  & 0.264  & 0.270  & 0.242  & 0.280  & 0.203  & 0.195  & 0.256  & 0.273  & 0.239  & 0.245  \\[0.2em]
\midrule \\[-1em]
\textbf{Study} & \textbf{J} &   & 30 & 16 & 14 & 27 & 17 & 27 & 24 & 15 & 19 & 25 & 30 & 28 & 14 & 24\\[0.2em]
\midrule\\[-1em]
\textbf{Total} & \textbf{n} &   & 4,188 & 1,874 & 968 & 5,629 & 3,911 &  6,464 & 3,937 & 3,304 & 1,261 & 1,366 & 6,102 & 5,015 & 866 & 2,138 \\[0.2em]
\bottomrule
\bottomrule
\end{tabular}%
}
\vspace{0.2cm}
\label{tab:addlabel}%
\end{table}%
\end{landscape}

 \begin{landscape}
 \begin{table}[ht]
\fontsize{9.5pt}{10.25pt}\selectfont
\centering
\caption{Estimated effect modification coefficients and standard errors for the intercept and six covariates. Each column gives the results for a given medication. In each cell, the upper number represents the estimates, and the lower (in parentheses) is the standard error. }
\resizebox{\columnwidth}{!}{%
\begin{tabular}{lrrrrrrrrrrrrrr}
\\[-2em]
 \toprule
 \toprule
 & \textbf{EMB} & \textbf{CAP} & \textbf{CIP} & \textbf{CS} & \textbf{ETO} & \textbf{OFX} & \textbf{PAS} & \textbf{PTO} & \textbf{RIF} & \textbf{SM} & \textbf{PZA} & \textbf{KM/AM} & \textbf{LgFQ} & \textbf{Gp5} \\ 
  \midrule
  \textbf{$\beta_0$} & 0.013 & 0.488 & 0.072 & 0.232 & 0.081 & 0.257 & -0.039 & -0.045 & -0.432 & 0.124 & 0.000 & 0.443 & 0.529 & 0.212 \\ 
   & (0.159) & (0.208) & (0.119) & (0.143) & (0.109) & (0.177) & (0.131) & (0.185) & (0.287) & (0.182) & (0.099) & (0.176) & (1.190) & (0.123) \\ 
\textbf{Age} & -0.015 & -0.020 & 0.001 & -0.002 & -0.025 & 0.022 & -0.013 & 0.008 & -0.011 & 0.005 & 0.009 & 0.014 & -0.105 & -0.013 \\ 
   & (0.022) & (0.017) & (0.025) & (0.014) & (0.021) & (0.016) & (0.012) & (0.022) & (0.032) & (0.031) & (0.011) & (0.033) & (0.213) & (0.032) \\ 
\textbf{Sex} & 0.095 & -0.012 & 0.052 & 0.026 & 0.020 & -0.014 & 0.017 & -0.037 & 0.137 & -0.133 & -0.001 & -0.031 & -0.114 & 0.010 \\ 
  & (0.032) & (0.047) & (0.036) & (0.017) & (0.042) & (0.032) & (0.027) & (0.021) & (0.185) & (0.059) & (0.034) & (0.055) & (0.386) & (0.039) \\ 
  \textbf{Afb} & -0.064 & -0.026 & -0.024 & -0.002 & -0.052 & -0.041 & 0.001 & 0.073 & 0.119 & 0.089 & -0.049 & 0.045 & -0.224 & 0.044 \\ 
   & (0.037) & (0.063) & (0.041) & (0.034) & (0.051) & (0.048) & (0.045) & (0.031) & (0.138) & (0.074) & (0.032) & (0.073) & (0.476) & (0.064) \\ 
\textbf{Cavity} &-0.085 & -0.145 & 0.023 & 0.013 & -0.066 & 0.077 & -0.004 & 0.067 & 0.013 & 0.059 & 0.020 & -0.108 & 0.081 & -0.070 \\ 
   & (0.049) & (0.069) & (0.071) & (0.052) & (0.056) & (0.056) & (0.038) & (0.063) & (0.086) & (0.060) & (0.031) & (0.086) & (0.616) & (0.058) \\ 
\textbf{HIV} & -0.007 & -0.243 & -0.270 & 0.025 & 0.130 & 0.050 & 0.016 & 0.104 & -0.125 & -0.151 & 0.130 & 0.091 & -0.208 & -0.135 \\ 
  & (0.046) & (0.095) & (0.116) & (0.046) & (0.183) & (0.198) & (0.095) & (0.105) & (0.078) & (0.120) & (0.102) & (0.132) & (0.915) & (0.127) \\ 
\textbf{pastTB} & 0.029 & 0.014 & 0.034 & -0.090 & 0.068 & 0.044 & 0.042 & -0.107 & 0.090 & 0.081 & 0.050 & 0.003 & 0.172 & -0.014 \\ 
   &(0.062) & (0.086) & (0.053) & (0.04) & (0.042) & (0.067) & (0.043) & (0.049) & (0.112) & (0.076) & (0.029) & (0.064) & (0.796) & (0.048) \\ 
\bottomrule
\end{tabular}
}
\end{table}
\end{landscape}

 \begin{landscape}
 \begin{table}[ht]
\fontsize{9.5pt}{10.25pt}\selectfont
\centering
\caption{Estimated effect modification coefficients and standard errors. Each column represents the effect modification of the named medication and each row represents the potential effect modifier. In each cell, the upper value represents the estimate, and the lower (in parentheses) is the standard error. }
\resizebox{\columnwidth}{!}{%
\begin{tabular}{lrrrrrrrrrrrrrr}
\\[-2em]
 \toprule
 \toprule
 & \textbf{EMB} & \textbf{CAP} & \textbf{CIP} & \textbf{CS} & \textbf{ETO} & \textbf{OFX} & \textbf{PAS} & \textbf{PTO} & \textbf{RIF} & \textbf{SM} & \textbf{PZA} & \textbf{KM/AM} & \textbf{LgFQ} & \textbf{Gp5} \\ 
  \midrule
\textbf{EMB} & - & -0.025 & 0.094 & -0.031 & -0.051 & -0.031 & 0.083 & 0.103 & 0.075 & 0.165 & -0.058 & -0.115 & -0.099 & 0.079 \\ 
   & - & (0.052) & (0.062) & (0.028) & (0.066) & (0.057) & (0.053) & (0.035) & (0.096) & (0.066) & (0.047) & (0.062) & (0.300) & (0.063) \\ 
\textbf{CAP} & -0.021 & - & -0.164 & -0.052 & -0.242 & -0.031 & -0.049 & -0.006 & 0.071 & -0.093 & -0.053 & -0.059 & -0.261 & -0.107 \\ 
 & (0.065) & - & (0.054) & (0.048) & (0.052) & (0.105) & (0.055) & (0.101) & (0.168) & (0.109) & (0.039) & (0.111) & (0.578) & (0.056) \\ 
\textbf{CIP} &-0.012 & -0.017 &- & 0.019 & 0.112 & 0.109 & 0.104 & 0.058 & -0.235 & -0.127 & 0.022 & -0.137 & 0.277 & -0.071 \\ 
   & (0.052) & (0.159) & - & (0.054) & (0.071) & (0.258) & (0.114) & (0.069) & (0.197) & (0.138) & (0.049) & (0.190) & (1.684) & (0.108) \\ 
 \textbf{CS} & -0.063 & -0.064 & 0.011 & - & 0.148 & -0.051 & -0.131 & -0.127 & -0.054 & -0.183 & 0.092 & -0.018 & -0.303 & -0.051 \\ 
  & (0.053) & (0.067) & (0.054) & - & (0.045) & (0.069) & (0.063) & (0.054) & (0.077) & (0.086) & (0.057) & (0.147) & (0.291) & (0.067) \\ 
\textbf{ETO} & 0.041 & -0.101 & 0.006 & 0.089 & - & -0.182 & 0.018 & 0.103 & -0.113 & 0.080 & -0.048 & -0.188 & -0.052 & -0.003 \\ 
  & (0.049) & (0.086) & (0.078) & (0.042) & - & (0.097) & (0.061) & (0.069) & (0.160) & (0.065) & (0.061) & (0.129) & (0.499) & (0.068) \\ 
\textbf{OFX} & -0.009 & -0.021 & 0.078 & -0.017 & 0.001 & - & 0.101 & -0.007 & 0.131 & -0.080 & -0.052 & -0.210 & -0.153 & 0.098 \\ 
  & (0.059) & (0.136) & (0.074) & (0.053) & (0.055) & - & (0.074) & (0.047) & (0.332) & (0.200) & (0.038) & (0.202) & (1.234) & (0.095) \\
\textbf{PAS} & 0.005 & -0.026 & 0.087 & -0.050 & 0.030 & 0.051 & - & 0.002 & 0.185 & 0.001 & -0.012 & 0.030 & 0.250 & -0.087 \\ 
  & (0.047) & (0.055) & (0.040) & (0.037) & (0.038) & (0.062) & - & (0.045) & (0.073) & (0.060) & (0.033) & (0.139) & (0.519) & (0.073) \\ 
\textbf{PTO} & 0.096 & 0.027 & 0.159 & -0.004 & 0.128 & -0.101 & 0.101 & - & -0.036 & -0.050 & -0.027 & -0.157 & 0.014 & 0.028 \\ 
   & (0.083) & (0.090) & (0.070) & (0.059) & (0.062) & (0.102) & (0.059) & - & (0.183) & (0.074) & (0.037) & (0.093) & (0.528) & (0.068) \\ 
   \textbf{RIF} &0.025 & 0.069 & -0.064 & -0.126 & -0.027 & -0.098 & 0.005 & -0.013 & -& 0.009 & 0.072 & -0.054 & 0.192 & -0.073 \\ 
   & (0.076) & (0.090) & (0.084) & (0.065) & (0.087) & (0.125) & (0.070) & (0.095) & - & (0.148) & (0.057) & (0.126) & (1.654) & (0.068) \\ 
   \textbf{SM} & 0.108 & -0.088 & -0.085 & -0.236 & -0.101 & -0.045 & -0.062 & -0.034 & -0.242 & - & 0.058 & -0.032 & -0.154 & -0.030 \\ 
   &(0.078) & (0.089) & (0.048) & (0.037) & (0.068) & (0.051) & (0.043) & (0.062) & (0.166) & - & (0.045) & (0.170) & (1.134) & (0.069) \\ 
   \textbf{PZA} & -0.036 & -0.066 & 0.101 & -0.044 & 0.029 & -0.069 & -0.047 & -0.080 & 0.174 & 0.116 & - & 0.043 & 0.000 & -0.013 \\ 
   & (0.065) & (0.092) & (0.047) & (0.036) & (0.051) & (0.055) & (0.054) & (0.059) & (0.115) & (0.052) & - & (0.131) & (0.291) & (0.041) \\ 
   \textbf{KM/AM} &-0.001 & -0.107 & -0.158 & -0.058 & -0.166 & -0.079 & -0.029 & 0.064 & -0.218 & -0.14 & -0.028 & - & -0.017 & -0.092 \\ 
   &  (0.064) & (0.083) & (0.061) & (0.031) & (0.044) & (0.079) & (0.046) & (0.042) & (0.302) & (0.100) & (0.042) & - & (0.783) & (0.078) \\ 
\textbf{LgFQ} & -0.026 & -0.273 & 0.016 & -0.103 & -0.013 & -0.011 & 0.036 & -0.001 & 0.036 & 0.019 & 0.051 & -0.232 & - & -0.127 \\ 
   & (0.059) & (0.272) & (0.075) & (0.064) & (0.059) & (0.327) & (0.056) & (0.076) & (0.674) & (0.16) & (0.033) & (0.173) & - & (0.088) \\ 
\textbf{Gp5} & 0.037 & -0.121 & -0.129 & 0.012 & -0.107 & 0.037 & -0.079 & 0.049 & 0.048 & -0.088 & -0.021 & 0.090 & -0.331 & - \\ 
  & (0.054) & (0.071) & (0.058) & (0.062) & (0.045) & (0.062) & (0.057) & (0.069) & (0.125) & (0.099) & (0.044) & (0.113) & (0.679) & - \\ 
\bottomrule
\end{tabular}
}
\end{table}
\end{landscape}

  \begin{table}[ht]
\centering
\caption{Summary of the estimated $Pr(A=1|\boldsymbol{X})$ for each of the 14 medications}
\resizebox{0.95\columnwidth}{!}{%
\begin{tabular}{lrrrrr}
\\[-0.7em]
\toprule
  \toprule
 & Minimum & 5\% Quantile & Median & 95\% Quantile & Maximum \\ 
  \midrule
\textbf{EMB} & 0.00576 & 0.02901 & 0.43321 & 0.92581 & 0.96728 \\ [0.2em]
  \textbf{CAP} & 0.00101 & 0.01462 & 0.09539 & 0.44997 & 0.51206 \\ [0.2em]
  \textbf{CIP} & 0.00022 & 0.00551 & 0.06477 & 0.38125 & 0.45060 \\ [0.2em]
  \textbf{CS} & 0.00525 & 0.08334 & 0.67773 & 0.86341 & 0.87079 \\ [0.2em]
  \textbf{ETO} & 0.00166 & 0.03193 & 0.32798 & 0.54188 & 0.54752 \\ [0.2em]
  \textbf{OFX} & 0.00058 & 0.06633 & 0.77979 & 0.85201 & 0.86978 \\ [0.2em]
  \textbf{PAS} & 0.00287 & 0.05702 & 0.42128 & 0.77329 & 0.86634 \\ [0.2em]
  \textbf{PTO} & 0.00003 & 0.00182 & 0.23199 & 0.47145 & 0.48373 \\ [0.2em]
  \textbf{RIF} & 0.00020 & 0.00333 & 0.05279 & 0.55159 & 0.74471 \\ [0.2em]
  \textbf{SM} & 0.00024 & 0.00680 & 0.06327 & 0.53681 & 0.80091 \\ [0.2em]
  \textbf{PZA} & 0.02318 & 0.11876 & 0.74804 & 0.95482 & 0.96586 \\ [0.2em]
  \textbf{KM/AM} & 0.00116 & 0.03656 & 0.54369 & 0.88310 & 0.89926 \\ [0.2em]
  \textbf{LgFQ} & 0.00001 & 0.00069 & 0.03262 & 0.35000 & 0.60993 \\ [0.2em]
  \textbf{Gp5} & 0.00182 & 0.02494 & 0.25323 & 0.65496 & 0.77199 \\ [0.2em]
   \bottomrule
   \bottomrule
\end{tabular}}
\vspace{0.3cm}
\end{table}

 \begin{figure}[!htbp]
 \centering
 \includegraphics[width=14cm,height=18cm]{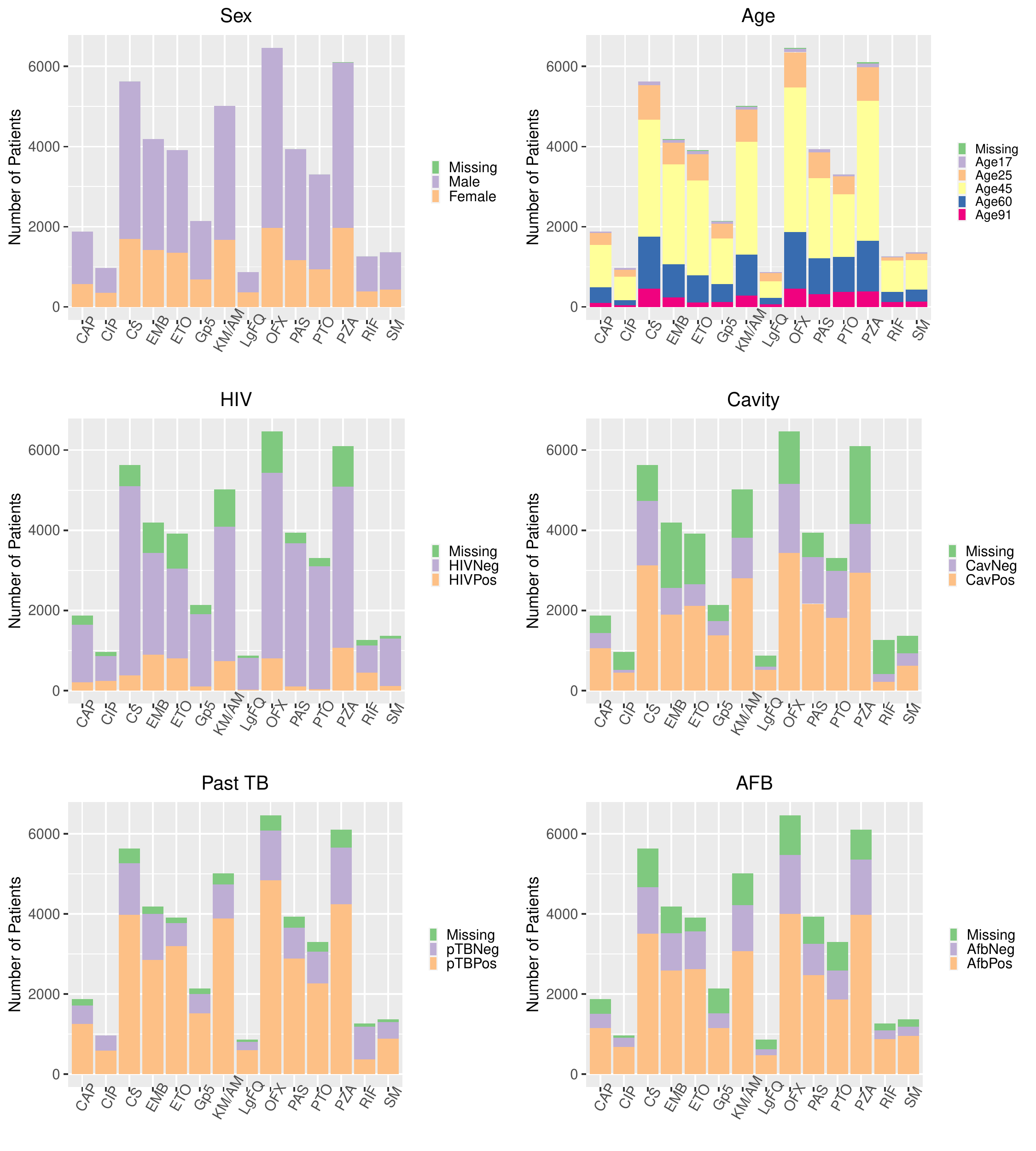}
 \caption{Summary of numbers of patients with combinations among 6 covariates and 14 medications. Age17 represents 0-17 year old group; Age25 represents the 18-25 year old group; Age45 represents the 26-45 year old group; Age60 represents the 46-60 year old group; Age91 represents the 61-91 year old group.}
 \end{figure}
 
 \begin{figure}[!htbp]
 \centering
 \includegraphics[width=16cm,height=20cm]{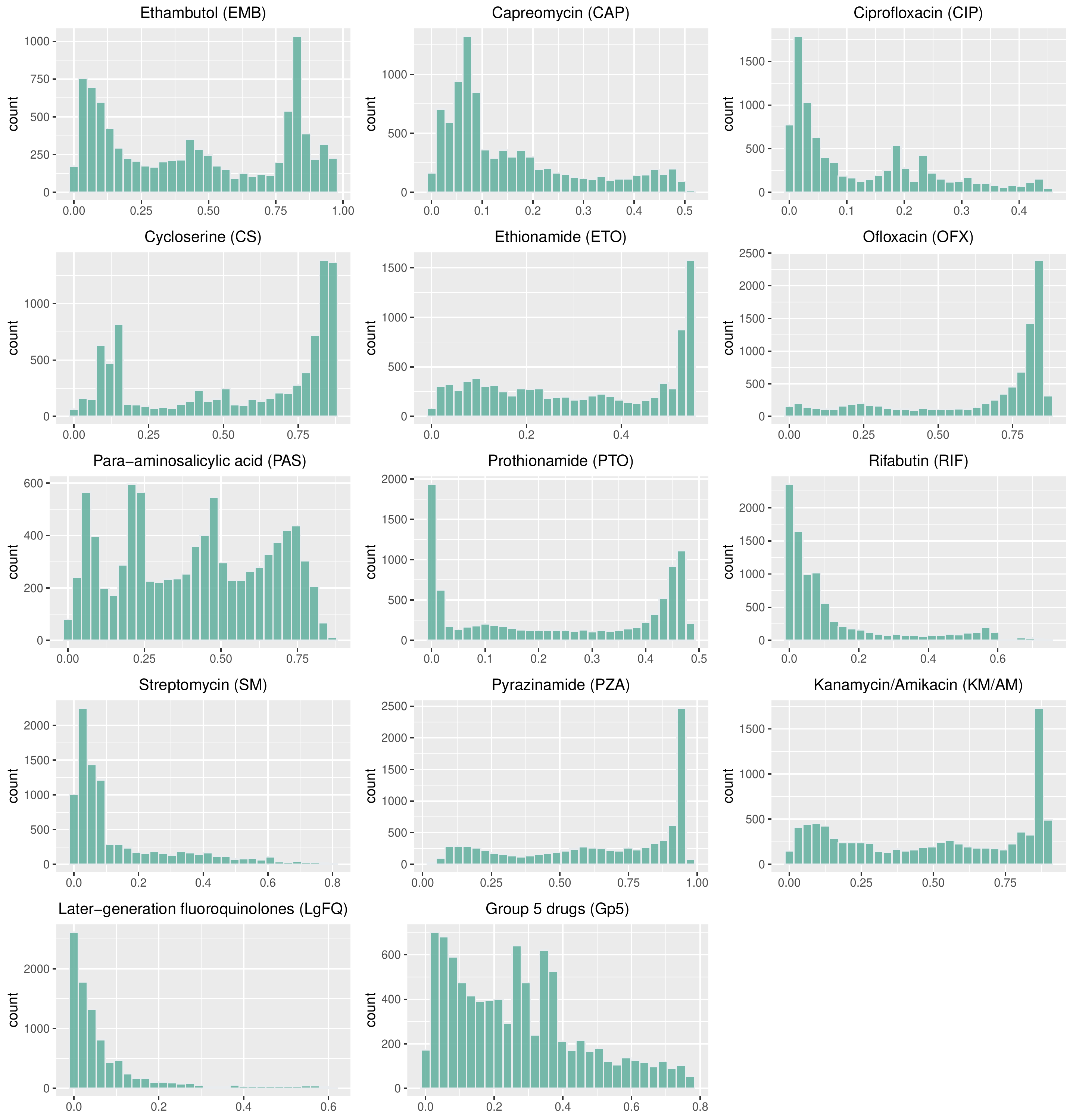}
 \caption{Distribution of propensity scores for 14 medications}
 \end{figure}
 
  \section{Appendix G: Multiple testing}
 
In a multiple testing situation, there are a variety of criteria that may be considered. The classic approach to the multiple comparison problem is to control the familywise error rate (FWER) which is the probability of making at least one false discovery. The Bonferroni correction is one of the most common ways to control the FWER. We can find the critical value $\alpha$ for an individual test through dividing the FWER (usually 0.05) by the number of tests. The Bonferroni correction tends to be a bit too conservative. It is mainly useful for a fairly small number of multiple comparisons.

An alternative approach is to control the expected proportion of falsely rejected hypotheses which is also called false discovery rate (FDR). It was developed and demonstrated to be a simple and powerful procedure by Benjamini and Hochberg (1995) \cite{fdr}. Compared to FWER, the error rate is equivalent when all hypotheses are true but is smaller otherwise. To perform a FDR test, we need to order the individual $p$ values from the smallest to the largest. If there are $n$ total number of tests, the smallest $p$ has a rank of $i=1$ and the largest has $i=n$. Then the next step is to compare each individual $p$ value to its Benjamini-Hochberg critical value, $(i/n)q$, where $q$ is the false discovery rate we choose. The largest $p$ value that has $p<(i/n)q$ is significant, and all of the $p$ values smaller than it are also significant, even those that are not less than their Benjamini-Hochberg critical value. There is another way to look at this method. We can calculate $q$ which is an estimate of FDR by $q=p(n/i)$ where $p$ is the $i$-th smallest $p$ value out of $n$ total $p$ values. The $q$ value equation can be treated as the expected false positives divided by the total number of positives actually accepted at that same $p$ value. Then we can use the $q$ value much like a $p$ value. For example, if we choose $q=0.05$, that means we expect that $5\%$ or less of our accepted results will be false. However, there is a problem when $q$ is not a monotonic function. To address this concern, Yekutieli and Benjamini introduced the FDR-adjustment method in which $q$ was replaced with the smallest value among all lower-rank $q$ values that we calculated then monotonicity is enforced \cite{fdr_adj}.\\

%% file: Sections/Introduction.tex
\textit{Precision medicine}, also known as personalized medicine, refers to treating patients according to their unique characteristics. \textit{Dynamic treatment regimes (DTRs)}, as a statistical framework for precision medicine, provide individualized treatment recommendations based on patients' individual information. Recently, the consideration of \textit{interference}, where one individual’s outcome is possibly affected by others’ treatment, has gained importance in estimating \textit{optimal} DTRs\cite{Jiang2022DTRint,su2019modelling, sherman2020general}, which are sequences of treatment rules that yield the best-expected health outcome across a population.

Recently, some researchers, such as Su et al. (2019)\cite{su2019modelling}, Jiang et al. (2022a)\cite{Jiang2022DTRint} and Park et
al. (2021)\cite{park2021optimal}, have focused on optimal DTR estimation in the presence of interference. In such cases, treatment-decision rules should involve others' information such as treatments and covariates. To conduct robust optimal DTR estimation with interference for continuous outcomes in the regression-based estimation framework,  Jiang et al. (2022a)\cite{Jiang2022DTRint} developed network balancing weights to extend the method of dynamic weighted ordinary least squares (dWOLS, Wallace and Moodie (2015) \cite{wallace2015doubly}). This method focused on a decision framework in cases where there is an ego (i.e., an individual of primary interest in a social network) and alters (i.e., those to whom the ego is linked). The covariates or treatments of the alters could affect the treatment or outcome of the ego, and the goal is to optimize the mean of the outcome of egos in the network. These recently developed interference-aware DTR estimation methods, and even many of the interference-unaware DTR estimation methods, however, focus primarily on continuous outcomes. Few publications have considered optimal DTR estimation for discrete outcomes, such as binary and ordinal outcomes, in the presence of interference. 

In order to estimate optimal DTRs with discrete outcomes, some methods have been developed without interference. Moodie et al. (2014)\cite{moodie2014q} first implemented more flexible modeling by adapting $Q$-learning to discrete utilities, such as Bernoulli and Poisson utilities.  Investigating discrete outcomes, Wallace et al. (2019)\cite{wallace2019model} introduced an extension of G-estimation to the case of non-additive treatment effects. Building on dWOLS, Simoneau et al. (2020)\cite{simoneau2020estimating} extended dWOLS to time-to-event data and developed DWSurv to determine the optimal DTR with right-censored survival outcomes. Further, focusing on discrete outcomes and particularly on binary outcomes, Jiang et al. (2022b)\cite{jiang2022doublyrobust} proposed a dynamic weighted generalized linear model in a multi-stage treatment decision analysis, employing two-step weighted logistic regression at each stage for binary outcomes. 

The methods developed in this paper are also motivated using data from the Population Assessment of Tobacco and Health (PATH) Study, a longitudinal study of smoking behaviours and cessation. Some studies have focused on the idea that a desire to quit smoking alone may not be sufficient motivation in itself \cite{hubbard2016systematic, foulstone2017partner}. Meanwhile, a  growing body of literature suggests that e-cigarettes, such as vaping, can be useful as a cessation aid\cite{benmarhnia2018can, hajek2019randomized}. Few studies have explored smoking cessation within couples or households, where interference may be present, and even fewer have examined the impact of participants' e-cigarette usage. However, the PATH study provides a unique chance to investigate these contexts. Motivated by this, in contrast to  Jiang et al.'s approaches to optimizing individual outcomes, our proposed framework for modeling household interference focuses on optimizing household utilities by making decisions for the household as a whole. In particular, we explore optimizing an ordinal utility across a household, combining a couple's two binary outcomes of quitting (or attempting to quit) smoking into a single ordinal outcome.

This paper is organized as follows. In Section \ref{Sec2}, we introduce the proposed approximately doubly robust regression-based DTR estimation framework for ordinal household utilities under household interference. Then, to achieve approximate double robustness in the face of model misspecification, we propose the estimation process of the joint propensity scores and construct the corresponding balancing weights. Through simulations of both single- and multi-stage treatment decisions, Section \ref{Sec3} demonstrates that our method is approximately doubly robust against misspecification of either the treatment-free or the joint propensity score model. Section \ref{Sec4} illustrates the implementation of our methods on PATH data. Section 5 concludes with a discussion of future research.

%% file: Sections/Methodology.tex
\subsection{Household Interference Modeling Framework with Ordinal Utilities} \label{C5sec3.1}
In the presence of household interference, we aim to estimate treatment decisions for both individuals in the same household, so the outcomes of interest should be related to both individuals of a couple in the same household \citep{lewis2006understanding}; thus, both covariates and treatments of individuals in the same household need to be considered in a household outcome model. First, we define the household utility function as a combination of the individuals' outcomes in the same household. For example, for a pair $(s, r)$, we may have the utility that $U(Y^{ s},\ Y^ {r})$ is equivalent to $\omega_sY^{s} +  \omega_rY^ {r}$, where the combination weights ($\omega_s$ and $\omega_r$) can be set based on the specific analytical goals.

For instance, we might set $\omega_s = \omega_r$ if outcomes of both $s$ and $r$ are considered equally of interest. Alternatively, we may instead consider a case where one individual of primary interest - the ego - is the sole focus of our optimization, but may be influenced by the treatments of their neighbour (the alter(s)). In this case, we would therefore set a weight of $1$ to the ego and $0$ to the alters.

For the binary outcome pairs $(Y^s, Y^r)$, where $(Y^s, Y^r) \in \{(0,0), (0,1), (1,0), (1,1) \}$, for simplicity and the goal of studying DTR with ordinal outcomes, we will specify that all the combination weights are equal to one ($\omega_s = \omega_r = 1$). Adding 1 to each sum, there are three possibilities 1, 2, or 3 of $U(Y^{s},\ Y^ {r})$ for a pair in the same household, and these can be considered ordinal outcomes for the household. That is, in this setting of a household's utility, the utilities of households, $U(Y^{s},\ Y^ {r}) = 1,2,3$, can be interpreted in an ordered way: for a pair in a household, (1) neither, (2) one, or (3) both of them incur a benefit such as smoking cessation, and the largest value (i.e., 3) is preferred. We consider that the model for such a household utility can be captured in the form of a function of
\begin{equation} \label{eq:intoutcome}
 f(\boldsymbol{x}^{\beta}) +  d_{\xi}(a^{s},\boldsymbol{x}^{\xi}) + d_{\psi}(a^{r},\boldsymbol{x}^{\psi}) + d_{int}(a^{s} a^{r},\boldsymbol{x}^{\phi}),
\end{equation}
where $\boldsymbol{x}^{\beta}$, often termed predictive variables, function to increase the precision of estimates, and $\boldsymbol{x}^{\xi}$, $\boldsymbol{x}^{\psi}$, $\boldsymbol{x}^{\phi}$, the so-called prescriptive or tailoring variables, are used to adapt treatment decisions to pairs in a household. That is, the model has a treatment-free function $f(\boldsymbol{x}^{\beta})$ and some decision functions $d_{\xi}(a^{s},\boldsymbol{x}^{\xi})$, $d_{\psi}(a^{r},\boldsymbol{x}^{\psi}) $, and $ d_{int}(a^{s} a^{r},\boldsymbol{x}^{\phi})$.

In practice, in the household-level model (\ref{eq:intoutcome}), covariates $\boldsymbol{x}^{\xi}$ and $\boldsymbol{x}^{\psi}$ can be individual-level covariates from each individual in the same household. These two covariates that contain individuals' characteristics indicate the ``personalized'' side of the model (\ref{eq:intoutcome}). Covariates $\boldsymbol{x}^{\phi}$ can be household-level covariates, and thus they represent households' characteristics. As such $\boldsymbol{x}^{\phi}$ are special tailoring variables for our household-interference treatment decisions case. Given the above utility model, the goal is to identify an \textit{optimal household treatment decision rule} $d^*(\boldsymbol{x}_s, \boldsymbol{x}_r)$ that maximizes the utility $U(Y^{s},\ Y^ {r})$, for binary outcomes $Y^s$ and $Y^r$. In our household case, the treatment decision rule $d(\boldsymbol{x}_s, \boldsymbol{x}_r)$ takes as input both individuals' covariates and outputs a treatment configuration for a couple in the same household.
\subsection{Proportional Odds Model and Target Decision Parameters} 
Let $U$ be an ordinal outcome with $C= 3$ categories. Then $\mathbb{P}(U \leq c)$ is the cumulative probability that $U$ is less than or equal to a specific category $c$. The log-odds of being less than or equal to a particular $c$ category can be defined as
\begin{equation*}
    log \frac{\mathbb{P}(U \leq c)}{\mathbb{P}(U > c)}= logit[\mathbb{P}(U \leq c)], \ \ \ \ \text{for}\  \ c = 1,..., C-1, 
\end{equation*}
where the logit link function is defined as $logit(p) = p/(1-p)$. Note that the denominator $\mathbb{P}(U > c)$ of the above equation will be zero if $c = C$; thus, $c = 1,..., C-1$. The \textit{proportional odds model} (POM) that specifies the cumulative log-odds for a particular category assumes that each explanatory variable exerts the same effect on each cumulative logit
regardless of the cutoff $c$, and is proposed by  McCullagh (1980)\cite{mccullagh1980regression} to be $logit[\mathbb{P}(U_h \leq c \mid \boldsymbol{x}_h)] = \zeta_c - \boldsymbol{\theta}^{\top}\boldsymbol{x}_h $, where coefficients $\zeta_c$ are category-specific intercepts and $\boldsymbol{\theta}$ are coefficients of covariates $\boldsymbol{x}_h$. The intercepts $\zeta_c$ are the only part that varies across the equations, and the effects of covariates $\boldsymbol{x}_h$ are assumed to be constant for all $c$, i.e., $\boldsymbol{\theta}_c = \boldsymbol{\theta}$. Building on the typical POM and our treatment decision set-up, we propose a proportional odds model for household ordinal utilities as follows, for $c = 1,2; \ \  h = 1,2,...H,$
\begin{flalign}\label{POMpro}
    logit[\mathbb{P}(U_h \leq c \mid a_h^{s} ,a_h^{r}, \boldsymbol{x}_h)] = \zeta_c - \boldsymbol{\beta}^{\top}\boldsymbol{x}^{\beta}_h -a_h^{s}  \boldsymbol{\xi}^{\top}\boldsymbol{x}^{\xi}_h - a_h^{r}\boldsymbol{\psi}^{\top}\boldsymbol{x}^{\psi}_h- a_h^{s} a_h^{r}\boldsymbol{\phi}^{\top}\boldsymbol{x}^{\phi}_h.
\end{flalign} 
According to the general utility model (\ref{eq:intoutcome}), we note that the treatment-free functions are identified in the above POM model as a linear form $f(\boldsymbol{x}^{\beta}) = \zeta_c - \boldsymbol{\beta}^{\top}\boldsymbol{x}^{\beta}$, and the decision functions are specified as $d_{\xi}(a^{s},\boldsymbol{x}^{\xi}) = -a^{s}  \boldsymbol{\xi}^{\top}\boldsymbol{x}^{\xi}$, $d_{\psi}(a^{r},\boldsymbol{x}^{\psi})= -a^{r}\boldsymbol{\psi}^{\top}\boldsymbol{x}^{\psi} $ and $ d_{int}(a^{s} a^{r},\boldsymbol{x}^{\phi}) = -a^{s} a^{r}\boldsymbol{\phi}^{\top}\boldsymbol{x}^{\phi}$, respectively. 

Focussing on the household ordinal outcome (\ref{POMpro}), we define the \textit{household blip function} as $\gamma[(A^{s} ,A^{r}),\boldsymbol{x}_h; \boldsymbol{\xi}, \boldsymbol{\psi}, \boldsymbol{\phi}] = A^{s}  \boldsymbol{\xi}^{\top}\boldsymbol{x}^{\xi}_h + A^{r}\boldsymbol{\psi}^{\top}\boldsymbol{x}^{\psi}_h+ A^{s} A^{r}\boldsymbol{\phi}^{\top}\boldsymbol{x}^{\phi}_h$, which represents the effects of the treatment configuration $(A^s, A^r)$ for a household compared with the null treatment configuration $(0, 0)$.  The estimation goal is to estimate target decision parameters, i.e., the blip parameters $\boldsymbol{\xi}$, $\boldsymbol{\psi}$, $\boldsymbol{\phi}$. From these blip-parameter estimates and given the household tailoring variables, the optimal treatment decisions for a pair in the household can be made. Given the four choices of $(A^s, A^r ) = (1,1),(1,0),(0,1)\ \text{or}\ (0,0)$, the corresponding blip value $\gamma[(A^{s} ,A^{r}),\boldsymbol{x}_h; \boldsymbol{\xi}, \boldsymbol{\psi}, \boldsymbol{\phi}]$ is $\boldsymbol{\xi}^{\top}\boldsymbol{x}^{\xi}_h + \boldsymbol{\psi}^{\top}\boldsymbol{x}^{\psi}_h + \boldsymbol{\phi}^{\top}\boldsymbol{x}^{\phi}_h$, $\boldsymbol{\xi}^{\top}\boldsymbol{x}^{\xi}_h$, $\boldsymbol{\psi}^{\top}\boldsymbol{x}^{\psi}_h$, and $0$, respectively. The decision goal is to maximize the outcome across a couple, which is equivalent to maximising the blip function.  Taking into account the blip values of all possible treatment configurations, an optimal treatment rule must choose the configuration that corresponds to the maximum blip value. Therefore, we have the following treatment decision rules for a household: \begin{decision}
The optimal household decision rules:\label{ruleHH}\\
 \textbf{Rule 1:} $d^*(\boldsymbol{x}^{\xi}, \boldsymbol{x}^{\psi}, \boldsymbol{x}^{\phi})= (1,1),$ if $\boldsymbol{\xi}^{\top}\boldsymbol{x}^{\xi}_h + \boldsymbol{\psi}^{\top}\boldsymbol{x}^{\psi}_h + \boldsymbol{\phi}^{\top}\boldsymbol{x}^{\phi}_h >0$ and $\boldsymbol{\psi}^{\top}\boldsymbol{x}^{\psi}_h + \boldsymbol{\phi}^{\top}\boldsymbol{x}^{\phi}_h > 0$, and $\boldsymbol{\xi}^{\top}\boldsymbol{x}^{\xi}_h + \boldsymbol{\phi}^{\top}\boldsymbol{x}^{\phi}_h > 0$.\\
 \textbf{Rule 2:} $d^*(\boldsymbol{x}^{\xi}, \boldsymbol{x}^{\psi}, \boldsymbol{x}^{\phi})= (1,0),$ if $\boldsymbol{\psi}^{\top}\boldsymbol{x}^{\psi}_h + \boldsymbol{\phi}^{\top}\boldsymbol{x}^{\phi}_h < 0$ and $\boldsymbol{\xi}^{\top}\boldsymbol{x}^{\xi}_h > \boldsymbol{\psi}^{\top}\boldsymbol{x}^{\psi}_h$ and $\boldsymbol{\xi}^{\top}\boldsymbol{x}^{\xi}_h > 0$.\\
 \textbf{Rule 3:} $d^*(\boldsymbol{x}^{\xi}, \boldsymbol{x}^{\psi}, \boldsymbol{x}^{\phi}) = (0,1),$ if $\boldsymbol{\xi}^{\top}\boldsymbol{x}^{\xi}_h + \boldsymbol{\phi}^{\top}\boldsymbol{x}^{\phi}_h < 0$ and $\boldsymbol{\psi}^{\top}\boldsymbol{x}^{\psi}_h > \boldsymbol{\xi}^{\top}\boldsymbol{x}^{\xi}_h$ and $\boldsymbol{\psi}^{\top}\boldsymbol{x}^{\psi}_h > 0$.\\
 \textbf{Rule 4:} $d^*(\boldsymbol{x}^{\xi}, \boldsymbol{x}^{\psi}, \boldsymbol{x}^{\phi}) = (0,0),$ if $\boldsymbol{\xi}^{\top}\boldsymbol{x}^{\xi}_h + \boldsymbol{\psi}^{\top}\boldsymbol{x}^{\psi}_h + \boldsymbol{\phi}^{\top}\boldsymbol{x}^{\phi}_h < 0$ and $\boldsymbol{\xi}^{\top}\boldsymbol{x}^{\xi}_h < 0$ and $\boldsymbol{\psi}^{\top}\boldsymbol{x}^{\psi}_h < 0$.
\end{decision}
Further, if we know the blip parameters $\boldsymbol{\xi}, \boldsymbol{\psi}$, and $\boldsymbol{\phi}$, then we have $\gamma^{*}[d^*(\boldsymbol{x}^{\xi}, \boldsymbol{x}^{\psi}, \boldsymbol{x}^{\phi}) ; \boldsymbol{\xi}, \boldsymbol{\psi}, \boldsymbol{\phi}] = A^{s*} \boldsymbol{\xi}^{\top}\boldsymbol{x}^{\xi} + A^{r*}\boldsymbol{\psi}^{\top}\boldsymbol{x}^{\psi}+ A^{s*}A^{r*}\boldsymbol{\phi}^{\top}\boldsymbol{x}^{\phi}$, where $\gamma^*$ means the arguments $(A^{s} ,A^{r})$ in the $\gamma$ function follow the optimal household decision rules Decision \ref{ruleHH}, and $d^*(\boldsymbol{x}^{\xi}, \boldsymbol{x}^{\psi}, \boldsymbol{x}^{\phi})$ and $(A^{s*},A^{r*})$ are the corresponding optimal treatments for the pair $(s, r)$. Therefore, to make decisions for the household, the estimates of blip parameters $\boldsymbol{\xi}, \boldsymbol{\psi}$, and $\boldsymbol{\phi}$ are necessary, and we present our approximately doubly robust methods in the following section. 
\subsection{Proposed Method and Approximate Double Robustness}
Under household interference, in a single-stage decision setting, we assume that the true ordinal-outcome model is, for $c = 1,2$, $$ logit[\mathbb{P}(U \leq c \mid a^{s} ,a^{r}, \boldsymbol{x})] = \zeta_c - f(\boldsymbol{x}^{\beta}; \boldsymbol{\beta}) -\gamma[(A^{s} ,A^{r}),\boldsymbol{x}; \boldsymbol{\xi}, \boldsymbol{\psi}, \boldsymbol{\phi}].$$ Our proposed method, the  Weighted Proportional Odds Model (WPOM), for a single-stage decision is applied by specifying three models: (1) Treatment-free model: $f(\boldsymbol{x}^{\beta} ;\boldsymbol{\beta});$ (2) Blip model: $\gamma[(A^{s} ,A^{r}),\boldsymbol{x}_h; \boldsymbol{\xi}, \boldsymbol{\psi}, \boldsymbol{\phi}]  = a^{s}  \boldsymbol{\xi}^{\top}\boldsymbol{x}^{\xi} + a^{r}\boldsymbol{\psi}^{\top}\boldsymbol{x}^{\psi} + a^{s} a^{r}\boldsymbol{\phi}^{\top}\boldsymbol{x}^{\phi};$ (3) Joint propensity score model\cite{Jiang2022DTRint}: $\pi^{a^sa^r}(\boldsymbol{x}_{s}, \boldsymbol{x}_{r}) = \mathbb{P} (A^{s} = a^s, A^{r} = a^r \mid \boldsymbol{x}_{s}, \boldsymbol{x}_{r}).$ Further, let $w^{std}$ denote ``standard'' interference-aware balancing weights\cite{Jiang2022DTRint}, which satisfy:
\begin{equation} \label{wetc}
 \pi^{0 0} w^{std}(0,0, \boldsymbol{x}) = \pi^{0 1} w^{std}(0,1, \boldsymbol{x}) =\pi^{1 0} w^{std}(1,0, \boldsymbol{x}) =\pi^{1 1} w^{std}(1,1, \boldsymbol{x}).
\end{equation}
In this context, we employ the term ``interference-aware balancing weights'' to differentiate them from those in non-interference settings, where the weights that are not ``interference-aware'' only involve the common propensity score (see the balancing weights in Wallace and Moodie (2015) \cite{wallace2015doubly}). The work by Jiang et al. (2023)\cite{Jiang2022DTRint}, where the focus is primarily on balancing weights considering network interference, outlined the criteria for balancing weights in network settings. Equation (\ref{wetc}) mentioned above represents a special case of their balancing weights criteria specifically adapted for household settings. For example, the inverse probability-based interference-aware balancing weight for the correlated treatments in a household is given by:
\begin{equation} \label{wetHH}
    w^{std}(a^s, a^r) \propto \frac{1}{\pi^{a^sa^r}} \times \frac{1}{\sum_{a^s, a^r}1/\pi^{a^sa^r}},\ \ \textrm{for}\ \  a^s = 0, 1; a^r = 0, 1.
\end{equation} 
The weight is proportional to the inverse of the joint propensities and divided by a ``normalization'' factor $\sum_{a^s, a^r}1/\pi^{a^sa^r}$. In particular, considering the balancing weights criterion in the form $\pi^{0 0} w(0,0, \boldsymbol{x}) = \pi^{0 1} w(0,1, \boldsymbol{x}) =\pi^{1 0} w(1,0, \boldsymbol{x}) =\pi^{1 1} w(1,1, \boldsymbol{x}) = \pi^{0 0}\pi^{1 0}\pi^{0 1}\pi^{1 1}$, we propose overlap-type balancing weights\citep{li2018balancing, li2019addressing}:
\begin{equation} \label{wetHH1}
    w^{std}(a^s, a^r) \propto  \frac{\pi^{0 0}\pi^{1 0}\pi^{0 1}\pi^{1 1}}{\pi^{a^sa^r}},\ \ \textrm{for}\ \ a^s = 0, 1; a^r = 0, 1. 
\end{equation}
The overlap-type weight for one treatment pair realization is proportional to the product of the joint propensities for the other possible realizations. Then, the WPOM for robust estimation of $\boldsymbol{\xi}$, $\boldsymbol{\psi}$, and $\boldsymbol{\phi}$ is applied:
\begin{algorithm}[!] 
 Compute the joint propensity scores and conduct a weighted POM with standard interference-aware overlap-type balancing weights in equation (\ref{wetHH1}) to obtain estimates $\hat{\boldsymbol{\zeta}}$, $\hat{\boldsymbol{\beta}}$, $\hat{\boldsymbol{\xi}}$, $\hat{\boldsymbol{\psi}}$, $\hat{\boldsymbol{\phi}}$.

 Construct the new weights, i.e., for $a^s,\ a^r = 0, 1$
\begin{equation} \label{wetimp}
    w(a^s, a^r) =  \frac{\pi^{0 0}\pi^{1 0}\pi^{0 1}\pi^{1 1}}{\pi^{a^sa^r}} \times \frac{\kappa(0,0, \boldsymbol{x})\kappa(1,0, \boldsymbol{x})\kappa(0,1, \boldsymbol{x})\kappa(1,1, \boldsymbol{x})}{\kappa(a^s, a^r, \boldsymbol{x})},
\end{equation}
based on $\kappa(a^s,a^r, \boldsymbol{x}) = \text{expit}(\hat{\eta}^{}_{2} )\left[1 - \text{expit}(\hat{\eta}^{}_{1})\right]\left[1 - \text{expit}(\hat{\eta}^{}_{2}) + \text{expit}(\hat{\eta}^{}_{1})\right],$ and \begin{equation*}
    \begin{cases}
\hat{\eta}^{}_{1}(a^s,a^r, \boldsymbol{x}) = \hat{\zeta}_1 + \hat{\boldsymbol{\beta}}^{\top} \boldsymbol{x}^{\beta} + \hat{\boldsymbol { \xi }}^{\top} a^s\boldsymbol{x}^{\xi}+ \hat{\boldsymbol { \psi }}^{\top} a^r\boldsymbol{x}^{\psi} + \hat{\boldsymbol { \phi }}^{\top} a^sa^r\boldsymbol{x}^{\phi},\\
\hat{\eta}^{}_{2}(a^s,a^r, \boldsymbol{x}) = \hat{\zeta}_2 + \hat{\boldsymbol{\beta}}^{\top} \boldsymbol{x}^{\beta} + \hat{\boldsymbol { \xi }}^{\top} a^s\boldsymbol{x}^{\xi}+ \hat{\boldsymbol { \psi }}^{\top} a^r\boldsymbol{x}^{\psi} + \hat{\boldsymbol { \phi }}^{\top} a^sa^r\boldsymbol{x}^{\phi},
\end{cases}
\end{equation*} which are calculated using Step 1 estimates $\hat{\boldsymbol{\zeta}}$, $\hat{\boldsymbol{\beta}}$, $\hat{\boldsymbol{\xi}}$, $\hat{\boldsymbol{\psi}}$, $\hat{\boldsymbol{\phi}}$.

 Use the new weights from Step 2, and conduct weighted POM again, to get new approximately consistent estimators $\boldsymbol{\widetilde{\xi}}$, $\boldsymbol{\widetilde{\psi}}$, $\boldsymbol{\widetilde{\phi}}$ for treatment decisions.
\caption{Weighted Proportional Odds Model (WPOM)}\label{AlgTNDDR}
\end{algorithm}
As stated in the following theorem, Step 2 in WPOM serves as the crucial key to ensuring approximate double robustness in consistently estimating the blip parameters, even when one of the treatment-free or joint propensity models is not correctly specified. 

\begin{theorem}{Approximate Double Robustness of WPOM:} \label{thmC5ordout} under the identifiability assumptions that (1) consistency \citep{rubin1980randomization}; (2) no unmeasured confounders; and (3) positivity \citep{robins2004optimal}, and suppose that the true ordinal-outcome model satisfies $$ logit[\mathbb{P}(U \leq c \mid a^{s} ,a^{r}, \boldsymbol{x})] = \zeta_c - f(\boldsymbol{x}^{\beta}; \boldsymbol{\beta}) -a^{s}  \boldsymbol{\xi}^{\top}\boldsymbol{x}^{\xi} - a^{r}\boldsymbol{\psi}^{\top}\boldsymbol{x}^{\psi}- a^{s} a^{r}\boldsymbol{\phi}^{\top}\boldsymbol{x}^{\phi}, $$ for $c = 1,2,$ and any treatment-free function $f(\boldsymbol{x}^{\beta}; \boldsymbol{\beta})$. Suppose we use weights that satisfy
\begin{equation} \label{wetcInt}
    \pi^{0 0} w(0,0) \kappa(0,0) = \pi^{0 1} w(0,1) \kappa(0,1)=\pi^{1 0} w(1,0) \kappa(1,0)=\pi^{1 1} w(1,1)\kappa(1,1),
\end{equation} where \begin{equation} \label{kappa}
    \kappa(a^s,a^r) = \text{expit}(\eta^{}_{2})\left[1 - \text{expit}(\eta^{}_{1})\right]\left[1 - \text{expit}(\eta^{}_{2}) + \text{expit}(\eta^{}_{1})\right].
\end{equation}
Then, a weighted proportional odds model based on the corresponding linear model  will yield approximately consistent estimators of $\boldsymbol{\xi}$, $\boldsymbol{\psi}$ as well as $\boldsymbol{\phi}$ if at least one of the joint propensity score and treatment-free models is correctly specified.
\end{theorem}
{\textbf{Proof:} See Appendix A of the Supplementary
materials.}
Note that the definitions of $\eta^{}_{1}, \eta^{}_{2}$ in (\ref{kappa}) are 
\begin{equation*}
    \begin{cases}
    \eta^{}_{1}(a^s,a^r, \boldsymbol{x}) := \zeta^*_1 + \boldsymbol {{\beta}^{*} }^{\top} \boldsymbol{x}^{\beta} + \boldsymbol { \xi^* }^{\top} a^s\boldsymbol{x}^{\xi}+ \boldsymbol { \psi^* }^{\top} a^r\boldsymbol{x}^{\psi} + \boldsymbol { \phi^* }^{\top} a^sa^r\boldsymbol{x}^{\phi},\\
      \eta^{}_{2}(a^s,a^r, \boldsymbol{x}) := \zeta^*_2 + \boldsymbol {{\beta}^{*} }^{\top} \boldsymbol{x}^{\beta} + \boldsymbol { \xi^* }^{\top} a^s\boldsymbol{x}^{\xi}+ \boldsymbol { \psi^* }^{\top} a^r\boldsymbol{x}^{\psi} + \boldsymbol { \phi^* }^{\top} a^sa^r\boldsymbol{x}^{\phi},
    \end{cases}\,
\end{equation*}
where $\boldsymbol { \xi^* }$, $\boldsymbol { \psi^* }$ and  $\boldsymbol { \phi^* }$ are the solutions of the estimation functions of the POM (\ref{POMpro}) with 
``standard'' interference-aware balancing weights that satisfy (\ref{wetc}). As a result, like dWOLS, a family of weights can be used if the criterion (\ref{wetcInt}) is satisfied.  Equation (\ref{wetimp}) in Step 2 provides an example of such weights because it is derived from equating (\ref{wetcInt}) to $\pi^{0 0}\pi^{1 0}\pi^{0 1}\pi^{1 1} \times \kappa(0,0, \boldsymbol{x})\kappa(1,0, \boldsymbol{x})\kappa(0,1, \boldsymbol{x})\kappa(1,1, \boldsymbol{x})$.

{\textbf{Remarks}} Similar to the balancing properties of dWOLS, the balancing properties of POM rely on the propensity score; however, the inference of household interference depends on the joint propensity functions, which will be discussed in the following subsection \ref{C5sec5.2}, in terms of estimation and construction of the balancing weights. The key factor of the balancing criterion (\ref{wetcInt}) is $\kappa$, called the ``adjustment factor'' \citep{jiang2022doublyrobust}. It adjusts for the nonlinearity of the link function, and it is special for the POM. Based on  (\ref{kappa}), we can conclude that the adjustment factor is the product of three terms: $\text{expit}(\eta^{}_2)$, $1 - \text{expit}(\eta^{}_1)$, and $1 - \text{expit}(\eta^{}_2) + \text{expit}(\eta^{}_1)$, where the first term $\text{expit}(\eta^{}_2)$ represents the estimated cumulative probabilities of categorical utilities $1$ and $2$, the second $1 - \text{expit}(\eta^{}_1)$ represents the estimated cumulative probabilities of categorical utilities $2$ and $3$, and the third term $1 - \text{expit}(\eta^{}_2) + \text{expit}(\eta^{}_1)$ represents the estimated cumulative probabilities of categorical utilities $1$ and $3$. Alternatively, the three terms of $\kappa$ in (8) can be expressed as $\text{expit}(\eta^{}_2) = 1 - \mathbb{P}(U = 3)$, $1 - \text{expit}(\eta^{}_1) = 1 - \mathbb{P}(U = 1)$, and $1 - \text{expit}(\eta^{}_2) + \text{expit}(\eta^{}_1) = 1 - \mathbb{P}(U = 2)$. Then the ``adjustment factor'' of  (\ref{kappa}) can be written as $\kappa(a^s,a^r, \boldsymbol{x}) = \Pi_{c=1}^{3}[1 - \mathbb{P}(U = c)]$. 

Regarding the approximate double robustness of WPOM, the ``approximate'' corresponds to ``approximately consistent'', which refers to a case where the estimators are derived from the estimating functions which are approximately unbiased with a small quantifiable bias (see proof of Theorem \ref{thmC5ordout} in Appendix A). We also use terms such as ``nearly unbiased'' or ``approximately unbiased'', and this quantifiable bias will be small when a linear predictor tends to vary in an interval where the $\text{expit}$ function is approximately linear\cite{Jiang2022essays}.

\subsection{Estimating Joint Propensity Score}  \label{C5sec5.2}
To maximize the assurance of approximate double robustness, we further present methods of estimating the joint propensity score that takes account of the correlations between treatments of individuals in the same household \citep{sherman2020general}. In a case where the treatments are correlated, the joint propensity functions are not equal to the product of the marginal propensities. To build accurate balancing weights and thus make robust estimations of optimal DTRs, we take into account the dependence among treatments observed in the same household. 

To estimate the joint propensity score,  letting $\boldsymbol{A}_h = (A^{s}_{h}, A^{r}_{h})^{\top}$ be the treatment vector for the $h^{th}$ household, we define $p_{hs}(\boldsymbol{\alpha}) := \mathbb{P}(A^{s}_{h} = 1 \mid \boldsymbol{x}_{hs}, \boldsymbol{\alpha})$ and $p_{hr}(\boldsymbol{\alpha}) := \mathbb{P}(A^{r}_{h} = 1 \mid \boldsymbol{x}_{hr}, \boldsymbol{\alpha})$ and $A_{hsr} := \mathbb{I}(A^{s}_{h} = 1, A^{r}_{h} = 1) = A^{s}_{h}A^{r}_{h}$, where $\mathbb{I}(x)$ is an indicator function. Also, we define $p_{hsr}:= \mathbb{P}(A_{hsr} = 1) = \mathbb{P}(A^{s}_{h} = 1, A^{r}_{h} = 1) $. Then, we provide a three-step estimation algorithm (i.e., Algorithm \ref{JointPS}). 
\begin{algorithm}[!] 
Estimate marginal propensity score models, for the $h^{th}$ household, i.e., $$p_{ht}(\boldsymbol{\alpha})= \mathbb{P} (A^{t}_{h} = 1 \mid \boldsymbol{x}_{ht}, \boldsymbol{\alpha})\ \  \text{for}\  t = s, r.$$

Model the association between pairs' treatments. For instance, denoting $\tau_{hsr}$ as the odds ratio for the pair of correlated binary variables $(A^{s}_{h},A^{r}_{h})$.

Calculate the joint propensity score based on Lipsitz et al. (1991)’s\cite{lipsitz1991generalized} formula 
\begin{equation} \label{C5jointpi}
    p_{hsr}= 
    \begin{dcases}
        \frac{b_{hsr}- \sqrt{b_{hsr}^{2}-4 \tau_{hsr}\left(\tau_{hsr}-1\right) p_{hs} p_{hr}} }{2\left(\tau_{hsr}-1\right)} & \left(\tau_{hsr} \neq 1\right), \\ p_{hs} p_{hr} & \left(\tau_{hsr}=1\right),
    \end{dcases}
\end{equation}
where $b_{hsr}=[1-\left(1-\tau_{hsr}\right)\left(p_{hr}+p_{hs}\right)]$. 
\caption{Estimating Joint Propensity Score}\label{JointPS}
\end{algorithm}
For the first step, where we estimate marginal propensity score models ($p_{ht}(\boldsymbol{\alpha})$) we employ Liang and Zeger (1986)’s\cite{liang1986longitudinal} first-order generalized estimating equation method for estimating parameter $\boldsymbol{\alpha}$. Regarding the second step, we model the association between pairs' treatments ($\tau_{hsr}$), we use Lipsitz et al. (1991)’s\cite{lipsitz1991generalized} pairwise odds ratios model, such that $log \tau_{hsr}(\boldsymbol{o}) = \boldsymbol{o}^{\top}\boldsymbol{x}_{hsr} $, where $\boldsymbol{x}_{sr}$ suppressing the $h$ are some pair-level covariates that may influence the odds-ratio between $A^{s}_{}$ and $A^{r}_{}$, and $\boldsymbol{o}$ represents the corresponding coefficients. Finally, in the third step, we calculate the joint propensity score based on Lipsitz et al. (1991)’s\cite{lipsitz1991generalized} formula, and the detailed instructions and techniques are outlined in Appendix B of the Supplementary Materials.  Therefore, building on estimators of both marginal probabilities ($p_{hs}(\hat{\boldsymbol{\alpha}})$ and $p_{hr}(\hat{\boldsymbol{\alpha}})$) and the odds ratios ($\hat{\tau}_{hsr}$), we can construct the estimator of joint propensity $\pi^{11}(\boldsymbol{x}_{hs}, \boldsymbol{x}_{hr}) = p_{hsr}$  by equation (\ref{C5jointpi}). Further, we have other estimators: 
$\hat{\pi}^{10}(\boldsymbol{x}_{hs}, \boldsymbol{x}_{hr})  = p_{hs}(\hat{\boldsymbol{\alpha}}) - \hat{\pi}^{11}(\boldsymbol{x}_{hs}, \boldsymbol{x}_{hr})$, $\hat{\pi}^{01}(\boldsymbol{x}_{hs}, \boldsymbol{x}_{hr})  = p_{hr}(\hat{\boldsymbol{\alpha}}) - \hat{\pi}^{11}(\boldsymbol{x}_{hs}, \boldsymbol{x}_{hr})$, and $\hat{\pi}^{00}(\boldsymbol{x}_{hs}, \boldsymbol{x}_{hr})  = 1-  p_{hs}(\hat{\boldsymbol{\alpha}}) - p_{hr}(\hat{\boldsymbol{\alpha}}) + \hat{\pi}^{11}(\boldsymbol{x}_{hs}, \boldsymbol{x}_{hr})$. Therefore, using equation (\ref{wetHH1}), we have overlap-type estimators of weights $\hat{w}(a^s, a^r) =  \frac{\hat{\pi}^{0 0}\hat{\pi}^{1 0}\hat{\pi}^{0 1}\hat{\pi}^{1 1}}{\hat{\pi}^{a^sa^r}}$, for $a^s = 0, 1; a^r = 0, 1.$ 

\subsection{Multiple-stage Decisions with Household Ordinal Utilities} \label{C5sec5.4}
For the multi-stage treatment decision setting, backward induction is utilized in most methods for sequential decision problems. Therefore, multi-stage treatment decision problems can be broken down into a group of single-stage decision problems. Then, for each stage, we employ a WPOM to consistently estimate the blip parameters, i.e., $\boldsymbol{\xi}$,  $\boldsymbol{\psi}$, and  $\boldsymbol{\phi}$.  Accordingly, we name our novel approach for DTR estimation with ordinal outcomes the dynamic weighted proportional odds model, namely the Dynamic Weighted Proportional Odds Model (dWPOM). 

If we acquire parameter estimates $\hat{\boldsymbol{\beta}}$, $\hat{\boldsymbol{\psi}}$, and $ \hat{\boldsymbol{\phi}}$, then we have the estimated optimal treatment blip $$\hat{\gamma}[\hat{d}^*(\boldsymbol{x}^{\xi}, \boldsymbol{x}^{\psi}, \boldsymbol{x}^{\phi}); \hat{\boldsymbol{\xi}}, \hat{\boldsymbol{\psi}}, \hat{\boldsymbol{\phi}}] = \hat{A}^{s*} \hat{\boldsymbol{\xi}}^{\top}\boldsymbol{x}^{\xi} + \hat{A}^{r*}\hat{\boldsymbol{\psi}}^{\top}\boldsymbol{x}^{\psi}+ \hat{A}^{s*}\hat{A}^{r*}\hat{\boldsymbol{\phi}}^{\top}\boldsymbol{x}^{\phi},$$ where the estimated optimal decisions $\hat{d}^*(\boldsymbol{x}^{\xi}, \boldsymbol{x}^{\psi}, \boldsymbol{x}^{\phi}) = (\hat{A}^{s*}, \hat{A}^{r*}) $ also depend on estimates $\hat{\boldsymbol{\beta}}$, $\hat{\boldsymbol{\psi}}$, and $ \hat{\boldsymbol{\phi}}$, and can be calculated by decision rules in Decision \ref{ruleHH}. Further, we can generate household level \textit{ordinal pseudo-utility} based on the ordinal pseudo-utility probability that:
\begin{equation}\label{Upseudo}
  \begin{split}
    \mathbb{P}(\widetilde{\mathcal{U}_h} =1 \mid \hat{d}^*_h, \boldsymbol{x}_h) =  & \text{expit}\left(\hat{\zeta}_1 - \hat{\boldsymbol{\beta}}^{\top}\boldsymbol{x}^{\beta}_h- \hat{\gamma}_h[\hat{d}^*_h ; \hat{\boldsymbol{\xi}}, \hat{\boldsymbol{\psi}}, \hat{\boldsymbol{\phi}}]\right), \\
    \mathbb{P}(\widetilde{\mathcal{U}_h} =2 \mid \hat{d}^*_h, \boldsymbol{x}_h) =  &\text{expit}\left(\hat{\zeta}_2 - \hat{\boldsymbol{\beta}}^{\top}\boldsymbol{x}^{\beta}_h- \hat{\gamma}_h[\hat{d}^*_h ; \hat{\boldsymbol{\xi}}, \hat{\boldsymbol{\psi}}, \hat{\boldsymbol{\phi}}]\right) \\ &-  \text{expit}\left(\hat{\zeta}_1 - \hat{\boldsymbol{\beta}}^{\top}\boldsymbol{x}^{\beta}_h- \hat{\gamma}_h[\hat{d}^*_h ; \hat{\boldsymbol{\xi}}, \hat{\boldsymbol{\psi}}, \hat{\boldsymbol{\phi}}]\right), \\
    \mathbb{P}(\widetilde{\mathcal{U}_h} =3 \mid \hat{d}^*_h, \boldsymbol{x}_h) =  &1 - \text{expit}\left(\hat{\zeta}_2 - \hat{\boldsymbol{\beta}}^{\top}\boldsymbol{x}^{\beta}_h- \hat{\gamma}_h[\hat{d}^*_h ; \hat{\boldsymbol{\xi}}, \hat{\boldsymbol{\psi}}, \hat{\boldsymbol{\phi}}]\right). 
  \end{split}
\end{equation}
Thus, building on equation (\ref{Upseudo}) and the estimates, we can compute the ordinal pseudo-utility probability, which is employed in the multiple-stage treatment decision settings.  This ordinal pseudo-utility probability represents the probability of the potential outcome that a household with the given history would have if they went on to receive the optimal treatment configuration in the current stage. 
 We incorporate the Brant-Wald test \cite{brant1990assessing} into the algorithm to evaluate whether the conditional expectations of the proposed ordinal pseudo-utility outcomes, specifically concerning the covariate values at the earlier stage, conform to a proportional odds model. The Brant-Wald test involves approximating a generalized ordinal logistic regression model and comparing it to the calculated proportional odds model, and the Wald test is then applied to assess the significance of the difference in model coefficients, generating a chi-square statistic. A low p-value (e.g., less than 0.05) in the Brant-Wald test suggests that the coefficients in the generalized model do not satisfy the proportional odds assumption. The \texttt{brant} package in R facilitates the implementation of the Brant-Wald test, and in our simulations supports the conclusion that the proportional odds assumption holds. In addition, in the multiple-stage decisions, to increase the estimation efficiency, we generate the ordinal pseudo-utility probability $\mathcal{R}$ times, and conduct $\mathcal{R}$ times estimation for the parameters of interest. Then, the final estimates of parameters are the averages of these $\mathcal{R}$ estimates. The detailed algorithm for a multiple-stage decision problem is outlined in Supplementary Materials Appendix C. As an illustrative demonstration, we present a two-stage setup in Algorithm \ref{2stage}. 
\begin{algorithm}[!] 
 Construct Stage $2$ ordinal \textit{pseudo-utility}: set $\widetilde{\mathcal{U}_2} =u_2$, where $u_2$ is the observed value of $U_2$. 
 
 Implement WPOM (i.e., Algorithm \ref{AlgTNDDR}) on ordinal pseudo-utility $\widetilde{\mathcal{U}_2}$ to acquire approximately consistent estimators $\boldsymbol{\widetilde{\xi}}_2$, $\boldsymbol{\widetilde{\psi}}_2$, $\boldsymbol{\widetilde{\phi}}_2$ for the second stage optimal treatment rule for the household, which is based on the rules in Decision \ref{ruleHH}. 
    
 Use Stage $2$ estimates $\boldsymbol{\hat{\beta}}_{2}$, $\boldsymbol{\hat{\xi}}_2$, $\boldsymbol{\hat{\psi}}_2$, $\boldsymbol{\hat{\phi}}_2$ to randomly generate ordinal pseudo-utility $\widetilde{\mathcal{U}_1}$, which takes the ordinal value $c$ with the ordinal probability $\mathbb{P}(\widetilde{\mathcal{U}_1}=c)$ (i.e., Equation \ref{Upseudo}), $\mathcal{R}$ times, to yield $\widetilde{\mathcal{U}_1^{1}}, \widetilde{\mathcal{U}_1^{2}}, ..., \widetilde{\mathcal{U}_1^{\mathcal{R}}}$.

Implement the Brant-Wald test for the pseudo-outcomes ($\widetilde{\mathcal{U}_1^{1}}, \widetilde{\mathcal{U}_1^{2}}, ..., \widetilde{\mathcal{U}_1^{\mathcal{R}}}$). (A warning message will be provided if the pseudo-outcomes fail the test.)

  For each $\mathfrak{r} = 1,..., \mathcal{R}$, implement WPOM (i.e., Algorithm \ref{AlgTNDDR}) on ordinal pseudo-utility $\widetilde{\mathcal{U}_1^{\mathfrak{r}}}$ to get revised estimates $\boldsymbol{\hat{\xi}}_{1}^{\mathfrak{r}}$, $\boldsymbol{\hat{\psi}}_{1}^{\mathfrak{r}}$, and $\boldsymbol{\hat{\phi}}_{1}^{\mathfrak{r}}$ for each $\mathfrak{r}$, and estimate $\boldsymbol{\xi}_1$, $\boldsymbol{\psi}_1$, and $\boldsymbol{\phi}_1$ by $\boldsymbol{\hat{\xi}}_{1} = \mathcal{R}^{-1}\sum_{\mathfrak{r}} \boldsymbol{\hat{\xi}}_{1}^{\mathfrak{r}}$, $\boldsymbol{\hat{\psi}}_{1} = \mathcal{R}^{-1}\sum_{\mathfrak{r}} \boldsymbol{\hat{\psi}}_{1}^{\mathfrak{r}}$, and $\boldsymbol{\hat{\phi}}_{1} = \mathcal{R}^{-1}\sum_{\mathfrak{r}} \boldsymbol{\hat{\phi}}_{1}^{\mathfrak{r}}$, respectively, then use parameter estimators  $\boldsymbol{\hat{\xi}}_{1}$,  $\boldsymbol{\hat{\psi}}_{1}$, and $\boldsymbol{\hat{\phi}}_{1}$ to construct the first stage optimal treatment rule (i.e., Decision \ref{ruleHH}). 
\caption{Two-stage Dynamic Weighted Proportional Odds Model}\label{2stage}
\end{algorithm}

%% file: Sections/Simulation.tex
In this section, we provide two simulation studies (Study 1 and 2) to illustrate our proposed methods for estimating optimal DTRs with ordinal outcomes under household interference. In each study, we first verify the approximate double robustness of our estimation method, and then check that the corresponding estimated optimal DTR outperforms those corresponding to other estimation methods.   In Study 1, we consider single-stage treatment decision problems, and in Study 2, we investigate a multi-stage decision problem in a two-stage case. 

To assess the performance of the methods, we construct three measures: (1) optimal treatment rate, (2) mean regret value, and (3) value functions for ordinal outcomes. First, based on the data-generating parameters, we can calculate the truly optimal treatments for each household.  Then, we can construct the recommended treatments from the estimated rules based on the estimated decision parameters. The optimal treatment rate (OTR) is then the percentage of the estimated recommended treatments that are in accord with the authentic optimal treatments. Second, the mean regret value (MRV) measures the difference between the blip value under the true optimal regime and under the estimated regime, and therefore measures the `loss' experienced by using the estimated regime instead of the truly optimal one. The detailed performance matrix outlining the definitions of the OTR and MRV is provided in Appendix D of the Supplementary Materials (subsection \ref{Performance Matrix}). Finally, we construct value functions for ordinal outcomes, which mainly compare the estimated optimal treatments with the observed treatments. We will give the formal definition of the value functions for ordinal outcomes building on the concept of the odds ratio. It is important to note that, because of the specific nature of ordinal outcomes, for the single-stage settings in Study 1, we compare the WPOM with methods that ignore interference (Study 1a), and focus on consistent estimation of the WPOM (Study 1b). For the multi-stage settings in Study 2, we primarily concentrate on the long-term treatment effects of the estimated DTRs. In that case, we examine value functions for ordinal outcomes to compare different methods.

\subsection{Single-stage Treatment Decision for a Couples Case} \label{Study1}

\subsubsection{Single-stage Treatment Decision for a Couples Case} \label{Study1a}

In \textit{Study 1a}, to evaluate the performance of the proposed WPOM, we compare the proposed approach with a simpler alternative that neglects interference. Specifically, we fit a standard logistic regression model separately for husbands and wives, each with their respective treatment, covariates, and household-level covariates. An example would be the model in Theorem H.2 in Jiang (2022) \cite{Jiang2022essays}, which is developed without considering the treatment of the spouse. We refer to this method as an \textit{interference-unaware} approach. We derive optimal treatment regimes by maximizing the conditional logistic probabilities associated with each individual, and compare their performance with that of the interference-aware method. Furthermore, we have developed a cross-validation variant of WPOM by partitioning the data into $K$ folds and have investigated the performance of $K-$fold cross-validated WPOM.

In the generation of ordinal outcomes for the households, based on the mixed cumulative logit model (\ref{POMpro}), the ordinal outcome is a random function of household treatment assignments and covariates $\boldsymbol{x}^{\beta}$, $\boldsymbol{x}^{\xi}$, $\boldsymbol{x}^{\psi}$ and $\boldsymbol{x}^{\phi}$. For simulation settings, in this part of the study, we take $B=500$ Monte Carlo replicates, generating the covariates, treatments and outcomes for each replicate. A comprehensive explanation of the steps involved in generating covariates and functions can be found in Section \ref{dgpS1} of Appendix D of the Supplementary Materials.

Table \ref{C5tb:Study1a} presents the three aforementioned performance metrics for the proposed WPOM method with the weights computed based on Algorithm (\ref{JointPS}), comparing it with the interference-unaware approach and the developed $20-$fold cross-validated WPOM. These performance metrics are (1) household OTR, (2) individual OTR, and (3) the mean regret value. Compared to the interference-unaware approach, both the proposed WPOM and the 20-fold cross-validated WPOM yielded higher values for household OTR and individual OTR, as well as a smaller value for the mean regret. We can conclude that in the context of household interference, decisions may be significantly compromised if the interference is disregarded. In comparison to the cross-validated WPOM, the standard WPOM demonstrates similar performance in these three performance measures. It is worth noting that $K-$fold validation is typically employed in a prediction context, but our approach focuses on estimation; this method serves as a means to assess the stability of the estimate and provides an error estimate that accounts for some of the model uncertainty.
\begin{table}[!]\centering
\setlength{\tabcolsep}{0.9pt}
\caption{Methods' performance measure estimates and their standard errors  (in parenthesis) in Study 1b. $H$ denotes the number of households. OTR-H: Household optimal treatment rate; OTR-I: Individual optimal treatment rate; MRV: Mean regret value.}\label{C5tb:Study1a} 
\setlength{\tabcolsep}{2pt} 
\renewcommand{\arraystretch}{1.2} 
\begin{tabular}{cl|l|llll}
\hline
\multirow{2}{*}{$H$} & \multirow{2}{*}{Performance} & \multicolumn{5}{c}{Method}                               \\ 
                  &                   & \multicolumn{2}{c}{Interference-unaware}    & \multicolumn{1}{c}{WPOM}  &\multicolumn{2}{c}{Cross-validated WPOM}  \\ \hline 
\multirow{3}{*}{$500$} 
                  & OTR-H               & \multicolumn{2}{c}{0.063 (0.005)}  & \multicolumn{1}{c}{0.395 (0.018)} & \multicolumn{2}{c}{0.393 (0.019)} \\ 
                  & OTR-I             & \multicolumn{2}{c}{0.751 (0.006)}  &\multicolumn{1}{c}{0.829 (0.016)} & \multicolumn{2}{c}{0.827 (0.016)}  \\ 
                  &  MRV            & \multicolumn{2}{c}{0.375 (0.008)}  & \multicolumn{1}{c}{0.055 (0.018)} & \multicolumn{2}{c}{0.060 (0.018)} \\ \hline 
\multirow{3}{*}{$1500$} 
                  & OTR-H               & \multicolumn{2}{c}{0.062 (0.003)}  & \multicolumn{1}{c}{0.530 (0.018)} & \multicolumn{2}{c}{0.522 (0.018)} \\ 
                  & OTR-I             & \multicolumn{2}{c}{0.750 (0.005)}  &\multicolumn{1}{c}{0.872 (0.014)} & \multicolumn{2}{c}{0.873 (0.014)}  \\ 
                  &  MRV            & \multicolumn{2}{c}{0.375 (0.006)}  & \multicolumn{1}{c}{0.032 (0.017)} & \multicolumn{2}{c}{0.037 (0.018)} \\ \hline 
\multirow{3}{*}{$3000$} 
                  & OTR-H               & \multicolumn{2}{c}{0.063 (0.003)}  & \multicolumn{1}{c}{0.622 (0.017)} & \multicolumn{2}{c}{0.618 (0.018)} \\ 
                  & OTR-I             & \multicolumn{2}{c}{0.750 (0.004)}  &\multicolumn{1}{c}{0.901 (0.012)} & \multicolumn{2}{c}{0.900 (0.013)}  \\ 
                  &  MRV            & \multicolumn{2}{c}{0.374 (0.005)}  & \multicolumn{1}{c}{0.024 (0.016)} & \multicolumn{2}{c}{0.028 (0.016)} \\ \hline 
\end{tabular}
\end{table}

\subsubsection{Approximate double robustness of WPOM} \label{Study1b}

In \textit{Study 1b}, to examine the approximate double robustness of the proposed method, we examine four scenarios. Scenario 1: neither the treatment-free model nor the treatment model is correctly specified. Scenario 2: the treatment-free model is correctly specified but the treatment model is misspecified. Scenario 3: the treatment model is correctly specified but the treatment-free model is misspecified.  Scenario 4: both treatment-free model and treatment model are correctly specified. 

Scenario 1 fails to specify a correct model, so consistent estimation of the blip parameters cannot be guaranteed. However, Scenarios 2, 3, and 4 correctly specify at least one of the treatment-free and treatment models, so the estimator of blip parameters should be close to consistent. In addition, note that we have only linear terms in our POM, while the true models can contain non-linear terms. If the true models contain non-linear terms, then we have misspecified the model. Moreover, in a real application, it is typically more challenging to correctly specify the treatment-free model than the treatment model, so we particularly highlight the results of Scenario 3. 

In each scenario, five different methods are investigated. Method 0 (M0) employs the proposed proportional odds model (\ref{POMpro}) without any balancing weights. Method 1 (M1) considers the same POM and uses the standard balancing weights,  but assumes independence between the treatments, like the weights in Jiang et al. (2022b)\cite{Jiang2022DTRint}. That is, $w = | A^s - \mathbb{P}(A^s = 1 \mid \boldsymbol{x}_s) | *  | A^r - \mathbb{P}(A^r = 1 \mid \boldsymbol{x}_r) |$. However, Methods 2 and 3 (M2 and M3) both consider the same POM, yet use the proposed interference balancing weights, which allow dependence between the treatments within the same household. In particular, M2 employs the inverse
probability-based weights (\ref{wetHH}) and M3 uses the overlap-type weights (\ref{wetHH1}). Furthermore, to contrast the performance of the weights in M2 and M3 with the weights that include the adjustment factor, i.e., (\ref{kappa}), we also consider Method 4 (M4), that is, using the same POM (\ref{POMpro}) with the adjusted overlap weights (\ref{wetimp}) that is based on proposed Algorithm \ref{JointPS}. 

Note that M0 is $Q$-learning in a single-stage decision setting, and M1, M2, M3, and M4 belong to our proposed WPOM yet with different balancing weights. M1 uses a no-treatment-association WPOM, but M2, M3, and M4 use treatment-association aware WPOMs. Methods M2 and M3 employ inverse probability type and overlap type weights, respectively. However, M4 utilizes adjusted overlap type weights. The adjusted weights (\ref{wetimp}) satisfy the weight criterion in Theorem \ref{thmC5ordout}, hence M4 is expected to provide close to consistent blip parameter estimators in Scenarios 2, 3, and 4. 

\begin{figure}[!]\centering
    \includegraphics[scale=0.45]{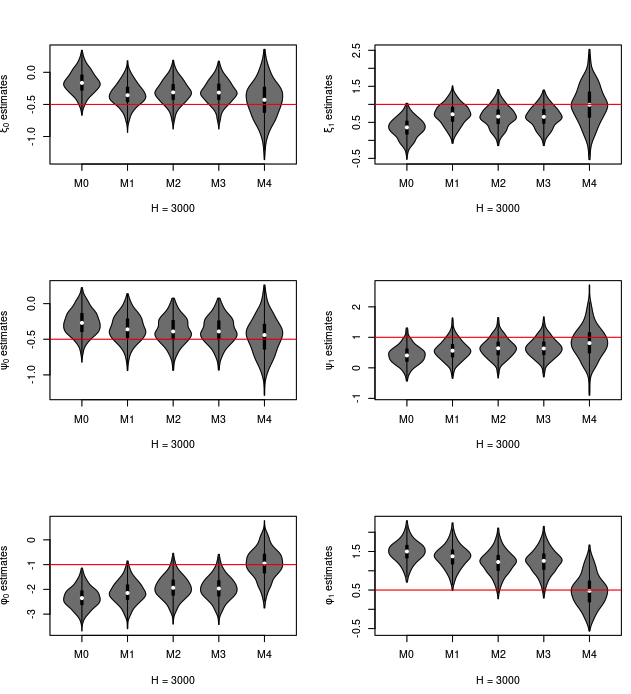} 
	\caption{Blip function parameter estimates, $\hat{\boldsymbol{\xi}}$ (top row), $\hat{\boldsymbol{\psi}}$ (middle row), and $\hat{\boldsymbol{\phi}}$ (bottom row) via Method 0 (M0, $Q$-learning), Method 1 (M1, no treatment-association WPOM), Method 2 (M2, treatment-association aware WPOM with IPW-type weights), Method 3 (M3, treatment-association aware WPOM with overlap-type weights) and Method 4 (M4, treatment-association aware WPOM with adjusted overlap-type weights), when the treatment model is correctly specified but the treatment-free model is misspecified (Scenario 3).}\label{P3study1a}
\end{figure}

The treatment decision rules in Decision \ref{ruleHH} rely on the estimates of blip parameters, that is, $\hat{\boldsymbol{\xi}}, \hat{\boldsymbol{\psi}}$, and $\hat{\boldsymbol{\phi}}$. Figure \ref{P3study1a} presents the distribution of blip parameter estimates from Methods 0, 1, 2, 3, and 4 in Scenario 3, where the treatment model is correctly specified but the treatment-free model is misspecified.  Moreover, the distributions of the blip parameter estimates in Scenarios 1, 2 and 4 are presented in Supplementary Materials Appendix D. From figures depicting these results, in particular Figure \ref{P3study1a}, the approximately consistent estimation of blip parameters (${\boldsymbol{\xi}}, {\boldsymbol{\psi}}$, and ${\boldsymbol{\phi}}$) from M4 is as expected. That is, the estimates of Method 4 from Scenarios 2, 3 and 4 appear consistent, and this verifies the approximate double robustness of our proposed adjusted weights (\ref{wetimp}) in the simulation setting. However, in Scenario 3, M0, M1, M2, and M3 offer biased blip parameter estimators. Even though the M1, M2, and M3 estimators are biased, the bias is smaller than for the M0 estimator which does not employ any balancing weights. Moreover, in this Scenario 3, compared with M1, where independence of the treatments is assumed, M2 and M3, which address the association between treatments, provide less biased estimators. This result confirms that if a correlation exists between treatments in the same household in truth, failing to take that into account will lead to biased estimation. Furthermore, as Figures \ref{P3study1aS2} and \ref{P3study1aS4} in Appendix D indicate, in both Scenarios 2 and 4, where treatment-free models are correctly specified, all the methods, even for M1, provide unbiased estimators of blip parameters. Thus, we find in the case where the treatment-free model is correctly specified that there is little distinction among the methods.

In this second part of Study 1b, focusing on Scenario 3, the settings are the same as those introduced above, except that we set different numbers of households $H$. Table \ref{C5tb:Study1} (subsection \ref{AddRes} of Appendix D) presents the three performance measures for all methods, i.e., household and individual OTR, and the mean regret value. As expected, in all the cases, compared with the $Q$-learning (M0), WPOM (M1 --- M4) methods provide higher values of both household and individual OTR, and lower mean regret value. From Table \ref{C5tb:Study1}, in the larger household sample cases, that is, $H \geq 3000$, compared with either M0 ($Q$-learning) or M1, M2, and M3 (WPOM with different types of weights), M4 which is WPOM with adjusted weights provides the highest in both household and individual OTRs, and the lowest MRV. As well, M0 which does not use any balancing weights outputs the lowest OTRs and the highest MRV. These results verify that the estimated treatment configuration from M4 is the closest to the optimal treatment configuration. Thus, in these large household sample cases, M4 performs best among all these methods, and M0 performs the worst.  

Simulation Study 2, a second simulation in a two-stage decision setting to demonstrate proposed dWPOM, is presented in Supplementary Materials Appendix E. These simulation results further demonstrate the robust estimation of the blip parameters resulting from the proposed method.

%% file: Sections/PATH.tex
\subsection{Data Source and Definition of Treatments and Outcome}
Investigating household ordinal outcomes, we now apply our approach, dWPOM with household interference, to longitudinal survey data in the Population Assessment of Tobacco and Health (PATH) study. We aim to estimate the optimal DTR for a pair in the same household, based on a sequence of rules of e-cigarette use or non-use, for achieving the smoking cessation of the pair in the household. Building on the PATH analysis in  Jiang et al. (2022a)\cite{Jiang2022DTRint}, we consider the subset of participant pairs both of whom smoke at the beginning of the study. In the PATH study, data were gathered in \textit{waves}, starting from 2011, with each subsequent wave beginning approximately one year after the previous one. Studying the first four waves, we formulate the PATH analysis as a three-stage decision problem by defining the $j^{th}$ stage, for $j = 1, 2, 3$, as the time from Wave $j$ to but not including Wave $j+1$.

In this analysis, the treatment variable is the use of e-cigarettes by cigarette smokers. Because waves were separated for approximately one year, we define e-cigarette use reported at the wave of the measured outcome as indicative of the pre-wave treatment. The e-cigarette usage variable is determined by the question “Do you now use e-cigarettes (a) Every day (b) Some days (c) Not at all.” Answers of either “Every day” or “Some days” are coded as $A = 1$, and answers of “Not at all” as $A = 0$. 

Further, our household ordinal utility is constructed by a combination of binary outcomes of individuals sharing a household, where the binary outcome variable is an indicator of whether participants have either given up smoking (traditional cigarettes) or have tried to quit smoking or using tobacco product(s). That is, the household utility is the sum of the final binary outcomes of a pair in the same household, which is interpreted, for a pair in a household, as (a) neither, (b) one, or (c) both of them incur a benefit such as smoking cessation. Jiang (2022)\cite{Jiang2022essays} provides a comprehensive discussion on the precise construction of binary outcomes using questionnaires.
\subsection{Household Covariates Choice and Model Settings}
As for the household covariates choice in our POM, for the $j^{th}$ stage, we first select the individual-level Wave $j$ variables: age (``less than 35” or ``35+”), education, non-Hispanic, race and ``plan to quit''. Then, building on these individual-level covariates, we can construct household or joint covariates for our household POM. For instance, the individual-level age variable is an indicator of ``less than 35”; for the household-level age variable, we thus have three possibilities for a pair in the same household:  both, one of them, or neither of them is less than 35. Therefore, we construct the single age variable with three categories for the household model. Similarly, with three possible values for each variable, we construct the non-Hispanic, race, and ``plan to quit'' variables at the household level. 

For the household covariates, we have denoted, age, education, non-Hispanic, race, and ``plan to quit'', as the covariates $\boldsymbol{x}_j^{\beta} = (x_{j1}, x_{j2}, x_{3}, x_{4}, x_{j5})^{\top}$ in the Stage $j$ treatment-free model. We note that, compared with the PATH analysis in  Jiang et al. (2022a)\cite{Jiang2022DTRint}, we omit the sex variable in the household covariates, because it has not been significant when we focus on the household-level model. Individuals $s$ and $r$ are respectively the first and second listed members of the household pair in the data set. In addition, building on  previous work studying  moderators in the relationships of prior wave predictors of quitting smoking (e.g.,  Le Grande et al. (2021)\cite{le2020predictive}), we select at each stage the variables age and ``plan to quit'' as tailoring variables, that is, $\boldsymbol{x}_j^{\xi} = (1, x^s_{j1})^{\top}$, $\boldsymbol{x}_j^{\psi} = (1, x^r_{j1})^{\top}$, and $\boldsymbol{x}_j^{\phi} = (1, x^s_{j5} + x^r_{j5})^{\top}$. Therefore, in estimation, the blip model is set up as $\gamma[(a_{j+1}^{s}, a_{j+1}^{r}),\boldsymbol{x}_{j}; \boldsymbol{\xi}_j, \boldsymbol{\psi}_j, \boldsymbol{\phi}_j] = a_{j+1}^{s}  \boldsymbol{\xi}_j^{\top}\boldsymbol{x}^{\xi}_j + a_{j+1}^{r}\boldsymbol{\psi}_j^{\top}\boldsymbol{x}^{\psi}_j+ a_{j+1}^{s} a_{j+1}^{r}\boldsymbol{\phi}_j^{\top}\boldsymbol{x}^{\phi}_j$, and the treatment-free model as $f(\boldsymbol{x}_{j}^{\beta}; \boldsymbol{\beta}_j) =  \boldsymbol{\beta}_{j}^{\top} \boldsymbol{x}_{j}^{\beta}$. Accordingly, solving the sequential decision problem by backward induction, for the Stage $j = 3,2,1$ and $c = 1,2$, we have the POM that $ 
    logit[\mathbb{P}(\widetilde{\mathcal{U}_j^r}  \leq c \mid a_j^{s} ,a_j^{r}, \boldsymbol{x}_j; \boldsymbol{\xi}_j, \boldsymbol{\psi}_j, \boldsymbol{\phi}_j)] = \zeta_{cj} - \boldsymbol{\beta}_j^{\top}\boldsymbol{x}_j^{\beta} -a_j^{s}  \boldsymbol{\xi}_j^{\top}\boldsymbol{x}_j^{\xi} - a_j^{r}\boldsymbol{\psi}_j^{\top}\boldsymbol{x}_j^{\psi}- a_j^{s} a_j^{r}\boldsymbol{\phi}_j^{\top}\boldsymbol{x}_j^{\phi}.$
    
To construct the balancing weights for the proposed POM, as introduced in Section \ref{C5sec5.2}, we estimate the marginal propensity scores, the pairwise odds ratios ($\tau_{sr}$), and the joint propensity scores. 
We first choose covariates at the individual level for the marginal treatment propensity models, based on the previous PATH studies of Benmarhnia et al. (2018)\cite{benmarhnia2018can}  and Jiang et al. (2022a)\cite{Jiang2022DTRint}, as
$\boldsymbol{x}_j^{\alpha}=  (x_{j1}, x_{j2}, x_{3},x_{4}, x_{5j}, x_{6})^{\top}$, namely, age, education, non-Hispanic, race, ``plan to quit'', and sex. Then, we employ logistic regression to acquire marginal treatment propensity scores. The pairwise odds ratios are modeled through a generalized linear model with the log link and covariates $\boldsymbol{x}_{sr} = [1,  (x^s_{j5} + x^r_{j5}) ]^{\top}$, which represents the number of individuals in the same household who have a plan to quit. That is, with parameter $\boldsymbol{o}$, $log \tau_{sr}(\boldsymbol{o}) = \boldsymbol{o}^{\top}\boldsymbol{x}_{sr}$. 

Following the methods that were introduced in Section \ref{Sec2}, we can further estimate the joint propensity score, and the corresponding weights. In particular, we compare four different weights, which are (I) no balancing weights (M0), (II) no-association overlap weights (M1), where the joint propensity functions are equal to the product of marginal propensities, (III) association-aware overlap weights (\ref{wetHH1}) (M3), and (IV) adjusted association-aware overlap weights (\ref{wetimp}) (M4). In this PATH analysis, we call them Methods I ($Q$-learning), II, III, and IV.  It is important to note that Method IV employs the adjusted balancing weights, and is our desired treatment-association aware dWPOM, which in theory guarantees approximately consistent estimators of the blip parameters.

\begin{table}[!]\centering
\setlength{\tabcolsep}{5pt}
\caption{Blip estimates and their replication standard errors (in parenthesis) from the analysis of PATH data. Optimal DTRs are  functions of blip parameter estimates based on decision rules in Decision \ref{ruleHH}. Est. stands for the blip parameters' estimates.}\label{C5t:pathhhRES} 
\setlength{\tabcolsep}{3pt} 
\renewcommand{\arraystretch}{1.2} 
\begin{tabular}{cllllll}
\hline
\multirow{2}{*}{Wave} & \multirow{2}{*}{Est.} & \multicolumn{4}{c}{ Methods }                               \\                   &                   & \multicolumn{1}{c}{I ($Q$-learning)}  & \multicolumn{1}{c}{II}  & \multicolumn{1}{c}{III}  &\multicolumn{1}{c}{IV} \\ \hline 
\multirow{6}{*}{$1 \sim 2$} & \multicolumn{1}{c}{$\hat{\xi}_0$ }               & \multicolumn{1}{c}{-0.100 (0.037)} & \multicolumn{1}{c}{-0.124 (0.044)} & \multicolumn{1}{c}{0.218 (0.072)} & \multicolumn{1}{c}{0.107 (0.047)} \\ 
                  & \multicolumn{1}{c}{$\hat{\xi}_1$ }             & \multicolumn{1}{c}{0.180 (0.043)} & \multicolumn{1}{c}{0.130 (0.055)} &\multicolumn{1}{c}{-0.188 (0.052)} & \multicolumn{1}{c}{-0.150 (0.068)}  \\ 
                  & \multicolumn{1}{c}{$\hat{\psi}_0$ }            & \multicolumn{1}{c}{-0.167 (0.040)} & \multicolumn{1}{c}{-0.321 (0.049)} & \multicolumn{1}{c}{-0.140 (0.047)} & \multicolumn{1}{c}{-0.070 (0.050)} \\ & \multicolumn{1}{c}{$\hat{\psi}_1$ }               & \multicolumn{1}{c}{0.251 (0.044)} & \multicolumn{1}{c}{0.257 (0.058)} & \multicolumn{1}{c}{0.017 (0.052)} & \multicolumn{1}{c}{-0.273 (0.059)} \\ 
                  & \multicolumn{1}{c}{$\hat{\phi}_0$ }             & \multicolumn{1}{c}{0.102 (0.079)} & \multicolumn{1}{c}{0.376 (0.078)} &\multicolumn{1}{c}{0.009 (0.081)} & \multicolumn{1}{c}{-0.001 (0.099)}  \\ 
                  & \multicolumn{1}{c}{$\hat{\phi}_1$ }            & \multicolumn{1}{c}{0.001 (0.047)} & \multicolumn{1}{c}{0.106 (0.048)} & \multicolumn{1}{c}{0.087 (0.050)} & \multicolumn{1}{c}{0.045 (0.058)} \\ \hline
\multirow{6}{*}{$2 \sim 3$} & \multicolumn{1}{c}{$\hat{\xi}_0$  }              & \multicolumn{1}{c}{0.341 (0.036)} & \multicolumn{1}{c}{0.116 (0.046)} & \multicolumn{1}{c}{0.081 (0.045)} & \multicolumn{1}{c}{-0.031 (0.045)} \\ 
                  & \multicolumn{1}{c}{$\hat{\xi}_1$  }            & \multicolumn{1}{c}{-0.276 (0.043)} & \multicolumn{1}{c}{0.188 (0.062)} &\multicolumn{1}{c}{-0.044 (0.056)} & \multicolumn{1}{c}{ 0.200 (0.058)} \\ 
                  & \multicolumn{1}{c}{$\hat{\psi}_0$ }            & \multicolumn{1}{c}{-0.078 (0.037)} & \multicolumn{1}{c}{0.052 (0.052)} & \multicolumn{1}{c}{0.205 (0.048)} & \multicolumn{1}{c}{0.004 (0.048)} \\
                  & \multicolumn{1}{c}{$\hat{\psi}_1$  }              & \multicolumn{1}{c}{0.068 (0.045)} & \multicolumn{1}{c}{0.101 (0.067)} & \multicolumn{1}{c}{0.054 (0.058)} & \multicolumn{1}{c}{0.064 (0.062)} \\ 
                  & \multicolumn{1}{c}{$\hat{\phi}_0$  }            & \multicolumn{1}{c}{0.377 (0.097)} & \multicolumn{1}{c}{0.663 (0.110)} &\multicolumn{1}{c}{0.360 (0.108)} & \multicolumn{1}{c}{ 0.358 (0.126)} \\ 
                  & \multicolumn{1}{c}{$\hat{\phi}_1$ }            & \multicolumn{1}{c}{-0.317 (0.060)} & \multicolumn{1}{c}{-0.568 (0.065)} & \multicolumn{1}{c}{-0.414 (0.067)} & \multicolumn{1}{c}{-0.507 (0.076)} \\ \hline
\multirow{6}{*}{$3 \sim 4$} & \multicolumn{1}{c}{$\hat{\xi}_0$}                & \multicolumn{1}{c}{0.966 (0.040)} & \multicolumn{1}{c}{1.067 (0.041)} & \multicolumn{1}{c}{1.233 (0.040)} & \multicolumn{1}{c}{0.785 (0.047)} \\ 
                  & \multicolumn{1}{c}{$\hat{\xi}_1$  }            & \multicolumn{1}{c}{-0.419 (0.055)} & \multicolumn{1}{c}{-0.268 (0.055)} &\multicolumn{1}{c}{-0.391 (0.054)} &  \multicolumn{1}{c}{0.304 (0.060)} \\ 
                  & \multicolumn{1}{c}{$\hat{\psi}_0$  }           & \multicolumn{1}{c}{0.808 (0.038)} & \multicolumn{1}{c}{1.527 (0.047)} & \multicolumn{1}{c}{1.217 (0.043)} & \multicolumn{1}{c}{0.690 (0.044)} \\ & \multicolumn{1}{c}{$\hat{\psi}_1$}                & \multicolumn{1}{c}{0.448 (0.044)} & \multicolumn{1}{c}{-0.507 (0.054)} & \multicolumn{1}{c}{0.138 (0.049)} & \multicolumn{1}{c}{0.987 (0.052)} \\ 
                  & \multicolumn{1}{c}{$\hat{\phi}_0$  }            & \multicolumn{1}{c}{-1.612 (0.137)} & \multicolumn{1}{c}{-0.150 (0.150)} &\multicolumn{1}{c}{-1.479 (0.147)} &  \multicolumn{1}{c}{-1.151 (0.228)} \\ 
                  & \multicolumn{1}{c}{$\hat{\phi}_1$  }           & \multicolumn{1}{c}{0.331 (0.069)} & \multicolumn{1}{c}{0.085 (0.076)} & \multicolumn{1}{c}{0.004 (0.076)} & \multicolumn{1}{c}{-0.091 (0.115)} \\\hline
\end{tabular}
\end{table}

\subsection{PATH Analysis Results}
Table \ref{C5t:pathhhRES} summarizes the blip estimates and their replication standard errors (in parenthesis) from Methods I, II, III, and IV in this PATH analysis. It is important to note that, for both members of a couple to either quit or attempt to quit smoking, the optimal DTRs for the household are functions of blip parameter estimates and the couple's tailoring variables, that is, the decision rules in Decision \ref{ruleHH}. Method IV, which employs the adjusted balancing weights, is expected to provide consistent estimation of these blip parameters. Thus, we particularly focus on the results from Method IV, while accounting for those from other methods.

Because the household case with four treatment configurations is more complicated than in the previous individual-level analysis, based on Rule \ref{ruleHH},  we give several examples of how the results may be interpreted. For Method IV, in Stage 3 (Wave $3 \sim 4$), for example, the blip estimate is $A^s(0.785 + 0.304*age_s) + A^r(0.690 + 0.98
7 * age_r) + A^s*A^r(- 1.151 - 0.091*PQ)$, where $PQ$ represents the plans of quitting for a couple in the same household, and $age_s$ and $age_r$ are ages of $s$ and $r$. When we plug in four possibilities of $(A^s, A^r) = {(1,1), (1,0), (0,1), (0,0)}$, the blip estimates are $0.785 + 0.304*age_s + 0.690 + 0.987 * age_r - 1.151 - 0.091*PQ$, $0.785 + 0.304*age_s$, $0.690 + 0.987 * age_r$, and $0$, respectively. Table \ref{tb:C5PATHblip} summaries the blip estimates for different treatment configurations $(A^{s}_{h},A^{r}_{h})$ from Method IV (Stage 3). To interpret these results, we provide the following examples.

\textbf{Example 1:} If we have household tailoring variables such that $age_s = 1$
$age_r = 0$, and $PQ = 0$, the blip estimates are $0.785 + 0.304 + 0.690 -1.151 = 0.62
8$, $0.785 + 0.304 = 1.089$, $0.690$, and $0$, respectively. The largest blip estimate is $1.089$, and corresponds to the treatment configuration $(A^s = 1, A^r = 0)$. Therefore, in Stage 3 (Wave $3 \sim 4$), if individual $s$ is less than $35$ but $r$ is not, and both of them have no plan to quit, then the treatment recommendation for this household should be $(A^s = 1, A^r = 0)$.

\begin{table}
\caption{Blip estimates for different treatment configurations $(A^{s}_{h},A^{r}_{h})$ from Method IV (Stage 3)}
\begin{center}
\begin{tabular}{l|r}
  ($A^{s}_{h}, A^{r}_{h}$) & Blip estimate from Method IV (Stage 3)\\
\hline
$ (1,1)$ & $0.785 + 0.304*age_s + 0.690 + 0.987 * age_r - 1.151 - 0.091*PQ$  \\
$ (1,0)$ & $0.785 + 0.304*age_s$  \\
$(0,1)$ & $0.690 + 0.987 * age_r$ \\
$(0,0)$ & 0 \\
\hline
\end{tabular}
\end{center}

\label{tb:C5PATHblip}
\end{table}

\textbf{Example 2:} If we have household tailoring variables such that $age_s = 0$,
$age_r = 1$, and $PQ = 2$ (individual $s$ is over $35$ but $r$ is not, and both of them have plans to quit), the blip estimates are $0.785 + 0.690 + 0.987 -1.151 - 0.09
1*2 = 1.192$, $0.785$, $0.690 + 0.997 = 1.687,$ and $0$, respectively; then the treatment recommendation should be $(A^s = 0, A^r = 1)$.

Finally, we note that an important aspect of rigorous real data analysis is implementing cross-validation to evaluate proposed methods properly. While the complexity of the PATH design may pose challenges, McConville (2011)\cite{mcconville2011improved}, Opsomer and Miller (2005)\cite{opsomer2005selecting} and You (2009)\cite{you2009cross} suggest that there are alternative cross-validation strategies and literature available to address these challenges, specifically in the context of complex survey data.  Users should carefully choose and adapt cross-validation methods to suit their specific data and research needs, ensuring the reliability and validity of their results. Consequently, a further research direction for our methods entails an examination of suitable cross-validation strategies for analyzing data from complex longitudinal surveys.

%% file: Sections/Conclusion.tex
In this paper, considering household interference and household utility, we proposed a robust DTR estimation method for ordinal outcomes to consistently estimate optimal DTRs. This method, namely dWPOM, uses sequential WPOM with adjusted balancing weights. We theoretically and empirically demonstrated the approximate double robustness property of our WPOM approach, which utilizes the proposed adjusted balancing weights. In the presence of household interference, our WPOM addresses household ordinal utility problems and provides optimal treatment recommendations for both individuals in the household. To address the ordinal outcomes challenge, we consider a POM because of its easy estimation and interpretation and note that any POM-related tools or techniques, such as those for variable selection or model diagnosis of POMs, can be employed in our method. Regarding inference, a single-stage decision can use standard errors from WPOM to create confidence intervals for blip parameters directly. However, for multi-stage decisions, non-regularity issues arise\cite{chakraborty2010inference}. Hence, future research is needed to develop methods like adaptive bootstrap and m-out-of-n bootstrap\cite{chakraborty2010inference} for constructing multi-stage decision confidence intervals.
 
We have also made a methodological contribution to the study of interference. In addition to considering the effects of neighbours' treatments on an individual's outcome, we considered a possible association between their treatments. Building on this, we presented the estimation process for joint propensity scores in the case where there exists an association between treatments of individuals in the same household, then estimated the corresponding balancing weights that satisfy the balancing criterion. Our simulation studies have revealed that if there exists an association between treatments but we fail to consider it, then the DTR estimation process will lead to bias. It would be straightforward to extend our household interference case to cases of partial interference, where treatments of individuals blocked by clusters can affect outcomes of the individuals in the same cluster, while also accounting for the association between these treatments of individuals in the same cluster. However, the association-aware estimation has an extra cost: modeling association between pairs of binary treatments, such as through a pairwise odds ratio model in our household case. For the cluster partial-interference case, we suggest considering the log-linear model to extend our work to estimate the ``higher-order'' odds ratio association \citep{yi2005marginal}. Note that in cases of association, the final goal is to estimate the joint propensity scores; therefore, we recommend employing machine learning methods, such as random forest or deep neural network, to directly train the model for the joint propensity scores.

We acknowledge that any misspecification in the association model (Step 2 of the Algorithm \ref{JointPS}) could impact the accuracy of our joint propensity score estimation, and thus affect the approximate double robustness of the proposed method. This step could be refined by using flexible data-adaptive approaches that accommodate correlated data.  Examples of such approaches include mixed-effect machine learning\cite{ngufor2019mixed} and smoothed kernel regression designed for dependent data\cite{park2022efficient}. Further investigation is needed to assess the robustness of the proposed methods in terms of association models and employing these data-driven approaches. Our formulation through the household utility function also allows for some association among the responses of the household members, conditional on their treatments.  In particular, when we define household utility as a function of the sum of individual utilities, which happens to be the sum of their response indicators, the utility distribution will imply an association between the responses of paired members.

We also note that individuals within the same household may experience varying magnitudes of interference effects. When the objective is to optimize the household's utility function, the fourth term in our model (Equation \ref{eq:intoutcome}) captures these interference effects, and distinguishing between the effects on individuals may not be necessary. However, if the goal is to optimize individual outcomes or understand the role of interference effects for each person (e.g., husband or wife), distinguishing between these interference effects becomes essential. To achieve this, we can examine individual outcome models, which involve modeling the outcomes of the husband and wife separately, such as using distinct logistic regression models. For a consistent estimation approach in logistic regression, we refer to Appendix H.1.2 in Jiang (2022)\cite{Jiang2022essays}, where the balancing property with household interference for binary outcomes was proposed for generalized linear models that consider interference. It would be of interest to explore further the estimation and development of optimal decision rules for this approach, in order to compare it with the approach based on household outcomes.

Extending the proposed method to households with varying numbers of individuals, including those with more than two individuals, presents a challenging yet important endeavor. This extension can be likened to addressing a partial interference\cite{sobel2006randomized, park2021optimal} problem, where treatments of individuals blocked by clusters can affect outcomes of the individuals in the same cluster, and households can be viewed as distinct clusters. 
To address the partial interference problem, two modeling are necessary for the investigation: (1) modeling the outcomes, this entails developing regression models for the outcomes. It is worth noting that dealing with high dimensionality (e.g., the treatment indicators of all study units in the cluster and cluster-level pre-treatment covariates) may necessitate additional assumptions, such as conditional stratified interference, to mitigate the challenges posed by the curse of dimensionality; see Section 2.2 of Park et al. (2021) \cite{park2021optimal} for additional discussions about these assumptions; (2) modeling the joint propensity scores: In the context of the joint propensity scores, it's important to consider the potential removal of the treatment-independent assumption as we investigated in Section \ref{C5sec5.2}. This can be achieved by accounting for the associations between treatments of individuals within the same cluster.


Through our analysis of the PATH study, we have demonstrated the practical applicability of our proposed methods.  We estimated a treatment decision function for household pairs to maximize the probability of achieving smoking cessation under the assumptions of a model for their treatment and success. In particular, we modeled the potential association of e-cigarette usage between members of a household pair and estimated the joint propensity scores that play a crucial role in approximately doubly robust estimation with interference. We acknowledge some limitations of our analysis: the PATH data points are a year apart, which is not an ideal spacing for treatment decisions, and the size and period of the PATH subsample provide insufficient data on some of the possible treatment sequences to make the findings easily interpretable or conclusive. In addition, we have implicitly assumed that there is meaning in having been the first of the two household members to be interviewed.  While this could well be the case in practice (e.g., the household head is interviewed first), it is important to recognize the semi-arbitrary nature of this labeling, and to assess its impact in future research. Due to these limitations, it is essential to note that the results of our PATH analysis are not intended as authentic treatment recommendations for smoking cessation. They nonetheless serve to demonstrate the underlying principles of household interference in such a context and the methodology we propose in this analysis.

%% file: Sections/appendix.tex
\title{\textbf{Supporting Information for ``Estimating dynamic treatment regimes for ordinal outcomes with household
interference: Application in household
smoking cessation''}}
\section*{Supplementary Materials Appendix} \label{AppendixC5}
\section*{Appendix A: Proof of Theorem \ref{thmC5ordout}} \label{C5PfThm}

In this Appendix section, we will prove Theorem \ref{thmC5ordout}. We assume that the ordinal outcome $U_h$ that takes the value $c$ follows a multinomial distribution that $U^{*}_h \sim multinom(\boldsymbol{\pi}_h, 1)$, where $U^{*}$ is ``one-hot'' encoded as a $C-$vector with a $1$ at the $c$th entry and 0 otherwise. Then, the likelihood function is:
\begin{equation*}
    \mathbb{P}\left(\boldsymbol{U}_{h}^{*}=\boldsymbol{u}_{h}^{*}\right)=\prod_{c} \pi_{h c}^{u_{h c}^{*}}, \ for\  c= 1,2,..., C,
\end{equation*}
where $\pi_{hc} = \mathbb{P}(U_h = c)$. In our $C=3$ case, building on the assumed POM (\ref{POMpro}) and denoting $\eta_{1h} = \zeta_1 - \boldsymbol{\beta}^{\top}\boldsymbol{x}^{\beta}_h -a^{s}  \boldsymbol{\xi}^{\top}\boldsymbol{x}^{\xi}_h - a^{r}\boldsymbol{\psi}^{\top}\boldsymbol{x}^{\psi}_h- a^{s} a^{r}\boldsymbol{\phi}^{\top}\boldsymbol{x}^{\phi}_h$ and $\eta_{2h} = \zeta_2 - \boldsymbol{\beta}^{\top}\boldsymbol{x}^{\beta}_h -a^{s}  \boldsymbol{\xi}^{\top}\boldsymbol{x}^{\xi}_h - a^{r}\boldsymbol{\psi}^{\top}\boldsymbol{x}^{\psi}_h- a^{s} a^{r}\boldsymbol{\phi}^{\top}\boldsymbol{x}^{\phi}_h$, then we have
\begin{align*}
   \pi_{h1} = \mathbb{P}(U_h =1 \mid a^{s} ,a^{r}, \boldsymbol{x}_h) &= {g^{-1}}^{}[\eta_{1h}] \\
\pi_{h2} = \mathbb{P}(U_h =2 \mid a^{s} ,a^{r}, \boldsymbol{x}_h) &= {g^{-1}}^{}[\eta_{2h}] -{g^{-1}}^{}[\eta_{1h}] \\
\pi_{h3} = \mathbb{P}(U_h =3 \mid a^{s} ,a^{r}, \boldsymbol{x}_h) &=1 - {g^{-1}}^{}[\eta_{2h}],
\end{align*}
where $g^{-1}$ is the inverse of the link function. However, we assume that  the true POM model has $\eta_{1h} = \zeta_1 - f(\boldsymbol{x}^{\beta}_h)  -a^{s}  \boldsymbol{\xi}^{\top}\boldsymbol{x}^{\xi}_h - a^{r}\boldsymbol{\psi}^{\top}\boldsymbol{x}^{\psi}_h- a^{s} a^{r}\boldsymbol{\phi}^{\top}\boldsymbol{x}^{\phi}_h$ and $\eta_{2h} = \zeta_2 - f(\boldsymbol{x}^{\beta}_h)  -a^{s}  \boldsymbol{\xi}^{\top}\boldsymbol{x}^{\xi}_h - a^{r}\boldsymbol{\psi}^{\top}\boldsymbol{x}^{\psi}_h- a^{s} a^{r}\boldsymbol{\phi}^{\top}\boldsymbol{x}^{\phi}_h$, for an arbitrary treatment-free function $f(\boldsymbol{x}^{\beta}_h)$.
Denoting the nuisance parameters as $\boldsymbol{\theta_1} = (\boldsymbol{\zeta} , \boldsymbol{\beta})$ and the parameters of interest as $\boldsymbol{\theta_2} = (\boldsymbol{\xi} , \boldsymbol{\psi}, \boldsymbol{\phi})$, the log-likelihood function is 
\begin{align*}
    \ell\left(\boldsymbol{\theta_1}, \boldsymbol{\theta_2} ; \boldsymbol{u}^{*}\right) &=\sum_{h} \sum_{c} u_{hc}^{*} \log \pi_{hc}\\
    &=\sum_{h} \sum_{c} \mathbb{I}(u_h = c) \log \pi_{hc}.
\end{align*}
Denoting $ \boldsymbol{\theta} = ( \boldsymbol{\theta_1},  \boldsymbol{\theta_2}) $ as the whole parameter in POM (\ref{POMpro}), then the corresponding score function system components are
\begin{align*}
    \sum_{h} \sum_{c} \mathbb{I}(u_h = c)  (\pi_{hc})^{-1} \partial \pi_{hc}/ \partial \boldsymbol{\theta}.
\end{align*}

Further, denoting $\mathcal{V}^{0}(A_h, X_h):= \left( 0, 1 , -X_h^{\beta} , -A^s_hX_h^{\xi} , -A^r_hX_h^{\psi} , -A^s_hA^r_hX_h^{\phi}\right)^{\top}$ and $\mathcal{V}^{1}(A_h, X_h):= \left(\begin{array}{c} 1, 0 , -X_h^{\beta} , -A^s_hX_h^{\xi} , -A^r_hX_h^{\psi} , -A^s_hA^r_hX_h^{\phi}\end{array}\right)^{\top},$ we give the unweighted score function system.  
In particular, in our case we have three components of the score estimation equation system, which are
\begin{align*}
    \sum_{h} \mathbb{I}(u_h = 1) ({g^{-1}}^{}[\eta_{1h}])^{-1} {g^{-1}}^{\prime}[\eta_{1h}] \mathcal{V}^{1}(A_h, X_h),
\end{align*}
\begin{align*}
   \sum_{h} \mathbb{I}(u_h = 2) ({g^{-1}}^{}[\eta_{2h}] - {g^{-1}}^{}[\eta_{1h}])^{-1} \left[ {g^{-1}}^{\prime}[\eta_{2h}]\mathcal{V}^{0}(A_h, X_h)
   - {g^{-1}}^{\prime}[\eta_{1h}] \mathcal{V}^{1}(A_h, X_h) \right], 
\end{align*}
\begin{align*}
    -\sum_{h} \mathbb{I}(u_h = 3) (1- {g^{-1}}^{}[\eta_{2h}])^{-1} {g^{-1}}^{\prime}[\eta_{2h}] \mathcal{V}^{0}(A_h, X_h),
\end{align*}
respectively. That is, the score estimation equation system is 
\begin{align} \label{POMscore}
    \sum_{h} S_{1h} \mathcal{V}^{1}(A_h, X_h) + \sum_{h} S_{2h} \mathcal{V}^{0}(A_h, X_h) = \boldsymbol{0},
\end{align} where $S_{1h}$ and $S_{2h}$ are defined as \begin{equation*} 
  \left\{
    \begin{aligned}
      &  S_{1h}:= \left[ \mathbb{I}(u_h = 1) ({g^{-1}}^{}[\eta_{1h}])^{-1} - \mathbb{I}(u_h = 2) ({g^{-1}}^{}[\eta_{2h}] - {g^{-1}}^{}[\eta_{1h}])^{-1} \right] {g^{-1}}^{\prime}[\eta_{1h}]\\
      & S_{2h} :=\left[ \mathbb{I}(u_h = 2) ({g^{-1}}^{}[\eta_{2h}] - {g^{-1}}^{}[\eta_{1h}])^{-1} -\mathbb{I}(u_h = 3) (1- {g^{-1}}^{}[\eta_{2h}])^{-1}\right]{g^{-1}}^{\prime}[\eta_{2h}].
    \end{aligned}
  \right.
\end{equation*}
There are two major components of the POM score equation system in (\ref{POMscore}). One of the components is the top two equations associated with $\zeta_1$ and $\zeta_2$, and the other is the bottom equations related to $(\boldsymbol{\beta}, \boldsymbol{\xi} , \boldsymbol{\psi}, \boldsymbol{\phi})$.
The top two rows yield $\sum_{h} S_{1h} = 0$ and $\sum_{h} S_{2h} = 0$.
Further, for convenience, we denote by $b_h = (b_{1h}, b_{2h})^\top$ the vector $(-S_{1h}/\left(S_{1h} + S_{2h} \right), -S_{2h}/\left(S_{1h} + S_{2h} \right))^\top$. In addition, considering the balancing weights, we can write the weighted analogues of (\ref{POMscore}) as:
\begin{align*}
    \sum_{h} \left(S_{1h} + S_{2h} \right)w(A^s_h, A^r_h, X_h) \left(\begin{array}{c} b_h \\ X_h^{\beta} \\ A^s_hX_h^{\xi} \\ A^r_hX_h^{\psi} \\ A^s_hA^r_hX_h^{\phi}\end{array}\right)  = \sum_{h} \left(\begin{array}{c}  V_{0h} \\ V_{1h} \\ V_{2h} \\ V_{3h}  \\ V_{4h} \end{array}\right) = \boldsymbol{0};
\end{align*}
where \begin{equation*} \label{forzeta}
    \sum_{h} V_{0h} = \sum_{h}\left(\begin{array}{c} S_{1h}\\ S_{2h} \end{array}\right) w(A^s_h, A^r_h, X_h)=  \boldsymbol{0}
\end{equation*} are additional equations that are used for solving $\zeta_1$, $\zeta_2$. Now, for simplicity, we assume that $\zeta_1$ and $\zeta_2$ are known. (In the implementation of the overall estimation procedure, these ``known'' values would be replaced by preliminary estimates.)
If we focus on the logit link, that is, $g^{-1}(t) = [1 + \mathrm{exp}(-t)]^{-1}$, then we have some properties, such as ${g^{-1}}^{\prime}(t) = g^{-1}(t)[1- g^{-1}(t)]$ or ${g^{-1}}^{\prime}(t) = g^{-1}(t) g^{-1}(- t)$, where $[1- g^{-1}(t)] = g^{-1}(-t)$. Thus, we have 
\begin{equation*}
    \begin{cases}
     \mathbb{I}(u_h = 1) ({g^{-1}}^{}[\eta_{1h}])^{-1}{g^{-1}}^{\prime}[\eta_{1h}] = \mathbb{I}(u_h = 1)(1 - {g^{-1}}^{}[\eta_{1h}]); \\
     \mathbb{I}(u_h = 3) (1 - {g^{-1}}^{}[\eta_{2h}])^{-1}{g^{-1}}^{\prime}[\eta_{2h}] = \mathbb{I}(u_h = 3){g^{-1}}^{}[\eta_{2h}].
    \end{cases}\,
\end{equation*}
In addition, we also have 
\begin{align*}
    & \mathbb{I}(u_h = 2) \left({g^{-1}}^{}[\eta_{2h}] - {g^{-1}}^{}[\eta_{1h}]\right)^{-1} \left( {g^{-1}}^{\prime}[\eta_{2h}] - {g^{-1}}^{\prime}[\eta_{1h}]\right)\\
    =& \mathbb{I}(u_h = 2) \left({g^{-1}}^{}[\eta_{2h}] - {g^{-1}}^{}[\eta_{1h}]\right)^{-1} \left( {g^{-1}}[\eta_{2h}] - {g^{-1}}^2[\eta_{2h}] - {g^{-1}}[\eta_{1h}] + {g^{-1}}^2[\eta_{1h}]\right)\\
    =& \mathbb{I}(u_h = 2) \left({g^{-1}}^{}[\eta_{2h}] - {g^{-1}}^{}[\eta_{1h}]\right)^{-1} \left( {g^{-1}}[\eta_{2h}] - {g^{-1}}[\eta_{1h}] + {g^{-1}}^2[\eta_{1h}] - {g^{-1}}^2[\eta_{2h}] \right)\\
    =& \mathbb{I}(u_h = 2) \left( 1 - {g^{-1}}[\eta_{1h}] - {g^{-1}}[\eta_{2h}] \right),
\end{align*}
where the first equality is based on the fact that $ {g^{-1}}^{\prime}[\eta_{2h}] - {g^{-1}}^{\prime}[\eta_{1h}]   =  {g^{-1}}[\eta_{2h}](1 - {g^{-1}}[\eta_{2h}]) - {g^{-1}}[\eta_{1h}](1- {g^{-1}}[\eta_{1h}]),$ and the last follows by cancelling the factor $\left({g^{-1}}^{}[\eta_{2h}] - {g^{-1}}^{}[\eta_{1h}]\right)^{-1}$.
Thus, we have 
\begin{align*}
    &S_{1h} + S_{2h}\\ = & \mathbb{I}(u_h = 1)(1 - {g^{-1}}^{}[\eta_{1h}]) + \mathbb{I}(u_h = 2) \left( 1 - {g^{-1}}[\eta_{1h}] - {g^{-1}}[\eta_{2h}] \right) - \mathbb{I}(u_h = 3){g^{-1}}^{}[\eta_{2h}] \nonumber \\  =& [\mathbb{I}(u_h = 1) + \mathbb{I}(u_h = 2)] (1 - {g^{-1}}^{}[\eta_{1h}])  - [\mathbb{I}(u_h = 2) + \mathbb{I}(u_h = 3)] {g^{-1}}^{}[\eta_{2h}].
\end{align*}
Therefore,  without
loss of generality, we assume that $X^{\beta} = X^{\xi} = X^{\psi} = X^{\phi} = X$; thus, we have 
\begin{align*}
& \sum_{h} \left[\begin{array}{c} V_{1h} - V_{2h} - V_{3h} + V_{4h} \\ V_{2h} - V_{4h} \\ V_{3h} - V_{4h} \\ V_{4h} \end{array}\right] \\ =& \sum_{h} \left[\left(\begin{array}{c} (1- A_h^s)(1-A_h^r)\\ (1-A_h^r)A_h^s \\ (1-A_h^s)A_h^r  \\ A_h^rA_h^s\end{array}\right) X w(A^s_h, A^r_h, X_h) \left(S_{1h} + S_{2h} \right)  \right] = \boldsymbol{0}.
\end{align*}
$ \sum_{h} (V_{1h} - V_{2h} - V_{3h} + V_{4h})$ uses the untreated part of the sample, and can be used to estimate  $\boldsymbol{\beta}$.  That is, $\hat{\boldsymbol { \beta }} $ can be solved by $ \sum_{h} (V_{1h} - V_{2h} - V_{3h} + V_{4h}) = \sum_h^H(1-A^s_h)(1- A^r_h)X_hw(A^s_h, A^r_h,X_h) (S^{00}_{1h} + S^{00}_{2h}) = \boldsymbol{0},$
where $S^{00}_{1h} + S^{00}_{2h}$ is defined as 
\begin{align*}
     S^{00}_{1h} + S^{00}_{2h} := & [\mathbb{I}(u_h = 1) + \mathbb{I}(u_h = 2)] (1 - {g^{-1}}^{}[\zeta_1 -\boldsymbol { {\beta}}^{\top} X_h])  \\ & - [\mathbb{I}(u_h = 2) + \mathbb{I}(u_h = 3)] {g^{-1}}^{}[\zeta_2 -\boldsymbol { {\beta}}^{\top} X_h].
\end{align*}
If the treatment-free model is not correct, the estimates  will actually estimate $\boldsymbol{\beta}^*$, which is defined as the solution of  $\mathbb{E}_{X}\left[\pi^{00}(X)Xw(0, 0,X) (\mathcal{S}^{00}_{1} + \mathcal{S}^{00}_{2} ) \right]= \boldsymbol{0},$ that is, $\mathbb{E}_{X}\left[\pi^{00}(X)Xw(0,0,X) (\mathcal{S}^{00}_{1} + \mathcal{S}^{00}_{2} ) \right] \Big|_{\boldsymbol{\beta} = \boldsymbol { {\beta}^*}}= \boldsymbol{0},$   where $\mathcal{S}^{00}_{1} + \mathcal{S}^{00}_{2}$ is defined as 
\begin{align*}
        &\mathcal{S}^{00}_{1} +  \mathcal{S}^{00}_{2} \\ = & {g^{-1}}^{}[ \zeta_2 -  f(X)] (1 - {g^{-1}}^{}[\zeta_1 -\boldsymbol{\beta}^{\top} X]) - (1 - {g^{-1}}^{}[ \zeta_1 -  f(X)]){g^{-1}}^{}[\zeta_2 -\boldsymbol{\beta}^{\top} X] \nonumber \\
        = & {g^{-1}}^{}[ \zeta_2 -  f(X)] - {g^{-1}}^{}[\zeta_2 -\boldsymbol{\beta}^{\top} X] +
        {g^{-1}}^{}[ \zeta_1 -  f(X)]{g^{-1}}^{}[\zeta_2 -\boldsymbol{\beta}^{\top} X] \\ & - {g^{-1}}^{}[ \zeta_2 -  f(X)] {g^{-1}}^{}[\zeta_1 -\boldsymbol{\beta}^{\top} X].
\end{align*}

Note that, conditional on treatments and covariates, $\mathbb{E}[I(u = 1)] = \pi_1 = g^{-1}[\eta_{1}] = {g^{-1}}^{}[ \zeta_1 -  f(X)]$, which depends on the true outcome models in terms of the treatment-free function $f(X)$.  Similar results are derived from $\mathbb{E}[I(u = 2)]$ and $\mathbb{E}[I(u = 3)]$. That is, $\mathbb{E}[I(u = 2)] = \pi_2 = g^{-1}[\eta_{2}] - g^{-1}[\eta_{1}] = {g^{-1}}^{}[ \zeta_2 -  f(X)] - {g^{-1}}^{}[ \zeta_1 -  f(X)]$, and $\mathbb{E}[I(u = 3)] = \pi_3 = 1 - g^{-1}[\eta_{2}] = 1- {g^{-1}}^{}[ \zeta_2 -  f(X)]$. Thus, conditional on treatments and covariates, we have  $\mathbb{E}[I(u = 1)] + \mathbb{E}[I(u = 2)] = {g^{-1}}^{}[ \zeta_2 -  f(X)]$, and $\mathbb{E}[I(u = 2)] + \mathbb{E}[I(u = 3)] = 1 - {g^{-1}}^{}[ \zeta_1 -  f(X)]$. 

To sum up, $\sum_{h} (V_{1h} - V_{2h} - V_{3h} + V_{4h}) = \boldsymbol{0}$ can be solved for $\hat{\boldsymbol { \beta }}$. According to large sample theory, $\hat{\boldsymbol { \beta }} $ tends to converge to $\boldsymbol { {\beta}^{*}}$, as $H \rightarrow \infty$\cite{white1982maximum}, where $\boldsymbol { {\beta}^{*}}$ is the solution of $H\mathbb{E}_{X}\left[\pi^{00}(X)Xw(0,0,X) (\mathcal{S}^{00}_{1} + \mathcal{S}^{00}_{2} ) \right]= \boldsymbol{0}$, with $\mathcal{S}^{00}_{1} +  \mathcal{S}^{00}_{2}:= {g^{-1}}^{}[ \zeta_2 -  f(X)] (1 - {g^{-1}}^{}[\zeta_1 -\boldsymbol{\beta}^{\top} X]) - (1 - {g^{-1}}^{}[ \zeta_1 -  f(X)]){g^{-1}}^{}[\zeta_2 -\boldsymbol{\beta}^{\top} X]$.


Then the expectation of $\sum_h^H (V_{1h} - V_{2h} - V_{3h} + V_{4h})$ conditional on $(X_1, ...X_H)$,  that is, 
\begin{equation} \label{C5eqtn00}
    \sum_h^H \pi^{00}X_hw(0, 0, X_h)\left[ (\mathcal{S}^{00}_{1h} +  \mathcal{S}^{00}_{2h})(\boldsymbol{\zeta}, \boldsymbol{\beta})\right],
\end{equation}
where, in terms of parameters of $\boldsymbol{\zeta}$, and $\boldsymbol{\beta}$, $(\mathcal{S}^{00}_{1h} +  \mathcal{S}^{00}_{2h})(\boldsymbol{\zeta}, \boldsymbol{\beta})$ is defined as:
\begin{align*}
    & (\mathcal{S}^{00}_{1h} +  \mathcal{S}^{00}_{2h})(\boldsymbol{\zeta}, \boldsymbol{\beta}):= \\ &g^{-1}\left[ \zeta_2 - f(X_h)\right]- g^{-1}\left(\zeta_2 - \boldsymbol { \beta }^{\top} X_h\right)  +
        {g^{-1}}^{}[ \zeta_1 -  f(X_h)]{g^{-1}}^{}[\zeta_2 -\boldsymbol { {\beta}}^{\top} X_h] \\ &- {g^{-1}}^{}[ \zeta_2 -  f(X_h)] {g^{-1}}^{}[\zeta_1 -\boldsymbol { {\beta}}^{\top} X_h],
\end{align*}
can be written using a Taylor series expansion, function $g^{-1}[\zeta - f(X_h)]$ at the points $ \zeta_1 - \boldsymbol { \beta }^{\top} X_h$ and $ \zeta_2 - \boldsymbol { \beta }^{\top} X_h$, respectively, as 
\begin{align} \label{C5eqtn00V1}
\sum_h^H \pi^{00}X_h w(0, 0, X_h) \left[\mathcal{K}\Delta+\mathcal{O}[\Delta^2]\right],
\end{align}
where \begin{equation*}
    \mathcal{K} := {g^{-1}}(\mathcal{Z}_2)\left[1 - g^{-1}(\mathcal{Z}_1)\right]\left[1 - g^{-1}(\mathcal{Z}_2) + g^{-1}(\mathcal{Z}_1)\right],
\end{equation*} where $\Delta :=  \boldsymbol { \beta}^{\top} X_h - f(X_h) $, $\mathcal{Z}_2 := \zeta_2 - \boldsymbol {\beta}^{\top} X_h$ and $\mathcal{Z}_1 := \zeta_1 - \boldsymbol {\beta}^{\top} X_h$, and the big $\mathcal{O}$ describes the error term in an approximation to the $g^{-1}$ function. To prove this, we initially focus on the first two terms of $\mathcal{S}^{00}_{1h} +  \mathcal{S}^{00}_{2h}$ in expression (\ref{C5eqtn00}), that is, $g^{-1}\left[ \zeta_2 - f(X_h)\right]- g^{-1}\left(\mathcal{Z}_2 \right) = {g^{-1}}^{\prime}(\mathcal{Z}_2)(\Delta)+\mathcal{O}[\Delta^2]$. Then the third and fourth terms are
 \begin{equation*} 
  \left\{
    \begin{aligned}
      &  {g^{-1}}^{}[ \zeta_1 -  f(X_h)]{g^{-1}}^{}[\mathcal{Z}_2 ]  = \left( g^{-1}\left(\mathcal{Z}_1\right) + {g^{-1}}^{\prime}(\mathcal{Z}_1)(\Delta)+\mathcal{O}[\Delta^2] \right)   {g^{-1}}^{}[\mathcal{Z}_2 ];\\
      &  {g^{-1}}^{}[ \zeta_2 -  f(X_h)] {g^{-1}}^{}[\mathcal{Z}_1] = \left( g^{-1}\left(\mathcal{Z}_2 \right) + {g^{-1}}^{\prime}(\mathcal{Z}_2)(\Delta)+\mathcal{O}[\Delta^2] \right)  {g^{-1}}^{}[\mathcal{Z}_1].
\    \end{aligned}
  \right.
\end{equation*} 
Thus, the Taylor series expansion of $\mathcal{S}^{00}_{1h} +  \mathcal{S}^{00}_{2h}$ in expression (\ref{C5eqtn00}) is
 \begin{align}\label{C5eqtn00mid}
     \left( {g^{-1}}^{\prime}(\mathcal{Z}_2) + {g^{-1}}^{\prime}(\mathcal{Z}_1) {g^{-1}}^{}(\mathcal{Z}_2) \nonumber  -  {g^{-1}}^{\prime}(\mathcal{Z}_2){g^{-1}}^{}(\mathcal{Z}_1)  \right) (\Delta)+\mathcal{O}[\Delta^2].  
 \end{align}
Further, using the property of the $g^{-1}$ that ${g^{-1}}^{\prime}(t) = g^{-1}(t)[1- g^{-1}(t)]$, we have the Taylor series expansion of $(\mathcal{S}^{00}_{1h} +  \mathcal{S}^{00}_{2h})(\boldsymbol{\zeta}, \boldsymbol{\beta})$ is
\begin{equation*}
    (\mathcal{S}^{00}_{1h} +  \mathcal{S}^{00}_{2h})(\boldsymbol{\zeta}, \boldsymbol{\beta}) = \mathcal{K}\Delta+\mathcal{O}[\Delta^2].
\end{equation*} Therefore, we finally have expression (\ref{C5eqtn00V1}). 
Then, we consider $\sum_h^H (V_{2h} - V_{4h})  = \boldsymbol{0}$ and $\sum_h^H (V_{3h} -V_{4h})  = \boldsymbol{0}$ which can be solved for $\hat{\boldsymbol{\xi}}$ and $\hat{\boldsymbol{\psi}}$, respectively, in terms of $\hat{\zeta}_1, \hat{\zeta}_2$, and $\hat{\boldsymbol {\beta }}$.  First, we study $\sum_h^H (V_{2h} - V_{4h}) = \boldsymbol{0}$, that is, $\sum_h^H (1-A_h^r)A_h^sX w(1, 0, X_h)\ \left( \mathcal{S}_{1h} +  \mathcal{S}_{2h} \right)  = \boldsymbol{0}$, which provides $\hat{\boldsymbol{\xi}}$ in terms of $\hat{\zeta}_1, \hat{\zeta}_2$, and $\hat{\boldsymbol {\beta }}$. To show that $\hat{\boldsymbol{\xi}}$ is (approximately) consistent, we would need to show that the expectation of $\sum_h^H(1-A^r_h)A^s_hX_h w(1,0, X_h)\ (\mathcal{S}^{10}_{1h} + \mathcal{S}^{10}_{2h})(\boldsymbol{\zeta}, \boldsymbol{\beta^*}, \boldsymbol{\xi})$ equals or is close to $\boldsymbol{0}$ for general $\boldsymbol { \xi }$, where $(\mathcal{S}^{10}_{1h} +  \mathcal{S}^{10}_{2h})(\boldsymbol{\zeta}, \boldsymbol{\beta^*}, \boldsymbol{\xi})$ is defined as 
\begin{align*}
    (\mathcal{S}^{10}_{1h} +  \mathcal{S}^{10}_{2h})(\boldsymbol{\zeta}, \boldsymbol{\beta^*}, \boldsymbol{\xi}) &:=  {g^{-1}}^{}[ \zeta_2 -  f(X_h) -  \boldsymbol { \xi }^{\top}  X_h] (1 - {g^{-1}}^{}[\zeta_1 -\boldsymbol { {\beta}^*}^{\top} X_h -  \boldsymbol { \xi }^{\top}  X_h]) \\ & - (1 - {g^{-1}}^{}[ \zeta_1 -  f(X_h) -  \boldsymbol { \xi }^{\top}  X_h]){g^{-1}}^{}[\zeta_2 -\boldsymbol { {\beta}^*}^{\top} X_h -  \boldsymbol { \xi }^{\top}  X_h] \nonumber \\
    & = {g^{-1}}^{}[ \zeta_2 -  f(X_h) -  \boldsymbol { \xi }^{\top}  X_h] - {g^{-1}}^{}[\zeta_2 -\boldsymbol { {\beta}^*}^{\top} X_h -  \boldsymbol { \xi }^{\top}  X_h] \nonumber \\
    & +  {g^{-1}}^{}[ \zeta_1 -  f(X_h) -  \boldsymbol { \xi }^{\top}  X_h]{g^{-1}}^{}[\zeta_2 -\boldsymbol { {\beta}^*}^{\top} X_h -  \boldsymbol { \xi }^{\top}  X_h] \\ & - {g^{-1}}^{}[ \zeta_2 -  f(X_h) -  \boldsymbol { \xi }^{\top}  X_h] {g^{-1}}^{}[\zeta_1 -\boldsymbol { {\beta}^*}^{\top} X_h -  \boldsymbol { \xi }^{\top}  X_h].
\end{align*}
Similar to the Taylor series expansion of $\mathcal{S}^{00}_{1h} +  \mathcal{S}^{00}_{2h}$ in expression (\ref{C5eqtn00}), we have 
\begin{align} \label{ss01}
    & (\mathcal{S}^{10}_{1h} +  \mathcal{S}^{10}_{2h})(\boldsymbol{\zeta}, \boldsymbol{\beta^*}, \boldsymbol{\xi})\nonumber \\ = & {g^{-1}}(\zeta_2 - \boldsymbol { {\beta}^*}^{\top} X_h  -  \boldsymbol { \xi }^{\top}  X_h) \left[1 - g^{-1}(\zeta_1 - \boldsymbol { {\beta}^*}^{\top} X_h -  \boldsymbol { \xi }^{\top}  X_h)\right] \nonumber \\ & \left[1 - g^{-1}(\zeta_2 - \boldsymbol { {\beta}^*}^{\top} X_h -  \boldsymbol { \xi }^{\top}  X_h)  + g^{-1}(\zeta_1 - \boldsymbol { {\beta}^*}^{\top} X_h -  \boldsymbol { \xi }^{\top}  X_h)\right](\Delta^*)+\mathcal{O}[ {\Delta^*}^2],
\end{align}
where $\Delta^* := \boldsymbol { \beta^*}^{\top} X_h - f(X_h)$. If we denote $\eta_1^{a^s,a^r}( \boldsymbol{\beta}^*) = \zeta_1 + \boldsymbol {{\beta} }^{\top} \boldsymbol{x}^{\beta} + \boldsymbol { \xi }^{\top} a^s\boldsymbol{x}^{\xi}+ \boldsymbol { \psi }^{\top} a^r\boldsymbol{x}^{\psi} + \boldsymbol { \phi }^{\top} a^sa^r\boldsymbol{x}^{\phi}$ and $\eta_2^{a^s,a^r}( \boldsymbol{\beta}^*) = \zeta_2 + \boldsymbol {{\beta} }^{\top} \boldsymbol{x}^{\beta} + \boldsymbol { \xi }^{\top} a^s\boldsymbol{x}^{\xi}+ \boldsymbol { \psi }^{\top} a^r\boldsymbol{x}^{\psi} + \boldsymbol { \phi }^{\top} a^sa^r\boldsymbol{x}^{\phi}$, then, from equation (\ref{ss01}), we have
\begin{equation*}
    \begin{split}
        (\mathcal{S}^{10}_{1h} +  \mathcal{S}^{10}_{2h})(\boldsymbol{\zeta}, \boldsymbol{\beta^*}, \boldsymbol{\xi})= {g^{-1}}[\eta_2^{1,0}( \boldsymbol{\beta}^*)]\left[1 - g^{-1}[\eta_1^{1,0}( \boldsymbol{\beta}^*)]\right]\left[1 - g^{-1}[\eta_2^{1,0}( \boldsymbol{\beta}^*)] \right. \\
    \left. +\  g^{-1}[\eta_1^{1,0}( \boldsymbol{\beta}^*)]\right](\Delta^*)+\mathcal{O}[ {\Delta^*}^2].
    \end{split}
\end{equation*}
If this is not the case, then the expectation of equation $\sum_i^n(V_{2h} - V_{4h}) = \boldsymbol{0}$  with $\boldsymbol{\beta} = \hat{\boldsymbol{\beta}}$ and $\boldsymbol{\xi} = \hat{\boldsymbol{\xi}}$ may approach the expectation of equation $\sum_h^n(V_{2h} - V_{4h})= \boldsymbol{0}$  with $\boldsymbol{\beta}$ set equal to $\boldsymbol { {\beta}^{*}}$ and $\boldsymbol{\xi}$ set equal to a similar limiting value $\boldsymbol { {\xi}^{*}}$ as $n \rightarrow \infty$.  The vector $\boldsymbol { {\xi}^{*}}$  will satisfy the condition that the expectation of  $$\sum_{h}^{n}A^{s}_h(1 - A^r_h)X_hw(1,0, X_h)\left[g^{-1}(-\boldsymbol { \xi^* }^{\top}  X_h^{}-\boldsymbol {{\beta }^{*}}^{\top} X_h^{}) - g^{-1}(-\boldsymbol { \xi^* }^{\top}  X_h^{}-f(X_h))\right]$$ equals or is  close to $\boldsymbol{0}$, but $\boldsymbol { {\xi}^{*}}$ will in general be different from the true $\boldsymbol { {\xi}}$.

Next, we study $\sum_h^H (V_{3h} - V_{4h}) = \boldsymbol{0}$, that is, $\sum_h^H (1-A^s)A^rX_hw(0, 1,X_h) \left( \mathcal{S}_{1h} +  \mathcal{S}_{2h}  \right)  = \boldsymbol{0}$, which offers $\hat{\boldsymbol{\psi}}$ in terms of $\hat{\zeta}_1, \hat{\zeta}_2$, and $\hat{\boldsymbol {\beta }}$. To show that $\hat{\boldsymbol{\psi}}$ is (approximately) consistent, we would need to show that the expectation of $\sum_h^H(1-A^s_h)A^r_hX_h w(0, 1,X_h) (\mathcal{S}^{01}_{1h} + \mathcal{S}^{01}_{2h})(\boldsymbol{\zeta}, \boldsymbol{\beta^*}, \boldsymbol{\psi})$ equals or is close to $\boldsymbol{0}$ for general $\boldsymbol { \psi }$, where $(\mathcal{S}^{01}_{1h} + \mathcal{S}^{01}_{2h})(\boldsymbol{\zeta}, \boldsymbol{\beta^*}, \boldsymbol{\psi})$ is defined as 
\begin{align*}
    (\mathcal{S}^{01}_{1h} + \mathcal{S}^{01}_{2h})(\boldsymbol{\zeta}, \boldsymbol{\beta^*}, \boldsymbol{\psi}) &:= {g^{-1}}^{}[ \zeta_2 -  f(X_h) -  \boldsymbol { \psi }^{\top}  X_h] (1 - {g^{-1}}^{}[\zeta_1 -\boldsymbol { {\beta}^*}^{\top} X_h -  \boldsymbol { \psi }^{\top}  X_h]) \nonumber \\ & - (1 - {g^{-1}}^{}[ \zeta_1 -  f(X_h) -  \boldsymbol { \psi }^{\top}  X_h]){g^{-1}}^{}[\zeta_2 -\boldsymbol { {\beta}^*}^{\top} X_h -  \boldsymbol { \psi }^{\top}  X_h] \nonumber \\
    &= {g^{-1}}^{}[ \zeta_2 -  f(X_h) -  \boldsymbol { \psi }^{\top}  X_h] - {g^{-1}}^{}[\zeta_2 -\boldsymbol { {\beta}^*}^{\top} X_h -  \boldsymbol { \psi }^{\top}  X_h] \nonumber\\
    &+ {g^{-1}}^{}[ \zeta_1 -  f(X_h) -  \boldsymbol { \psi }^{\top}  X_h]{g^{-1}}^{}[\zeta_2 -\boldsymbol { {\beta}^*}^{\top} X_h -  \boldsymbol { \psi }^{\top}  X_h] \nonumber \\ & - {g^{-1}}^{}[ \zeta_2 -  f(X_h) -  \boldsymbol { \psi }^{\top}  X_h] {g^{-1}}^{}[\zeta_1 -\boldsymbol { {\beta}^*}^{\top} X_h -  \boldsymbol { \psi }^{\top}  X_h].
\end{align*}
Again, consistent with the Taylor series expansion of $(\mathcal{S}^{00}_{1h} +  \mathcal{S}^{00}_{2h})(\boldsymbol{\zeta}, \boldsymbol{\beta})$ in expression (\ref{C5eqtn00}), we have
\begin{align*}
    & (\mathcal{S}^{01}_{1h} + \mathcal{S}^{01}_{2h})(\boldsymbol{\zeta}, \boldsymbol{\beta^*}, \boldsymbol{\psi})\\  = & {g^{-1}}(\zeta_2 - \boldsymbol { {\beta}^*}^{\top} X_h -  \boldsymbol { \psi }^{\top}  X_h)\left[1 - g^{-1}(\zeta_1 - \boldsymbol { {\beta}^*}^{\top} X_h -  \boldsymbol { \psi }^{\top}  X_h)\right]\\ & \left[1 - g^{-1}(\zeta_2 - \boldsymbol { {\beta}^*}^{\top} X_h -  \boldsymbol { \psi }^{\top}  X_h) + g^{-1}(\zeta_1 - \boldsymbol { {\beta}^*}^{\top} X_h -  \boldsymbol { \psi }^{\top}  X_h)\right](\Delta^*)+\mathcal{O}[ {\Delta^*}^2],
\end{align*}
or equivalently, 
\begin{align*} 
    (\mathcal{S}^{01}_{1h} + \mathcal{S}^{01}_{2h})(\boldsymbol{\zeta}, \boldsymbol{\beta^*}, \boldsymbol{\psi})  = & {g^{-1}}[\eta_2^{0,1}( \boldsymbol{\beta}^*)]\left[1 - g^{-1}[\eta_1^{0,1}( \boldsymbol{\beta}^*)]\right]\\ &\left[1 - g^{-1}[\eta_2^{0,1}( \boldsymbol{\beta}^*)] + g^{-1}[\eta_1^{0,1}( \boldsymbol{\beta}^*)]\right](\Delta^*)+\mathcal{O}[ {\Delta^*}^2],
\end{align*}
Similar arguments are applied to $\sum_h^H  V_{4h} = \boldsymbol{0}$, that is, $\sum_h^H A^sA^rX_h w(1, 1,X_h)  \left( \mathcal{S}_{1h} +  \mathcal{S}_{2h}  \right)  = \boldsymbol{0}$, which offers $\hat{\boldsymbol{\phi}}$ in terms of $\hat{\zeta}_1, \hat{\zeta}_2$, $\hat{\boldsymbol {\beta }}$, $\hat{\boldsymbol{\xi}}$, and $\hat{\boldsymbol{\phi}}$.  To show that $\hat{\boldsymbol{\phi}}$ is (approximately) consistent, we would need to show that the expectation of $\sum_h^HA^s_hA^r_hX_h w(1, 1,X_h) (\mathcal{S}^{11}_{1h} + \mathcal{S}^{11}_{2h})(\boldsymbol{\zeta}, \boldsymbol{\beta^*}, \boldsymbol{\theta}_2)$ equals or is close to $\boldsymbol{0}$ for general $\boldsymbol { \phi }$, where $(\mathcal{S}^{11}_{1h}+ \mathcal{S}^{11}_{2h})(\boldsymbol{\zeta}, \boldsymbol{\beta^*}, \boldsymbol{\theta}_2)$ can be written as by Taylor series expansion 
\begin{equation*}
    \begin{split}
        (\mathcal{S}^{11}_{1h}+ \mathcal{S}^{11}_{2h})(\boldsymbol{\zeta}, \boldsymbol{\beta^*}, \boldsymbol{\theta}_2):= {g^{-1}}[\eta_2^{1,1}( \boldsymbol{\beta}^*)]\left[1 - g^{-1}[\eta_1^{1,1}( \boldsymbol{\beta}^*)]\right]\left[1 - g^{-1}[\eta_2^{1,1}( \boldsymbol{\beta}^*)] \right. \\
    \left. +\  g^{-1}[\eta_1^{1,1}( \boldsymbol{\beta}^*)]\right](\Delta^*)  +\mathcal{O}[ {\Delta^*}^2].
    \end{split}
\end{equation*}

Therefore, the expectation of $\sum_{h} \left[\begin{array}{c} V_{1h} - V_{2h} - V_{3h} + V_{4h} \\ V_{2h} - V_{4h} \\ V_{3h} - V_{4h} \\ V_{4h} \end{array}\right]$ conditional on $(X_1, ...X_n)$,  that is, $$\sum_{h} \left\{\left(\begin{array}{c} \pi^{00}\\ \pi^{10} \\ \pi^{01}\\ \pi^{11}\end{array}\right) X_h \left(\begin{array}{c} w(0,0, X_h)\\ w(1,0, X_h) \\ w(0,1, X_h)\\ w(1,1, X_h)\end{array}\right) \left[\begin{array}{c} (\mathcal{S}^{00}_{1h} +  \mathcal{S}^{00}_{2h})(\boldsymbol{\zeta}, \boldsymbol{\beta})\\ (\mathcal{S}^{10}_{1h} +  \mathcal{S}^{10}_{2h})(\boldsymbol{\zeta}, \boldsymbol{\beta}, \boldsymbol{\xi}) \\ (\mathcal{S}^{01}_{1h} +  \mathcal{S}^{01}_{2h})(\boldsymbol{\zeta}, \boldsymbol{\beta}, \boldsymbol{\psi}) \\ (\mathcal{S}^{11}_{1h} + \mathcal{S}^{11}_{2h})(\boldsymbol{\zeta}, \boldsymbol{\beta}, \boldsymbol{\theta}_2)\end{array}\right] \right\},$$ can be written using a Taylor series expansion as 
\begin{equation}\label{tayloreqtn}
    \sum_{h} \Pi X\Omega \left[\begin{array}{c} {g^{-1}}(\eta_2^{00} )\left[1 - g^{-1}(\eta_1^{00} )\right]\left[1 - g^{-1}(\eta_2^{00} ) + g^{-1}(\eta_1^{00} )\right](\Delta)+\mathcal{O}[\Delta^2]\\ {g^{-1}}(\eta_2^{10} )\left[1 - g^{-1}(\eta_1^{10} )\right]\left[1 - g^{-1}(\eta_2^{10} ) + g^{-1}(\eta_1^{10} )\right](\Delta)+\mathcal{O}[\Delta^2]\\ {g^{-1}}(\eta_2^{01} )\left[1 - g^{-1}(\eta_1^{01} )\right]\left[1 - g^{-1}(\eta_2^{01} ) + g^{-1}(\eta_1^{01} )\right](\Delta)+\mathcal{O}[\Delta^2]\\ {g^{-1}}(\eta_2^{11} )\left[1 - g^{-1}(\eta_1^{11} )\right]\left[1 - g^{-1}(\eta_2^{11} ) + g^{-1}(\eta_1^{11} )\right](\Delta)+\mathcal{O}[\Delta^2] \end{array}\right],
\end{equation}
where $\Pi X\Omega := \left(\begin{array}{c} \pi^{00}\\ \pi^{10} \\ \pi^{01}\\ \pi^{11}\end{array}\right) X_h \left(\begin{array}{c} w(0,0)\\ w(1,0) \\ w(0,1)\\ w(1,1)\end{array}\right)$,  $\eta_1^{a^sa^r} = \zeta_1 + \boldsymbol {{\beta} }^{\top} \boldsymbol{x}^{\beta} + \boldsymbol { \xi }^{\top} a^s\boldsymbol{x}^{\xi}+ \boldsymbol { \psi }^{\top} a^r\boldsymbol{x}^{\psi} + \boldsymbol { \phi }^{\top} a^sa^r\boldsymbol{x}^{\phi}$ and $\eta_2^{a^sa^r} = \zeta_2 + \boldsymbol {{\beta} }^{\top} \boldsymbol{x}^{\beta} + \boldsymbol { \xi }^{\top} a^s\boldsymbol{x}^{\xi}+ \boldsymbol { \psi }^{\top} a^r\boldsymbol{x}^{\psi} + \boldsymbol { \phi }^{\top} a^sa^r\boldsymbol{x}^{\phi}$.

Define $\kappa^{*}(A^s,A^r,X) = {g^{-1}}\left(\eta_{2} \right)\left[1 - g^{-1}\left(\eta_{1}\right)\right]\left[1 - g^{-1}\left(\eta_{2}\right) + g^{-1}\left(\eta_{1}\right) \right]$ with $\eta_{1}(a^s,a^r, X) = \zeta^*_1 + \boldsymbol {{\beta}^{*} }^{\top} X + \boldsymbol { \xi^* }^{\top} a^sX+ \boldsymbol { \psi^* }^{\top} a^rX + \boldsymbol { \phi^* }^{\top} a^sa^rX$, $\eta_{2}(a^s,a^r, X) = \zeta^*_2 + \boldsymbol {{\beta}^{*} }^{\top} X + \boldsymbol { \xi^* }^{\top} a^sX+ \boldsymbol { \psi^* }^{\top} a^rX + \boldsymbol { \phi^* }^{\top} a^sa^rX$,  and $\boldsymbol{ \xi^*}$, $\boldsymbol{ \psi^*}$, and $\boldsymbol{ \phi^*}$ are assumed to be limiting values for $\hat{\boldsymbol{ \xi}}$, $\hat{\boldsymbol{ \psi}}$, and $\hat{\boldsymbol{ \phi}}$, respectively.  Then if weights are defined to satisfy a new balancing criterion \begin{equation*}
    \pi^{0 0} w(0,0) \kappa(0,0) = \pi^{0 1} w(0,1) \kappa(0,1)=\pi^{1 0} w(1,0) \kappa(1,0)=\pi^{1 1} w(1,1)\kappa(1,1),
\end{equation*} and if the distribution of $X$ is such that the inverse link function is close to linear for the range of $f(X) - \boldsymbol{{\beta}^{*}}^{\top}X$ (so that the Taylor expansion error term is small),  the fact that the expectation of the first (top) equation in (\ref{tayloreqtn}) is $\boldsymbol{0}$ for $\boldsymbol{\beta} = \boldsymbol{{\beta}^{*}}$ means that the expectation of the remaining (the second to the forth) equations (\ref{tayloreqtn}) are close to $\boldsymbol{0}$ too for $\boldsymbol{\beta} = \boldsymbol{{\beta}^{*}}$. Again, this argument establishes the approximate consistency of  the corresponding  new estimators of $\boldsymbol{\xi}$, $\boldsymbol{\psi}$, and $\boldsymbol{\phi}$.

To conclude, the corresponding overlap-type weights for POM with ordinal outcomes are:
\begin{equation*} 
    w(a^s, a^r) \propto  \frac{\pi^{0 0}\pi^{1 0}\pi^{0 1}\pi^{1 1}}{\pi^{a^sa^r}} \times \frac{\kappa(0,0, \boldsymbol{x})\kappa(1,0, \boldsymbol{x})\kappa(0,1, \boldsymbol{x})\kappa(1,1, \boldsymbol{x})}{\kappa(a^s, a^r, \boldsymbol{x})},\ \ \text{for}\ a^s,\ a^r = 0, 1.
\end{equation*}
where $\kappa(a^s,a^r, \boldsymbol{x}) = {g^{-1}}\left(\eta_{2} \right)\left[1 - g^{-1}\left(\eta_{1}\right)\right] \left[1 - g^{-1}\left(\eta_{2}\right) + g^{-1}\left(\eta_{1}\right)\right]$ with $\eta_{1}(a^s,a^r, \boldsymbol{x}) = \zeta^*_1 + \boldsymbol {{\beta}^{*} }^{\top} \boldsymbol{x}^{\beta} + \boldsymbol { \xi^* }^{\top} a^s\boldsymbol{x}^{\xi}+ \boldsymbol { \psi^* }^{\top} a^r\boldsymbol{x}^{\psi} + \boldsymbol { \phi^* }^{\top} a^sa^r\boldsymbol{x}^{\phi}$, $\eta_{2}(a^s,a^r, \boldsymbol{x}) = \zeta^*_2 + \boldsymbol {{\beta}^{*} }^{\top} \boldsymbol{x}^{\beta} + \boldsymbol { \xi^* }^{\top} a^s\boldsymbol{x}^{\xi}+ \boldsymbol { \psi^* }^{\top} a^r\boldsymbol{x}^{\psi} + \boldsymbol { \phi^* }^{\top} a^sa^r\boldsymbol{x}^{\phi}$, and $\zeta^*_1$, $\zeta^*_2$, $\boldsymbol { \beta^* }$, $\boldsymbol { \xi^* }$, $\boldsymbol { \psi^* }$ and  $\boldsymbol { \phi^* }$ are the solutions of the estimation functions of POM (\ref{POMpro}) with standard overlap weights (\ref{wetHH1}). 

Thus, to conduct robust estimation for blip parameters in POMs, we need three steps. First, construct ``adjustment factor'' from weighted POMs with balancing weights from either (\ref{wetHH}) or (\ref{wetHH1}). Second, construct new balancing weights that satisfy equation (\ref{wetcInt}). Third, conduct weighted POMs again to acquire approximate double robust estimators of the blip parameters. It is important to note that to estimate the balancing weights for ordinal outcomes in POMs, the key is to precisely estimate the joint propensity function $\pi^{a^sa^r}(\boldsymbol{x}_{s}, \boldsymbol{x}_{r})$.

\section*{Appendix B: Details of estimating the joint propensity score} \label{detailsofJPS} 
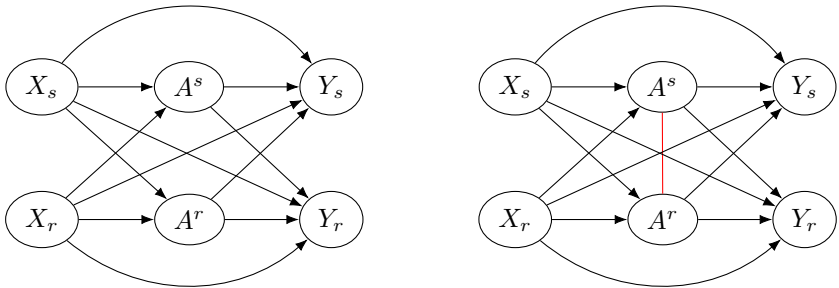
\begin{figure}	
	\begin{subfigure}[t]{.5\textwidth}\centering
	\begin{tikzpicture}[scale=1]
    \node[state] (x) at (0,0) {$X_{s}$};
    \node[state] (a) [right =of x] {$A^{s}$};
    \node[state] (y) [right =of a] {$Y_{s}$};

    \node[state] (x_al) [below =of x] {$X_{r}$};
        \node[state] (h) [right =of x_al] {$A^{r}$};
    \node[state] (yr) [below =of y] {$Y_{r}$};
    
   \path (x_al) edge (y);
    \path (x) edge (a);
    \path (a) edge (y);
      \path (a) edge (yr);
        \path (h) edge (yr);
    \path (x) edge[bend left=50] (y);
        \path (x) edge (h);
    \path (x_al) edge (a);
    \path (h) edge (y);
    \path (x) edge (yr);
    \path (x_al) edge (h);
    \path (x_al) edge[bend right=40] (yr);
    \end{tikzpicture}
		\caption{DAG 1}\label{C5fig:1a}		
	\end{subfigure}
	\begin{subfigure}[t]{.5\textwidth}\centering
	\begin{tikzpicture}[scale=0.5]
    \node[state] (x) at (0,0) {$X_{s}$};
    \node[state] (a) [right =of x] {$A^{s}$};
    \node[state] (y) [right =of a] {$Y_{s}$};

    \node[state] (x_al) [below =of x] {$X_{r}$};
        \node[state] (h) [right =of x_al] {$A^{r}$};
    \node[state] (yr) [below =of y] {$Y_{r}$};
        \draw[red] (h) -- (a)  -- cycle;
    
   \path (x_al) edge (y);
    \path (x) edge (a);
    \path (a) edge (y);
      \path (a) edge (yr);
        \path (h) edge (yr);
    \path (x) edge[bend left=50] (y);
    \path (h) edge (y);
    \path (x) edge (yr);
    \path (x) edge (h);
    \path (x_al) edge (a);
    \path (x_al) edge (h);
    \path (x_al) edge[bend right=40] (yr);
    \end{tikzpicture}
	\caption{DAG 2}\label{C5fig:1b}
	\end{subfigure}
	\caption{Directed acyclic graphs of interference analysis, without (DAG 1) and with (DAG 2) association between $A^s$ and $A^r$ conditional on $(\boldsymbol{x}_s, \boldsymbol{x}_r)$.}\label{fig:P3DAG1}
\end{figure}

In this appendix section, we provide details about estimating joint propensity scores, including data-generating processes and estimation methods. As presented in Figure \ref{fig:P3DAG1}, the directed acyclic graphs for the data-generating process, the red path between $A^s$ and $A^r$ in Figure \ref{C5fig:1b} indicates an association (conditional on $\boldsymbol{x}_s$ and $\boldsymbol{x}_r$) of the treatments between the individuals who are ``interfering with'' each other in the same household. 

With regard to estimation methods for joint propensity, for our pairs case, one straightforward way is to model three probabilities \citep{baker1995marginal}; for example, those for $(A^s, A^r) = (0,1), (1,0)$ and $(1,1)$. Thus, the remaining probability (i.e., $\mathbb{P}(A^s = 0, A^r =0)$) equals to one minus the sum of the probabilities of the three other possibilities.  However, this method would not easily be extended to a larger number of individuals in the same cluster. Therefore, we focus on an alternative general way to model the marginal probabilities and the associated structures. For these structures, the odds ratio, which has some desirable properties and is employed as the correlation coefficient, is studied as a measure of association between pairs of binary variables \cite{lipsitz1991generalized} \cite{yi2005marginal}. To generate the correlated treatments for a household,  the idea is to specify the joint distribution (or joint propensity) of the two binary variables by specifying the marginal distributions of those two binary variables and the odds ratio. Then, building on the joint propensity, we can generate the treatment configuration as $(0,0), (0,1), (1,0)$ or $(1,1)$. Formally, suppose there are $h = 1,...H$ households, and one pair is in each household, that is, the pair $(s,r)^{h}$ belongs to the $h^{th}$ household. Based on Lipsitz et al. (1991)’s\cite{lipsitz1991generalized} parameterization of correlated binary data, letting $\boldsymbol{A}_h = (A^{s}_{h}, A^{r}_{h})^{\top}$ be the treatment vector for the $h^{th}$ household, we define $p_{hs}(\boldsymbol{\alpha}) := \mathbb{P}(A^{s}_{h} = 1 \mid \boldsymbol{x}_{hs}, \boldsymbol{\alpha})$ and $p_{hr}(\boldsymbol{\alpha}) := \mathbb{P}(A^{r}_{h} = 1 \mid \boldsymbol{x}_{hr}, \boldsymbol{\alpha})$ and $A_{hsr} := \mathbb{I}(A^{s}_{h} = 1, A^{r}_{h} = 1) = A^{s}_{h}A^{r}_{h}$, where $\mathbb{I}(x)$ is an indicator function. Also, we define $p_{hsr}:= \mathbb{P}(A_{hsr} = 1) = \mathbb{P}(A^{s}_{h} = 1, A^{r}_{h} = 1) $. Through cross-classifying each individual's treatment at household $h$, we can form a $2 \times 2$ contingency  table (Table \ref{tb:odds}), and the odds ratio for the pair of correlated binary variables $(A^{s}_{h},A^{r}_{h})$ is
\begin{table}
\begin{center}
\begin{tabular}{l|r|r}

  \backslashbox{$A^{s}_{h}$}{$A^{r}_{h}$}& 1 & 0 \\
\hline
1 & $p_{hsr}$ & $p_{hs} - p_{hsr}$ \\

0 &  $p_{hr} - p_{hsr}$ &    $1 - p_{hs} - p_{hr} + p_{hsr}$ \\
\hline
\end{tabular}
\end{center}
\caption{Contingency table for binary treatment variables $(A^{s}_{h},A^{r}_{h})$ in the $h^{th}$ household.}
\label{tb:odds}
\end{table}
\begin{equation} \label{C5odds}
    \tau_{hsr} := \frac{\mathbb{P}\left(A^{s}_{h}=1, A^{r}_{h}=1\right) \mathbb{P}\left(A^{s}_{h}=0, A^{r}_{h}=0\right)}{\mathbb{P}\left(A^{s}_{h}=1, A^{r}_{h}=0\right) \mathbb{P}\left(A^{s}_{h}=0, A^{r}_{h}=1\right)} = \frac{p_{hsr}\left(1-p_{hs}-p_{hr}+p_{hsr}\right)}{\left(p_{hs}-p_{hsr}\right)\left(p_{hr}-p_{hsr}\right)}. 
\end{equation}
If the odds ratio is known, based on equation (\ref{C5odds}), then we can then solve for $p_{hsr}$ in terms of the two marginal probabilities ($p_{hs}$ and $p_{hr}$) and the odds ratio ($\tau_{hsr}$) such that:
\begin{equation}
    p_{hsr}= 
    \begin{dcases}
        \frac{b_{hsr}- \sqrt{b_{hsr}^{2}-4 \tau_{hsr}\left(\tau_{hsr}-1\right) p_{hs} p_{hr}} }{2\left(\tau_{hsr}-1\right)} & \left(\tau_{hsr} \neq 1\right), \\ \tau_{hsr} p_{hs} p_{hr} & \left(\tau_{hsr}=1\right),
    \end{dcases}
\end{equation}
where $b_{hsr}=[1-\left(1-\tau_{hsr}\right)\left(p_{hr}+p_{hs}\right)]$. Note that $p_{hsr}$ is always bounded in $[0, 1]$, so there is only one feasible choice between the two solutions of the quadratic equation leading to (\ref{C5jointpi})\cite{mardia1967some}. Therefore, in the dependent binary treatment-generating process, first, we generate the marginal propensity based on the individual covariates. Second, we generate the odds ratio, which is typically based on a log-linear regression model \citep{lipsitz1991generalized}, such that $log \tau_{hsr}(\boldsymbol{o}) = \boldsymbol{o}^{\top}\boldsymbol{x}_{hsr} $, where $\boldsymbol{x}_{sr}$ suppressing the $h$ are some pair-level covariates that may influence the odds-ratio between $A^{s}_{}$ and $A^{r}_{}$, and $\boldsymbol{o}$ represents the  corresponding coefficients. We can consider these pair-level covariates $\boldsymbol{x}_{sr}$ to be functions of $\boldsymbol{x}_{s}$, $\boldsymbol{x}_{r}$.  Third, based on the marginal propensity and odds ratio, we generate the joint probability (joint propensity) of the binary treatments configuration. Finally, we generate a treatment configuration for each household according to its joint propensity. 

After seeing the data-generating process for the correlated treatments, let us now review some notable methods for estimating joint propensity score. In the estimation process, the first step is to identify marginal models for multivariate binary data; \cite{liang1986longitudinal} proposed a first-order generalized estimating equation method to provide efficient estimates of regression coefficients. Suppose that there are $h = 1,2,.., H$ households and household $h$ contains a treatment record of a pair $\boldsymbol{A}_h = (A^{s}_{h}, A^{r}_{h})^{\top}$, and we are interested in inference about the parameters of the marginal probabilities such that $p_{ht}(\boldsymbol{\alpha})= \mathbb{P} (A^{t}_{h} = 1 \mid \boldsymbol{x}_{ht}, \boldsymbol{\alpha})$ in terms of covariates $\boldsymbol{x}_h$ and marginal parameter $\boldsymbol{\alpha}$ for $t = s, r$. The optimal estimating equation for parameter $\boldsymbol{\alpha}$ is
\begin{equation} \label{marginGEE}
    U (\boldsymbol{\alpha})=\sum_{h=1}^{H} \boldsymbol{D}_{h}^{\top} \boldsymbol{V}_{h}^{-1}\left\{\boldsymbol{A}_{h}-p_{h}(\boldsymbol{\alpha})\right\}=0
\end{equation}
where $\boldsymbol{D}_h = \partial p_{h}(\boldsymbol{\alpha})/ \partial \boldsymbol{\alpha}^{\top}$, and $\boldsymbol{V}_h$ is the working covariance matrix of $\boldsymbol{A}_h$. The working covariance matrix $\boldsymbol{V}_h$ has the form as $\boldsymbol{V}_h = \Lambda_h^{1/2} corr(\boldsymbol{A}_h) \Lambda_h^{1/2}$, where the diagonal matrix $\Lambda_h = diag\{p_{ht} (1-p_{ht})\}$, and correlation matrix $corr(\boldsymbol{A}_h)$ is the working correlation matrix of $\boldsymbol{A}_h$. 
Then the second step is to model the association between pairs of responses. Prentice (1988)\cite{prentice1988correlated} utilized second-order estimating equations, which also offered efficient estimates of association parameters; however, as the cluster size grows, the computation cost becomes huge. Lipsitz et al. (1991)\cite{lipsitz1991generalized} considered the odds ratio to model the association between binary responses, and then modified the estimating equations of Prentice to estimate the pairwise odds ratios. Define $\tau_{sr} := \frac{\mathbb{P}\left(A^{s}=1, A^{r}=1\right) \mathbb{P}\left(A^{s}=0, A^{r}=0\right)}{\mathbb{P}\left(A^{s}=1, A^{r}=0\right) \mathbb{P}\left(A^{s}=0, A^{r}=1\right)}$, as in Lipsitz et al.'s odds ratio model mentioned in the above generating process, i.e., $log \tau_{sr}(\boldsymbol{o}) = \boldsymbol{o}^{\top}\boldsymbol{x}_{sr}$. The pairwise odds ratios are assumed to be non-negative and are modeled through a generalized linear model with the log link and parameters $\boldsymbol{o}$.
In addition, Carey et al. (1993)\cite{carey1993modelling} proposed alternating logistic regressions (ALR), simultaneously regressing the response on covariates and modeling the association among responses in terms of odds ratios. Two logistic regressions are iterated: one to estimate regression coefficients using Liang and Zeger (1986)'s\cite{liang1986longitudinal} first-order generalized estimating equations, the other to update the odds ratio parameters $\boldsymbol{o}$ using an offset. The R software package \texttt{mets} \citep{mets} provides functions that output results from ALR estimation. 

\section*{Appendix C: Dynamic Weighted Proportional Odds Model (dWPOM)} \label{dWPOM}
In this section, we present our Dynamic Weighted Proportional Odds Model (dWPOM) for multiple-stage decisions with household ordinal utilities. 

For the multi-stage decision analysis, our  dWPOM procedure could be implemented by the following steps at each stage of the analysis, starting from the last stage $K$ and working backwards towards the first stage (subscript $j$ indicates the number of stages):

 \textbf{Step 1:} Construct the stage $j$ ordinal \textit{pseudo-utility}: set $\widetilde{\mathcal{U}_j} =u_K$, where $u_K$ is the observed value of $U_K$, if $j = K$. Otherwise, use prior estimates $\boldsymbol{\hat{\beta}}_{K}$, $\boldsymbol{\hat{\underline{\xi}}}_{j +1} = (\boldsymbol{\hat{\xi}}_{j+1},..., \boldsymbol{\hat{\xi}}_{K})$, $\boldsymbol{\hat{\underline{\psi}}}_{j +1} = (\boldsymbol{\hat{\psi}}_{j+1},..., \boldsymbol{\hat{\psi}}_{K})$, and $\boldsymbol{\hat{\underline{\phi}}}_{j +1} = (\boldsymbol{\hat{\phi}}_{j+1},..., \boldsymbol{\hat{\phi}}_{K})$ to randomly generate $\widetilde{\mathcal{U}_j}$, which takes the ordinal value $c$ with the ordinal probability $\mathbb{P}(\widetilde{\mathcal{U}_j}=c)$, $\mathcal{R}$ times, to yield $\widetilde{\mathcal{U}_j^{1}}, \widetilde{\mathcal{U}_j^{2}}, ..., \widetilde{\mathcal{U}_j^{\mathcal{R}}}$.

\textbf{Step 2:} Implement the Brant-Wald test for these pseudo-outcomes $\widetilde{\mathcal{U}_j^{1}}, \widetilde{\mathcal{U}_j^{2}}, ..., \widetilde{\mathcal{U}_j^{\mathcal{R}}}$. A warning message will be provided if the pseudo-outcomes fail the test.
 
\textbf{Step 3:} Estimate the stage $j$ joint propensity model $\pi^{a^sa^r}(\boldsymbol{h}_{js}, \boldsymbol{h}_{jr})$ (e.g., via alternating logistic regression), then compute the corresponding the association-concerned interference balancing weights, such as $$  w(a_j^s, a_j^r) =  \frac{\pi^{0 0}\pi^{1 0}\pi^{0 1}\pi^{1 1}}{\pi^{a_j^sa_j^r}}\ \ \text{for}\ a_j^s = 0, 1; a_j^r = 0, 1.$$
    
\textbf{Step 4:} Specify the stage $j$ treatment-free and blip models, and perform a weighted cumulative link mixed model of $\widetilde{\mathcal{U}_j^{\mathfrak{r}}}$ on the terms in the treatment-free and blip models, using weights from Step 2 to get estimates $\hat{\zeta^*_{1j}}^{\mathfrak{r}}$, $\hat{\zeta^*_{2j}}^{\mathfrak{r}}$, $\boldsymbol{\hat{\beta^*}}_{j}^{\mathfrak{r}}$, $\boldsymbol{\hat{\xi^*}}_{j}^{\mathfrak{r}}$, $\boldsymbol{\hat{\psi^*}}_{j}^{\mathfrak{r}}$, and $\boldsymbol{\hat{\phi^*}}_{j}^{\mathfrak{r}}$ for $\mathfrak{r} = 1,..., \mathcal{R}$; for example, for each $\mathfrak{r} = 1,..., \mathcal{R}$, use the POM, for $c = 1,2$, \begin{align*} \label{dynamicPOM}
    logit[\mathbb{P}(\widetilde{\mathcal{U}_j^\mathfrak{r}}  \leq c \mid a_j^{s} ,a_j^{r}, \boldsymbol{h}_j; \boldsymbol{\xi}_j, \boldsymbol{\psi}_j, \boldsymbol{\phi}_j)] = \zeta_{cj} - \boldsymbol{\beta}_j^{\top}\boldsymbol{h}_j^{\beta} -a_j^{s}  \boldsymbol{\xi}_j^{\top}\boldsymbol{h}_j^{\xi} - a_j^{r}\boldsymbol{\psi}_j^{\top}\boldsymbol{h}_j^{\psi}- a_j^{s} a_j^{r}\boldsymbol{\phi}_j^{\top}\boldsymbol{h}_j^{\phi}.
    \end{align*}
    
\textbf{Step 5:} Use estimates from Step 3 (i.e., $\hat{\zeta^*_{1j}}^{\mathfrak{r}}$, $\hat{\zeta^*_{2j}}^{\mathfrak{r}}$, $\boldsymbol{\hat{\beta^*}}_{j}^{\mathfrak{r}}$, $\boldsymbol{\hat{\xi^*}}_{j}^{\mathfrak{r}}$, $\boldsymbol{\hat{\psi^*}}_{j}^{\mathfrak{r}}$, and $\boldsymbol{\hat{\phi^*}}_{j}^{\mathfrak{r}}$) to compute $$\kappa^{\mathfrak{r}}(a_j^s,a_j^r, \boldsymbol{h}_{j}) = \text{expit}(\hat{\eta}^{}_{2} )\left[1 - \text{expit}(\hat{\eta}^{}_{1})\right]\left[1 - \text{expit}(\hat{\eta}^{}_{2}) + \text{expit}(\hat{\eta}^{}_{1})\right],$$ where $\hat{\eta}^{}_{c} = \hat{\eta}^{}_{c}(a_j^s,a_j^r, \boldsymbol{h}_j) = \hat{\zeta_{c}} - \hat{\boldsymbol{\beta}}^{\top} \boldsymbol{h}_j^{\beta} - \hat{\boldsymbol { \xi }}^{\top} a_j^s\boldsymbol{h}_j^{\xi}- \hat{\boldsymbol { \psi }}^{\top} a_j^r\boldsymbol{h}_j^{\psi} - \hat{\boldsymbol { \phi }}^{\top} a_j^sa_j^r\boldsymbol{h}_j^{\phi}$, with values $$(\hat{\zeta_{c}}, \hat{\boldsymbol{\beta}}, \hat{\boldsymbol{\xi}}, \hat{\boldsymbol{\psi}}, \hat{\boldsymbol{\phi}}) = (\hat{\zeta^*_{cj}}^{\mathfrak{r}}, \boldsymbol{\hat{\beta^*}}_{j}^{\mathfrak{r}} , \boldsymbol{\hat{\xi^*}}_{j}^{\mathfrak{r}}, \boldsymbol{\hat{\psi^*}}_{j}^{\mathfrak{r}}, \boldsymbol{\hat{\phi^*}}_{j}^{\mathfrak{r}}), \ \text{for}\ \ c=1,2.$$ Then, for $a_j^s,\ a_j^r = 0, 1,$ construct the new weights 
    \begin{equation}\label{C5neww}
        w^{new, \mathfrak{r}}(a_j^s, a_j^r) =  \frac{\pi^{0 0}\pi^{1 0}\pi^{0 1}\pi^{1 1}}{\pi^{a_j^sa_j^r}} \times \frac{\kappa^{\mathfrak{r}}(0,0, \boldsymbol{h}_j)\kappa^{\mathfrak{r}}(1,0, \boldsymbol{h}_j)\kappa^{\mathfrak{r}}(0,1, \boldsymbol{h}_j)\kappa^{\mathfrak{r}}(1,1, \boldsymbol{h}_j)}{\kappa^{\mathfrak{r}}(a_j^s, a_j^r, \boldsymbol{h}_j)}.
    \end{equation}
    
\textbf{Step 6:} Perform a weighted POM with the new weights (i.e., $w^{new, \mathfrak{r}}(a^s_j, a^r_j)$) to get revised estimates $\boldsymbol{\hat{\xi}}_{j}^{\mathfrak{r}}$, $\boldsymbol{\hat{\psi}}_{j}^{\mathfrak{r}}$, and $\boldsymbol{\hat{\phi}}_{j}^{\mathfrak{r}}$ for each $\mathfrak{r}$, and estimate $\boldsymbol{\xi}_j$, $\boldsymbol{\psi}_j$, and $\boldsymbol{\phi}_j$ by $\boldsymbol{\hat{\xi}}_{j} = \mathcal{R}^{-1}\sum_{r} \boldsymbol{\hat{\xi}}_{j}^{\mathfrak{r}}$, $\boldsymbol{\hat{\psi}}_{j} = \mathcal{R}^{-1}\sum_{r} \boldsymbol{\hat{\psi}}_{j}^{\mathfrak{r}}$, and $\boldsymbol{\hat{\phi}}_{j} = \mathcal{R}^{-1}\sum_{r} \boldsymbol{\hat{\phi}}_{j}^{\mathfrak{r}}$, respectively, then use parameter estimators  $\boldsymbol{\hat{\xi}}_{j}$,  $\boldsymbol{\hat{\psi}}_{j}$, and $\boldsymbol{\hat{\phi}}_{j}$ to construct the $j^{th}$ stage optimal treatment rule, which is based on Decision \ref{ruleHH}. 

\textbf{Step 7:} Return to Step 1 and analyze stage $j - 1$ if there are more stages to analyze.

\section*{Appendix D: Performance Matrix and Simulation Study 1 data generation process and some results} \label{C5SimStudy1}

In this section, we outline the performance matrix concerning the definitions of the Optimal Treatment Rate and Mean Regret Value (refer to subsection \ref{Performance Matrix}). Additionally, we delve into the data generation process for Study 1 (refer to subsection \ref{dgpS1}).
\subsection{Optimal treatment rate and mean regret value}\label{Performance Matrix}
To evaluate the performance of the methods that correspond to different balancing weights, we construct two main measures, the optimal treatment rate (OTR) and value functions for ordinal outcomes.

First, we consider the optimal treatment rate. The estimated recommended treatment configuration for a pair could be that both, one or neither of the treatments are the same as the true optimal treatments; therefore, there are two different quantities: one is the optimal treatment rate for the household, and the other is the optimal treatment rate for the individual. The household OTR represents the average decision accuracy for the pair in the same household. Formally, if the true treatment decision for $h^{th}$ household is $(a_h^{s*}, a_h^{r*})$, the optimal treatment rate for the household can be expressed as ${H}^{-1} \sum_{h=1}^{H} \mathbb{I}\left(a_h^{s*} = \hat{a}^s_{h}, a^{r*}_h = \hat{a}^r_{h}\right)$. In this case, decision accuracy requires estimated pairs' treatment configurations to be consistent with the true optimal treatment configuration. Alternatively, the individual OTR describes the average decision accuracy for individuals, and can be denoted as ${(2H)}^{-1} \sum_{h=1}^{H} \left[\mathbb{I}\left(a_h^{s*} = \hat{a}^s_{h}\right) + \mathbb{I}\left( a^{r*}_h = \hat{a}^r_{h}\right) \right]$. In both cases, the higher OTR value indicates greater decision accuracy and thus the superiority of the corresponding method. 

Second, in a single-stage setting, to compare the effects of the optimal treatment configuration with those of the estimated one, we consider the value of the regret function when the estimated treatment configuration is implemented.  We first define the regret function, which is the expected loss in outcome if we prescribe a non-optimal treatment $(A^{s} ,A^{r})$ compared to if we prescribe an optimal one ($(A^{s*} ,A^{r*})$), as 
\begin{align*}
    \mu[d^*, (A^{s} ,A^{r})] &=  \gamma^{*}[d^*(\boldsymbol{x}^{\xi}, \boldsymbol{x}^{\psi}, \boldsymbol{x}^{\phi}) ; \boldsymbol{\xi}, \boldsymbol{\psi}, \boldsymbol{\phi}] - \gamma[(A^{s} ,A^{r}),\boldsymbol{x}; \boldsymbol{\xi}, \boldsymbol{\psi}, \boldsymbol{\phi}] \\ &= (A^{s*} - A^{s} ) \boldsymbol{\xi}^{\top}\boldsymbol{x}^{\xi} + (A^{r*} - A^{r}) \boldsymbol{\psi}^{\top}\boldsymbol{x}^{\psi}+ (A^{s*}A^{r*} - A^{s} A^{r})\boldsymbol{\phi}^{\top}\boldsymbol{x}^{\phi}.
\end{align*}
Then, according to the definition of the regret function, with the estimated treatment configuration input, the mean regret value is $\mu[d^*, \hat{d}^*],$ where are $\hat{d}^* = \hat{d}^*(\boldsymbol{x}^{\xi}, \boldsymbol{x}^{\psi}, \boldsymbol{x}^{\phi})$ can be computed based on blip estimates and rules in Decision \ref{ruleHH}. The mean regret value measures the difference between the value under the optimal regime, i.e., $d^*(\boldsymbol{x}^{\xi}, \boldsymbol{x}^{\psi}, \boldsymbol{x}^{\phi}) = (A^{s*}, A^{r*})$ and that under the estimated optimal regime, i.e., $\hat{d}^*(\boldsymbol{x}^{\xi}, \boldsymbol{x}^{\psi}, \boldsymbol{x}^{\phi}) = (\hat{A}^{s*} ,\hat{A}^{r*})$. A smaller mean regret value indicates that the estimated regime is closer to the optimal regime; therefore, the smaller the mean regret value corresponds to the better method.

\subsection{Study 1 data generation process} \label{dgpS1}
In this subsection, we present the detailed generation process of covariates and functions, and the simulation (Study 1) figures, which depict the distribution of the blip parameter estimates in Scenarios 1, 2, and 4.

First, covariates and functions are generated as follows. The \textbf{individual-level and household-level covariates} are generated as follows. For each individual, covariates are generated as $x_1 \sim U[0, 1]$ that is uniformly distributed on $[0, 1]$, $x_2 \sim N(0,1)$ that is normally distributed with mean $0$ and variance $1$, and $x_3 \sim Ber(0.5)$, $x_4 \sim Ber(0.75)$ that are Bernoulli distributed with success probability $0.5$ and $0.75$, respectively.  Supposing each household contains two individuals denoted as $(s, r)$, we let $\boldsymbol{x}^{\beta}$ and $\boldsymbol{x}^{\phi}$ include both individuals' covariates, but $\boldsymbol{x}^{\xi}$ and $\boldsymbol{x}^{\psi}$ each include only the single individual's covariates. For example, in this simulation, denoting $\boldsymbol{x}^s$ and $\boldsymbol{x}^r$ as the covariates of $(s,r)$, respectively, we let $\boldsymbol{x}^{\beta} = (1, x_1^s, x_2^s, x_3^s, x_4^s, 1, x_1^r, x_2^r, x_3^r, x_4^r)$, $\boldsymbol{x}^{\xi} = (1, x_1^s)$, $\boldsymbol{x}^{\psi} =  (1, x_1^r)$, and $\boldsymbol{x}^{\phi} = (1, x_3^s + x_3^r)$, where $x^s + x^r$ is the household-level value which is the sum of two individuals' information. 

The \textbf{true treatment-free function} is set as $f(\boldsymbol{x}^{\beta}) =  \mathrm{cos}(x_1^s +  x_1^r) - (x_1^s +  x_1^r)^3 - \mathrm{log}|1/x^s_1| -2*(x^r_1)^2 + (x_3^s + x_3^r)^3$. It is important to note that this true treatment-free setting relies on various non-linear functions, including both even and odd functions. It also consists of both household-level information, such as $x_1^s +  x_1^r$ and $x_3^s +  x_3^r$, and individual-level information, including $x^s_1$ and $x^r_1$.  In addition, the \textbf{true blip function} is $\gamma[(A^{s} ,A^{r}),\boldsymbol{x}; \boldsymbol{\xi}, \boldsymbol{\psi}, \boldsymbol{\phi}] = A^{s}*(-0.5 + x_1^s) + A^{r}*(-0.5 + x_1^r) + A^{s}*A^{r}*[-1 + 0.5*(x_3^s + x_3^r)]$, where the tailoring variables are set as $\boldsymbol{x}^{\xi} = (1, x^s_1)^{\top}$, $\boldsymbol{x}^{\psi} = (1, x^r_1)^{\top}$, and $\boldsymbol{x}^{\phi} = (1, x^s_3 + x^r_3 )^{\top}$. The corresponding parameters of interest, that is, the true tailoring parameters, are set as follows: $\boldsymbol{\xi} = (-0.5, 1)^{\top}$, $\boldsymbol{\psi} = (-0.5, 1)^{\top}$, and $\boldsymbol{\phi} = (-1, 0.5)^{\top}$. Finally, the \textbf{true mean of the latent household outcome} is thus $
    \mu = f(\boldsymbol{x}^{\beta}) + A^{s}*(-0.5 + x_1^s) + A^{r}*(-0.5 + x_1^r) + A^{s}*A^{r}*[-1 + 0.5*(x_3^s + x_3^r)].$

Our present case needs two \textbf{thresholds} ($\zeta_1$ and $\zeta_2$) to classify the latent continuous outcomes into three categories.  The following is the process to \textbf{generate the three categorical outcomes} in our study. In Step 1, we set reference probabilities for the three-category outcome $\boldsymbol{p} = (p_1, p_2, p_3) = (0.65, 0.25, 0.1)$, and compute the cumulative reference probabilities. In Step 2, with the logistic distribution, we map cumulative reference probabilities to thresholds, which are $\zeta_1 = 0.619, \zeta_2 = 2.197$. In Step 3, based on individual and household covariates and treatments, we calculate household-level thresholds such that $(\zeta_1 - \mu, \zeta_2 - \mu)$. In Step 4, for each household, based on household-level thresholds, we can compute cumulative probabilities and thus acquire the category probabilities. Finally, in Step 5, we generate household-level category outcomes based on the category probabilities in Step 4. 

The \textbf{treatments generation process} follows the methods that are introduced in Section \ref{C5sec5.2}. The marginal propensity model for each individual is set as $\mathbb{P}(A = 1 \mid \boldsymbol{x}) = \mathrm{expit}(-1.15 + 0.5*\mathrm{exp}(x_1) - 0.25*x_2^2 + 0.25*x_3 + 0.6*x_4)$, and the odds ratio model is set as $\tau = \mathrm{exp}[-0.25 + 0.25*(x_{1}^s + x_1^r) + 0.5*(x_3^s +x_3^r)]$. Based on the marginal propensity and odds ratio models, joint probabilities can be generated. Therefore, the treatment configuration $(A^s, A^r)$ can be generated by the corresponding joint propensities.

\subsection{Additional Study 1 results} \label{AddRes}
This subsection presents additional results from Study 1b.
First, for the second part of study 1b, centered on Scenario 3, the parameters remain consistent with those outlined earlier, except for the variation in the number of households, denoted as $H$, which is set to 500, 1000, 3000, 5000, and 10000. Table \ref{C5tb:Study1} showcases the three performance metrics for all methods: (1) household OTR, (2) individual OTR, and (3) the mean regret value.

In the smaller household sample cases, $H < 3000$, M0 still provides the worst OTRs and MRV. M4 does not always provide the highest OTRs and the lowest MRV; however, in these cases, either M2 or M3, which are treatment-association aware WPOM, offer the best OTR or MRV, compared with M1 which assumes independent treatment. Note that M2, M3, and M4 all belong to treatment-association aware WPOM; therefore, these results indicate that treatment-association aware WPOM performs better than no treatment-association WPOM if an association exists between treatments.
\begin{table}[!]\centering
\setlength{\tabcolsep}{0.9pt}
\caption{Methods' performance measure estimates and their standard errors  (in parenthesis) from Methods 0, 1, 2,  3, and 4, when the treatment model is correctly
specified but the treatment-free model is misspecified (Scenario 3) in Study 1b. $H$ denotes the number of households. OTR-H: Household optimal treatment rate; OTR-I: Individual optimal treatment rate; MRV: Mean regret value.}\label{C5tb:Study1} 
\setlength{\tabcolsep}{2pt} 
\renewcommand{\arraystretch}{1.2} 
\begin{tabular}{cl|l|llll}
\hline
\multirow{2}{*}{$H$} & \multirow{2}{*}{Performance} & \multicolumn{5}{c}{Method}                               \\ 
                  &                   & \multicolumn{1}{c}{M0}  & \multicolumn{1}{c}{M1}  & \multicolumn{1}{c}{M2}  &\multicolumn{1}{c}{M3} &\multicolumn{1}{c}{M4} \\ \hline 
\multirow{3}{*}{$500$} 
                  & OTR-H               & \multicolumn{1}{c}{0.38 (0.18)} & \multicolumn{1}{c}{0.44 (0.18)} & \multicolumn{1}{c}{0.45 (0.18)} & \multicolumn{1}{c}{0.45 (0.18)} & \multicolumn{1}{c}{0.42 (0.17)}\\ 
                  & OTR-I             & \multicolumn{1}{c}{0.58 (0.12)} & \multicolumn{1}{c}{0.63 (0.11)} &\multicolumn{1}{c}{0.64 (0.11)} & \multicolumn{1}{c}{0.63 (0.11)} & \multicolumn{1}{c}{0.63 (0.10)} \\ 
                  &  MRV            & \multicolumn{1}{c}{0.19 (0.09)} & \multicolumn{1}{c}{0.16 (0.09)} & \multicolumn{1}{c}{0.17 (0.10)} & \multicolumn{1}{c}{0.17 (0.10)}& \multicolumn{1}{c}{0.17 (0.14)} \\ \hline 
\multirow{3}{*}{$1000$} 
                  & OTR-H               & \multicolumn{1}{c}{0.43 (0.17)} & \multicolumn{1}{c}{0.52 (0.17)} & \multicolumn{1}{c}{0.52 (0.17)} & \multicolumn{1}{c}{0.52 (0.17)} & \multicolumn{1}{c}{0.51 (0.18)}\\ 
                  & OTR-I             & \multicolumn{1}{c}{0.62 (0.11)} & \multicolumn{1}{c}{0.68 (0.10)} &\multicolumn{1}{c}{0.69 (0.10)} & \multicolumn{1}{c}{0.69 (0.10)} & \multicolumn{1}{c}{0.68 (0.10)} \\ 
                  & MRV            & \multicolumn{1}{c}{0.16 (0.10)} & \multicolumn{1}{c}{0.13 (0.10)} & \multicolumn{1}{c}{0.13 (0.10)} & \multicolumn{1}{c}{0.13 (0.10)}& \multicolumn{1}{c}{0.14 (0.12)} \\ \hline
\multirow{3}{*}{$3000$} 
                  & OTR-H               & \multicolumn{1}{c}{0.48 (0.14)} & \multicolumn{1}{c}{0.59 (0.13)} & \multicolumn{1}{c}{0.61 (0.14)} & \multicolumn{1}{c}{0.61 (0.14)} & \multicolumn{1}{c}{0.61 (0.16)}\\ 
                  & OTR-I             & \multicolumn{1}{c}{0.65 (0.08)} & \multicolumn{1}{c}{0.74 (0.07)} &\multicolumn{1}{c}{0.75 (0.07)} & \multicolumn{1}{c}{0.75 (0.07)} & \multicolumn{1}{c}{0.75 (0.07)} \\ 
                  & MRV            & \multicolumn{1}{c}{0.14 (0.06)} & \multicolumn{1}{c}{0.09 (0.05)} & \multicolumn{1}{c}{0.09 (0.05)} & \multicolumn{1}{c}{0.09 (0.05)}& \multicolumn{1}{c}{0.10 (0.07)} \\ \hline
\multirow{3}{*}{$5000$} 
                  & OTR-H               & \multicolumn{1}{c}{0.52 (0.13)} & \multicolumn{1}{c}{0.63 (0.11)} & \multicolumn{1}{c}{0.65 (0.11)} & \multicolumn{1}{c}{0.65 (0.10)} & \multicolumn{1}{c}{0.68 (0.12)}\\ 
                  & OTR-I             & \multicolumn{1}{c}{0.68 (0.07)} & \multicolumn{1}{c}{0.76 (0.06)} &\multicolumn{1}{c}{0.77 (0.06)} & \multicolumn{1}{c}{0.77 (0.05)} & \multicolumn{1}{c}{0.80 (0.06)} \\ 
                  & MRV            & \multicolumn{1}{c}{0.13 (0.05)} & \multicolumn{1}{c}{0.08 (0.03)} & \multicolumn{1}{c}{0.07 (0.03)} & \multicolumn{1}{c}{0.07 (0.03)}& \multicolumn{1}{c}{0.06 (0.04)} \\ \hline
\multirow{3}{*}{$10000$} 
                  & OTR-H               & \multicolumn{1}{c}{0.56 (0.13)} & \multicolumn{1}{c}{0.67 (0.09)} & \multicolumn{1}{c}{0.69 (0.09)} & \multicolumn{1}{c}{0.69 (0.09)} & \multicolumn{1}{c}{0.75 (0.11)}\\ 
                  & OTR-I             & \multicolumn{1}{c}{0.71 (0.07)} & \multicolumn{1}{c}{0.78 (0.05)} &\multicolumn{1}{c}{0.80 (0.05)} & \multicolumn{1}{c}{0.80 (0.05)} & \multicolumn{1}{c}{0.84 (0.06)} \\ 
                  & MRV            & \multicolumn{1}{c}{0.11 (0.04)} & \multicolumn{1}{c}{0.07 (0.03)} & \multicolumn{1}{c}{0.06 (0.03)} & \multicolumn{1}{c}{0.06 (0.03)}& \multicolumn{1}{c}{0.04 (0.03)} \\ \hline
\end{tabular}
\end{table}

\begin{figure}[!]\centering
    \includegraphics[scale=0.55]{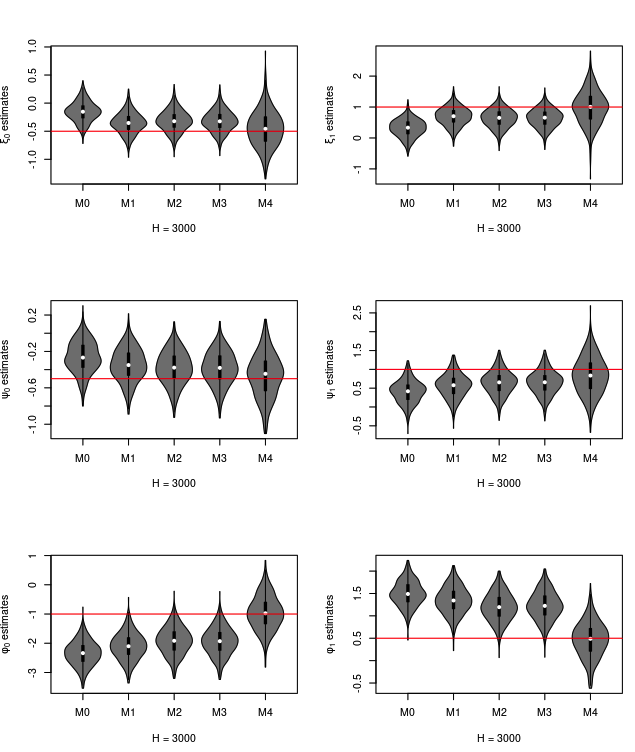} 
	\caption{Blip function parameter estimates, $\hat{\boldsymbol{\xi}}$ (top row), $\hat{\boldsymbol{\psi}}$ (middle row), and $\hat{\boldsymbol{\phi}}$ (bottom row) via Method 0 (M0, $Q$-learning), Method 1 (M1, no treatment-association dWPOM), Method 2 (M2, treatment-association aware dWPOM with IPW-type weights), Method 3 (M3, treatment-association aware dWPOM with overlap-type weights) and Method 4 (M4, treatment-association aware dWPOM with adjusted overlap-type weights), when both treatment model and treatment-free model are misspecified (Scenario 1).}\label{P3study1aS1}
\end{figure}

\begin{figure}[!]\centering
    \includegraphics[scale=0.55]{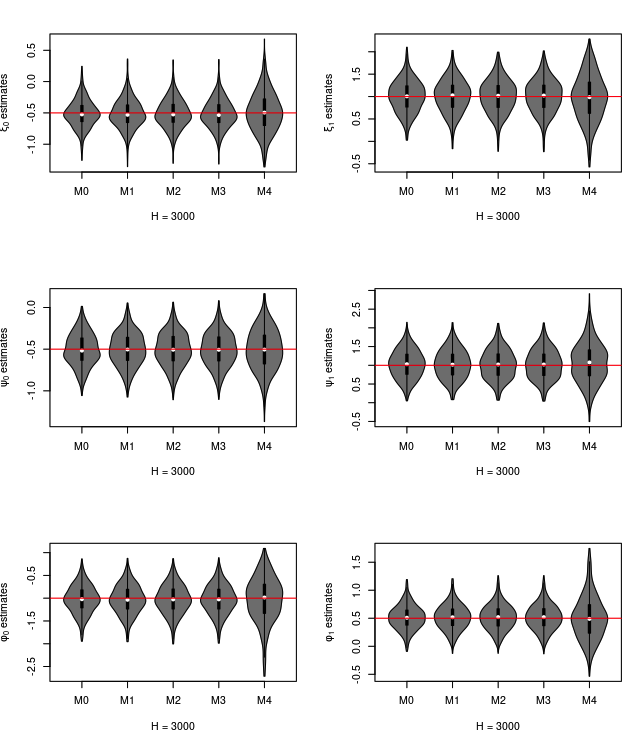} 
	\caption{Blip function parameter estimates, $\hat{\boldsymbol{\xi}}$ (top row), $\hat{\boldsymbol{\psi}}$ (middle row), and $\hat{\boldsymbol{\phi}}$ (bottom row) via Method 0 (M0, $Q$-learning), Method 1 (M1, no treatment-association dWPOM), Method 2 (M2, treatment-association aware dWPOM with IPW-type weights), Method 3 (M3, treatment-association aware dWPOM with overlap-type weights) and Method 4 (M4, treatment-association aware dWPOM with adjusted overlap-type weights), when the treatment-free model is correctly specified but the treatment model is misspecified (Scenario 2).}\label{P3study1aS2}
\end{figure}

Second, regarding the simulation figures in Study 1, Figure \ref{P3study1aS1} corresponds to Scenario 1 where both the treatment model and treatment-free model are misspecified; Figure \ref{P3study1aS2} correspond to Scenario 2 where the treatment-free model is correctly specified but the treatment model is misspecified, and Figure \ref{P3study1aS2} present the result from Scenario 4 in which both treatment model and treatment-free model are correctly specified.

\begin{figure}[!]\centering
    \includegraphics[scale=0.55]{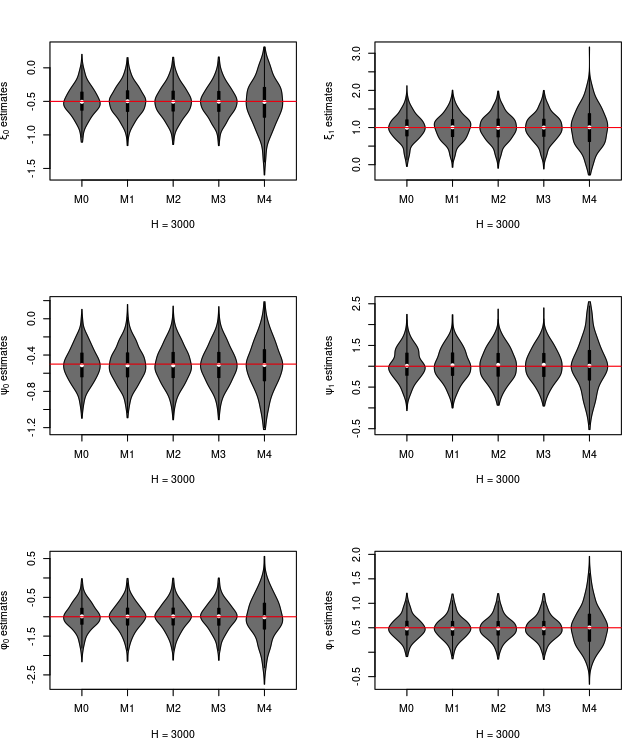} 
	\caption{Blip function parameter estimates, $\hat{\boldsymbol{\xi}}$ (top row), $\hat{\boldsymbol{\psi}}$ (middle row), and $\hat{\boldsymbol{\phi}}$ (bottom row) via Method 0 (M0, $Q$-learning), Method 1 (M1, no treatment-association dWPOM), Method 2 (M2, treatment-association aware dWPOM with IPW-type weights), Method 3 (M3, treatment-association aware dWPOM with overlap-type weights) and Method 4 (M4, treatment-association aware dWPOM with adjusted overlap-type weights), when both treatment model and treatment-free model are correctly specified (Scenario 4).}\label{P3study1aS4}
\end{figure}

From Figure \ref{P3study1aS1}, even though both the treatment model and treatment-free model are incorrectly specified, M4 which utilizes adjusted overlap-type weights can still provide less biased blip parameters' estimators. Except for estimator $\hat{\boldsymbol{\psi}}$, which displays little bias, estimators $\hat{\boldsymbol{\xi}}$ and $\hat{\boldsymbol{\phi}}$ appear unbiased. From Figures \ref{P3study1aS2} and \ref{P3study1aS4}, because of the correct identification of the treatment-free model, all the methods provide consistent estimators of the blip parameters. 
\section*{Appendix E: Multiple-stage Treatment Decision for a Couples Case} \label{C5SimRes}
\subsubsection{Approximate robust estimation of dWPOM}
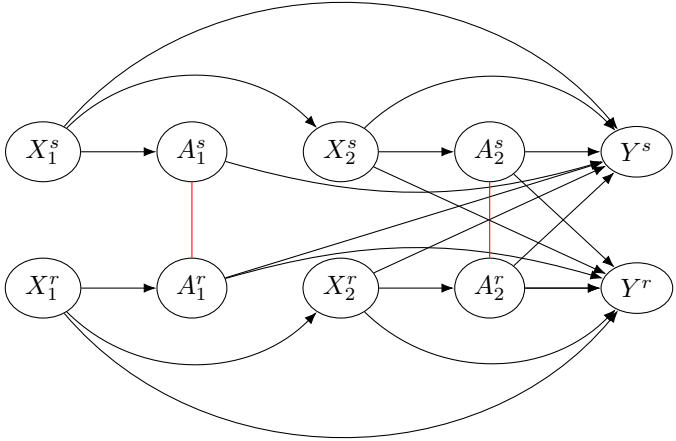
\begin{figure}\centering
	\begin{tikzpicture}
    \node[state] (a_1) at (0,0) {$A^{s}_{1}$};
    \node[state] (o_2) [right =of a_1] {$X^{s}_{2}$};
    \node[state] (a_2) [right =of o_2] {$A^{s}_{2}$};
    \node[state] (o_1) [left =of a_1] {$X^{s}_{1}$};
        \node[state] (xal1) [below = of o_1] {$X^{r}_{1}$};
    \node[state] (aal1) [below = of a_1] {$A^{r}_{1}$};
            \node[state] (xal2) [below = of o_2] {$X^{r}_{2}$};
    \node[state] (aal2) [below = of a_2] {$A^{r}_{2}$};
    \node[state] (y) [right =of a_2] {$Y^{s}$};
    \node[state] (yr) [right =of aal2] {$Y^{r}$};
    \path (o_1) edge (a_1);
    \path (o_1) edge[bend left=45] node[below, el]{} (o_2);
    \path (o_2) edge (a_2);
    \path (a_2) edge (y);
    \path (o_1) edge[bend left=50] node[below, el]{} (y);
    \path (o_2) edge[bend left=45] node[below, el]{} (y);
    \path (a_1) edge[bend right=16] node[below, el]{} (y);
    \path (xal1) edge (aal1);
    \path (xal2) edge (aal2);
    \path (aal2) edge (y);
    \path (aal1) edge (y);
    \path (xal1) edge[bend right=45] node[below, el]{} (xal2);
    \path (xal1) edge[bend right=50] node[below, el]{} (yr);
        \path (xal2) edge[bend right=45] node[below, el]{} (yr);
                \path (aal1) edge[bend left=16] node[below, el]{} (yr);
    \path (xal2) edge (y);
      \draw[red] (aal1) -- (a_1) -- cycle;
        \draw[red] (aal2) -- (a_2) -- cycle;
    \path (a_2) edge (yr);
        \path (o_2) edge (yr);
            \path (aal2) edge (yr);
    \path (aal2) edge (yr);

    \end{tikzpicture}
\caption{The DAG of interference analysis: dWPOM for household ordinal outcomes two-stage decision problems.  }\label{C5fig:multistage_h}
\end{figure}

For multi-stage settings with ordinal outcomes, the technique of dWPOM is explained in Section \ref{C5sec5.4}. We now conduct a simulation study (Study 2) to illustrate our estimation of a two-stage treatment decision problem with ordinal outcomes under household interference. The causal diagram for two-stage treatment decisions with household ordinal outcomes in the presence of interference is presented in Figure \ref{C5fig:multistage_h}, and the simulation’s data-generating process is according to this causal diagram. Again, the red path between $A^s$ and $A^r$ in DAG \ref{C5fig:multistage_h} indicates that the individuals who are interfering with each other in the same household receive correlated treatments. 
 
To generate data, taking $H = 1000$ households and $B = 500$ replicates, we denote covariates as $x_{jpt}$, where $j = 1,2$ indicates the $j^{th}$ stage, $p =1,2$ represents the dimension of covariates, and $t = s, r$ corresponds to individual $s$ or $r$ in the same household. In Stage 1, the covariates of $s$ and $r$ are $x_{11s} \sim N(0,1)$, $x_{11r} \sim N(0,1)$, $x_{12s} \sim Ber(0.5)$, $x_{12r} \sim Ber(0.5)$, that is, $x_{11}$ of $s$ and $r$ are normally distributed with mean 0 and variance 1, and $x_{12}$ of $s$ and $r$ are Bernoulli distributed with success probability 0.5 in Stage 1. Similarly, in Stage 2,  the covariates of $s$ and $r$ are $x_{21s} \sim 0.5* N(x_{11s},1)$, $x_{21r} \sim 0.5*N(x_{11r},1)$, $x_{22s} \sim Ber(0.1 + 0.5*x_{12s})$, $x_{22r} \sim Ber(0.1 + 0.5*x_{12r})$. To generate correlated treatments of a pair, for stage $j=1,2$, the marginal propensity model for each individual is set as $\mathbb{P}(A_j = 1 \mid \boldsymbol{x}_j) = \mathrm{expit}(-1 + 1.15*\mathrm{exp}(x_{j1}) - 0.5*x_{j2})$, and the odds ratio model is set as $\tau = \mathrm{exp}[-0.15 + 0.25*(x_{j2s} + x_{j2r})]$. The true treatment-free function is set as $f(\boldsymbol{x}^{\beta}) = \sum_{j=1}^2f_j(\boldsymbol{x}_j^{\beta}) =  \mathrm{cos}[\pi*(x_{11s} +  x_{11r})] +0.5*\mathrm{exp}(x_{21s} +  x_{21r}) +0.2*(x_{12s} + x_{12r})^3$, and for $j = 1,2$, the true blip functions ($\gamma_j[(A_j^{s} ,A_j^{r}),\boldsymbol{x}_j; \boldsymbol{\xi}, \boldsymbol{\psi}, \boldsymbol{\phi}]$) are set as: $A_j^{s}*(-0.25 + 0.5*x_{j1s}) + A_j^{r}*(-0.25 + 0.5*x_{j1r}) + A_j^{s}*A_j^{r}*[-0.5 + 0.25*(x_{j2s} + x_{j2r})],$ where the true blip parameters are $\boldsymbol{\xi} = (-0.25, 0.5)^{\top}$, $\boldsymbol{\psi} = (-0.25, 0.5)^{\top}$, and $\boldsymbol{\phi} = (-0.5, 0.25)^{\top}$. To generate household-level ordinal outcomes, following the same outcome-generating procedure in Study 1, we set $\zeta_1 = 0.405, \zeta_2 = 1.735$. Building on individual and household covariates and treatments, we calculate household-level thresholds $(\zeta_1 - \mu, \zeta_2 - \mu)$, where $\mu = f(\boldsymbol{x}^{\beta}) + \sum_{j=1}^2\gamma_j[(A_j^{s} ,A_j^{r}),\boldsymbol{x}_j; \boldsymbol{\xi}, \boldsymbol{\psi}, \boldsymbol{\phi}]$. Then, from these we can compute cumulative probabilities and thus acquire the category probabilities.
\begin{figure}[!]\centering
    \includegraphics[scale=0.35]{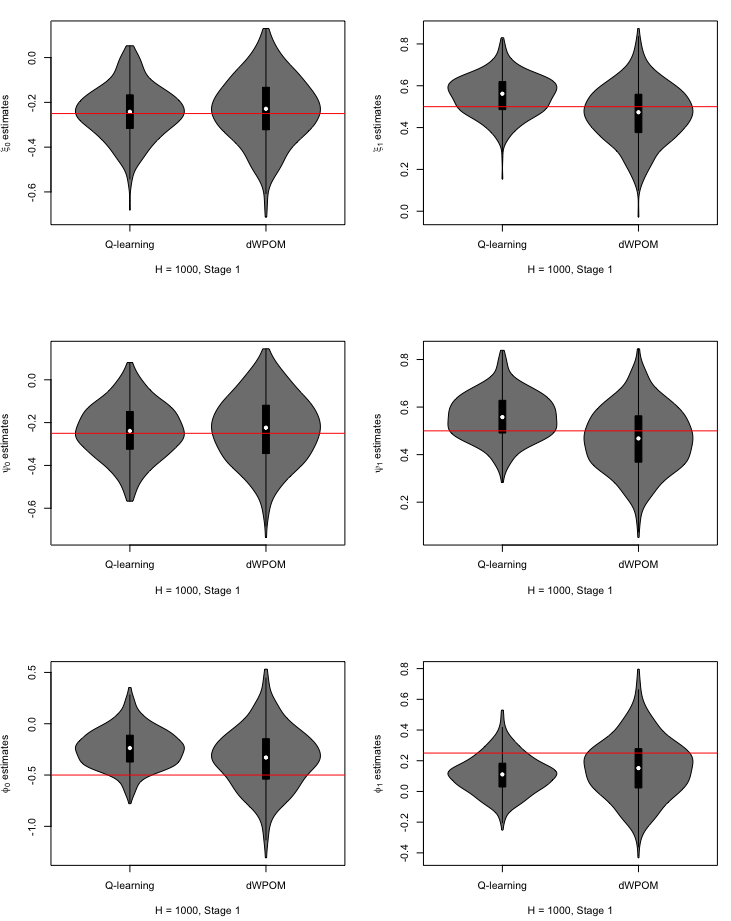}
	\caption{Blip function parameter estimates in Stage 1 of Study 2, $\hat{\boldsymbol{\xi}}$ (top row), $\hat{\boldsymbol{\psi}}$ (middle row), and $\hat{\boldsymbol{\phi}}$ (bottom row) via $Q$-learning and treatment-association aware dWPOM with adjusted overlap-type weights in Case (1), where the treatment-free models are misspecified, but the treatment models are correctly specified in both Stages 2 and 1.}\label{P3study2S1}
\end{figure}

\begin{figure}[!]\centering
    \includegraphics[scale=0.35]{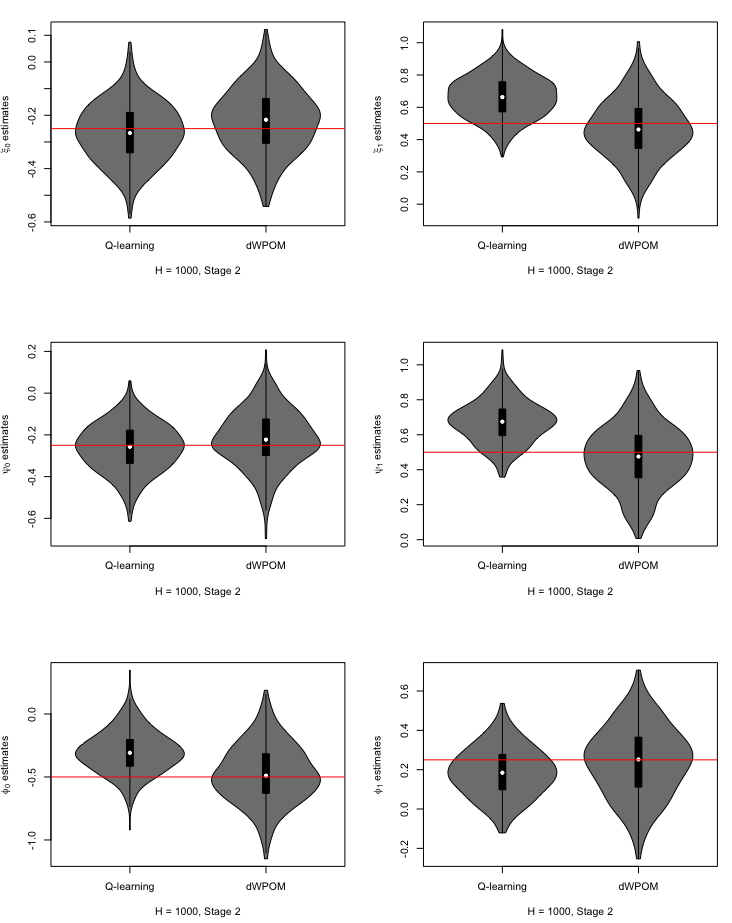} 
	\caption{Blip function parameter estimates in Stage 2 of Study 2, $\hat{\boldsymbol{\xi}}$ (top row), $\hat{\boldsymbol{\psi}}$ (middle row), and $\hat{\boldsymbol{\phi}}$ (bottom row) via $Q$-learning and treatment-association aware dWPOM with adjusted overlap-type weights in Case (1), where the treatment-free models are misspecified, but the treatment models are correctly specified in both Stages 2 and 1.}\label{P3study2S2}
\end{figure}
In this simulation study, examining two-person household interference, we focus on interference-aware $Q$-learning and our proposed dWPOM with adjusted weights. We highlight the approximate consistent estimation of the proposed dWPOM method if at least either the treatment-free or treatment model is correct, by considering various types of model misspecification at the two stages. The treatment-free model is harder to correctly specify; thus, in this study, we specifically consider two cases: (1) in both Stages 2 and 1, the treatment-free models are misspecified, but the treatment models are correctly specified; (2) in Stage 2, the treatment-free model is misspecified,  but the treatment model is correctly specified, while in Stage 1, alternatively the treatment model is misspecified, but the treatment-free model is correctly specified. Note that the true treatment-free models and treatment models contain non-linear covariate terms, and if we consider only the linear covariate terms in a model, we will mis-specify that model. 

For Case (1), the distributions of the blip estimates (i.e., $\hat{\boldsymbol{\xi}}, \hat{\boldsymbol{\psi}}, \hat{\boldsymbol{\phi}}$) are presented in Figures \ref{P3study2S1} and \ref{P3study2S2}, which correspond to Stage 1 and Stage 2, respectively. The distributions of the blip estimates from Case (2) are presented in Appendix \ref{C5SimRes} Figures \ref{P3study2S1C2} and \ref{P3study2S2C2}.  In Figure \ref{P3study2S2}, which depicts blip estimates from Case (1) in Stage 2, all the blip estimates from our dWPOM appear to be symmetrically distributed and centred at the true blip parameter values, but $Q$-learning provides biased estimators. From Figure \ref{P3study2S1}, which corresponds to Case (1) in Stage 1, for our dWPOM, blip estimates $ \hat{\boldsymbol{\xi}}, \hat{\boldsymbol{\phi}}$ are also symmetrically distributed and centred at the true parameters’ values, but the blip estimates $\hat{\phi}_0, \hat{\phi}_1$ appear to be slightly off the true values. We suspect that this misalignment results from the fact that the estimation is only approximately consistent, because of the omission of remainder terms in the Taylor expansion (see proof of Theorem \ref{thmC5ordout} in Appendix \ref{C5PfThm}). 
 
\subsubsection{Value function and  performance of identified DTR}
 As previously stated, our value functions for ordinal outcomes are evaluated by computing the odds ratio. We now define value functions for ordinal outcomes by first introducing the odds of a particular outcome. 
From the proposed POM, we can predict ordinal outcomes when the estimated optimal treatments have been implemented. Building on these predicted ordinal outcomes, we can compute the odds:  the probability that the preferred outcome will occur is divided by the probability that the preferred outcome will not occur. That is, we calculate the ratio of the number of the preferred outcome to the number of the outcome that is not preferred. For instance, in our three ordinal outcome case ($U = 1, 2, 3$), the preferred outcome is $3$, and the odds of $U = 3$ being among the predicted ordinal outcomes is the ratio of the number of $U=3$ to the number of $U=1$ or $U=2$, i.e., $Odd_{pred} = \sum^{H}_h\mathbb{I}(\hat{U}_h = 3)/ (H - \sum^{H}_h\mathbb{I}(\hat{U}_h = 3))$. Moreover, from among the observed ordinal outcomes, we can compute the other odds of $U = 3$, which correspond to implementing the observed DTRs, i.e., $Odd_{obs} = \sum^{H}_h\mathbb{I}(U_h = 3)/ (H - \sum^{H}_h\mathbb{I}(U_h = 3))$. Finally, we can define the value functions for ordinal outcomes: 
the ratio of the odds of $U = 3$ among the predicted ordinal outcomes to the odds of $U = 3$ among the observed ordinal outcomes (i.e., $OR(U = 3) = Odd_{pred}/Odd_{obs}$). Similarly, if our preferred outcome is ``$U=2$ or $U=3$'', meaning that at least one individual quits smoking, we can also compute the corresponding value functions. In the simulation study, we estimate the mean value function as the average of $B=500$ replicates ($B^{-1}\sum^{B}_{b}OR_{b}(U=3)$). 
 
The average value function for ordinal outcomes, which is the odds ratio of outcome $U=3$ from interference-aware $Q$-learning, in Case (1), is 1.25, but that from interference-aware dWPOM is 1.36, which is greater than $Q$-learning’s. For Case (2), the estimated value function for ordinal outcomes from $Q$-learning is 1.16, which is smaller than those (1.28) from dWPOM. These results indicate that both methods provide better treatment regimes than the observed treatment regime, but the treatment regime from the proposed interference-aware dWPOM performs better than interference-aware $Q$-learning.  In both cases, the true models are the same, so that the optimal treatment regime would be the same,  and the data are the same. Interference-aware dWPOM is estimating the model parameters correctly, and thus estimating the optimal treatment regime, while because of its biases $Q$-learning is effectively estimating a sub-optimal treatment regime.

In this section, we present the additional simulation results from simulation Study 2, a two-stage DTR estimation with ordinal outcomes under interference. Consistent with the same data-generating process in Case (1) Study 2,  Case (2), in Stage 2, misspecifies the treatment-free model but correctly specifies the treatment model. Stage 1, Case (2) misspecifies the treatment model, but correctly specifies the treatment-free model.

\begin{figure}[!]\centering
    \includegraphics[scale=0.34]{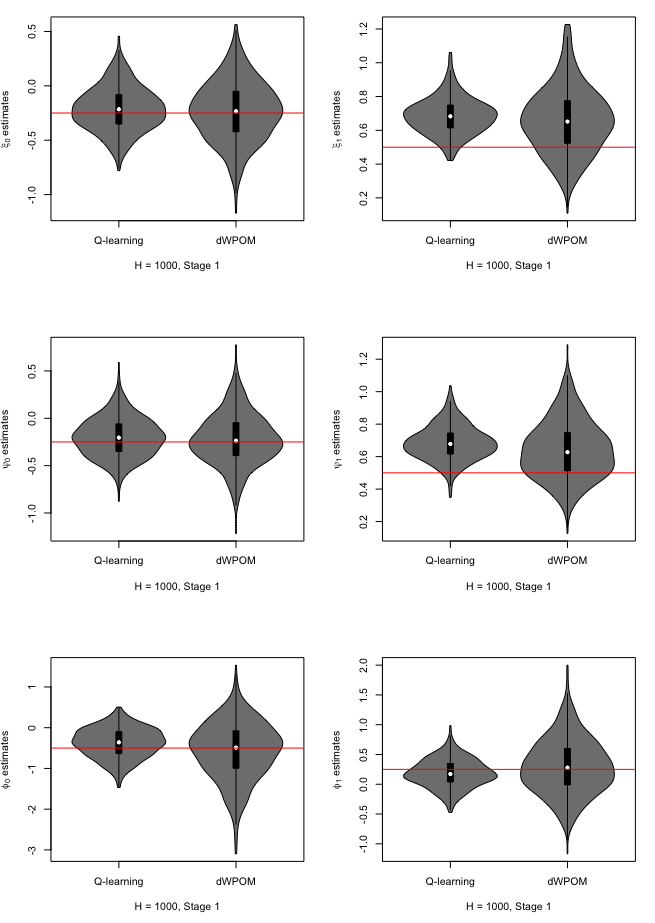} 
	\caption{Blip function parameter estimates in Stage 1 of Study 2, $\hat{\boldsymbol{\xi}}$ (top row), $\hat{\boldsymbol{\psi}}$ (middle row), and $\hat{\boldsymbol{\phi}}$ (bottom row) via $Q$-learning and treatment-association aware dWPOM with adjusted overlap-type weights in Case (2), where the treatment-free model is misspecified,  but the treatment model is correctly specified in Stage 2, and the treatment model is misspecified, but the treatment-free model is correctly specified in Stage 1.}\label{P3study2S1C2}
\end{figure}

\begin{figure}[!]\centering
    \includegraphics[scale=0.35]{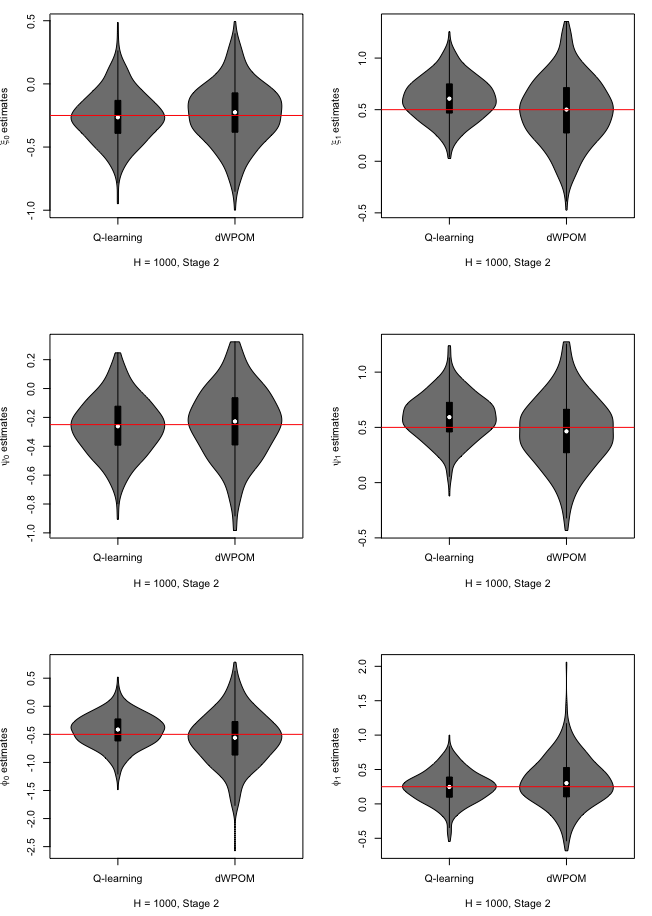} 
	\caption{Blip function parameter estimates in Stage 2 of Study 2, $\hat{\boldsymbol{\xi}}$ (top row), $\hat{\boldsymbol{\psi}}$ (middle row), and $\hat{\boldsymbol{\phi}}$ (bottom row) via $Q$-learning and treatment-association aware dWPOM with adjusted overlap-type weights in Case (2), where the treatment-free model is misspecified,  but the treatment model is correctly specified in Stage 2, and the treatment model is misspecified, but the treatment-free model is correctly specified in Stage 1.}\label{P3study2S2C2}
\end{figure}

The distributions of the blip estimates (i.e., $\hat{\boldsymbol{\xi}}, \hat{\boldsymbol{\psi}}, \hat{\boldsymbol{\phi}}$)  from Case (2) are presented in Figures \ref{P3study2S1C2} and \ref{P3study2S2C2}, which correspond to Stage 1 and Stage 2, respectively. Similar to the results for Case (1), in Figure \ref{P3study2S2C2}, which depicts blip estimates from Cases (2) in Stage 2, all the blip estimates from our dWPOM appear to be normally distributed and centred by the true blip parameters’ values, but $Q$-learning provides biased estimators. From Figure \ref{P3study2S1C2}, which corresponds to Case (1) in Stage 1, for our dWPOM, blip estimates $\hat{\xi}_0, \hat{\psi}_0,  \hat{\boldsymbol{\phi}}$ are also normally distributed and centred by the true parameters’ values, but the blip estimates $\hat{\xi}_1, \hat{\psi}_1$ appear to be slightly off the true values. Again, we suspect that this misalignment results from the approximately consistent estimation, essentially caused by the omission of remainder terms in the Taylor expansion. 

In our two-stage simulations generating data from the proportional odds model, the Brant-Wald test does support the use of the proportional odds assumption at the earlier stage. Around $3.11\%$ of the 7500 simulation replications indicate a test failure.
 